\documentclass[twocolumn]{aastex63}
\usepackage{lineno}

\newcommand{\gaia}{\it Gaia}
\newcommand{\kepler}{\it Kepler}
\newcommand{\ktwo}{\it K2}
\newcommand{\tess}{\it TESS}
\newcommand{\Teff}{$T_{\text{eff}}$}

\newcommand{\clusterone}{NGC 2547}
\newcommand{\clustertwo}{Pleiades}
\newcommand{\clusterthree}{M50}
\newcommand{\clusterfour}{NGC 2516}
\newcommand{\clusterfive}{M37}
\newcommand{\clustersix}{Praesepe}
\newcommand{\clusterseven}{NGC 6811}

\newcommand{\editfinal}[1]{#1}

\received{}
\revised{}
\accepted{}
\submitjournal{-}

\shorttitle{Rotation of Open Clusters Stars with Gaia}
\shortauthors{Godoy-Rivera et al.}

\begin{document}

\title{Stellar Rotation in the Gaia Era: Revised Open Clusters Sequences}
\correspondingauthor{Diego Godoy-Rivera}
\email{godoyrivera.1@osu.edu}

\author[0000-0003-4556-1277]{Diego Godoy-Rivera}
\affiliation{Department of Astronomy, The Ohio State University, 140 West 18th Avenue, Columbus, OH 43210, USA}

\author[0000-0002-7549-7766]{Marc H. Pinsonneault}
\affiliation{Department of Astronomy, The Ohio State University, 140 West 18th Avenue, Columbus, OH 43210, USA}

\author[0000-0001-6381-515X]{Luisa M. Rebull}
\affiliation{Infrared Science Archive (IRSA), IPAC, 1200 E. California Blvd., California Institute of Technology, Pasadena, CA 91125, USA}

\begin{abstract}
The period versus mass diagrams (i.e., rotational sequences) of open clusters provide crucial constraints for angular momentum evolution studies. However, their memberships are often heavily contaminated by field stars, which could potentially bias the interpretations. In this paper, we use data from {\gaia} DR2 to re-assess the memberships of seven open clusters with \editfinal{ground- and space-based} rotational data, and present an updated view of stellar rotation as a function of mass and age. We use the {\gaia} astrometry to identify the cluster members in phase-space, and the photometry to derive revised ages and place the stars on a consistent mass scale. Applying our membership analysis to the rotational sequences reveals that: 1) the contamination in clusters observed from the ground can reach up to $\sim 35\%$; 2) the overall fraction of rotational outliers decreases substantially when the field contaminants are removed, but some outliers persist; 3) there is a sharp upper edge in the rotation periods at young ages; 4) \editfinal{at young ages,} stars in the $1.0\textendash0.6 M_{\odot}$ range inhabit a global maximum of rotation periods, potentially providing an optimal window for habitable planets. Additionally, we see clear evidence for a strongly mass-dependent spin-down process. In the regime where rapid rotators are leaving the saturated domain, the rotational distributions broaden (in contradiction with popular models), which we interpret as evidence that the torque must be lower for rapid rotators than for intermediate ones. The cleaned rotational sequences from ground-based observations can be as constraining as those obtained from space.
\end{abstract}

\section{Introduction}
\label{sec:intro}

Together with mass and metallicity, rotation is one of the defining properties of stars. Stellar rotation is strongly mass-dependent \citep{kraft67b,garcia14}, and  while stars with radiative envelopes ($M\gtrsim 1.4 M_{\odot}$) live their lives as rapid rotators, stars with convective envelopes spin down as they age \citep{skumanich72}. The overall framework of the rotational evolution of low mass stars is that they are born with a range of rotation rates ($\sim 1\textendash10$ days; \citealt{herbst02}) and star-disk coupling timescales (from $< 1$ and up to $\sim 10$ Myr; \citealt{williams11}), and once they arrive onto the zero-age main-sequence (ZAMS), they experience a decrease in their rotational velocities due to angular momentum losses caused by magnetized winds \citep{parker58,schatzman62,weber67,kawaler88,pinsonneault89}.

For lower main-sequence stars, the observable stellar properties do not appreciably change with time, which makes the determination of ages of individual stars challenging \citep{soderblom10}. In this regard, the empirical spin-down of low mass stars offers a valuable age diagnostic, and this idea is known as gyrochronology \citep{barnes03b,barnes07,mamajek08}. 

If properly validated, gyrochronology can provide valuable age information for Galactic and stellar population studies \citep{mcquillan13,mcquillan14,davenport17,davenport18,vansaders19}, as well as for the characterization of exoplanet hosts (e.g., \citealt{gallet19,gallet20,zhou20,carmichael21,david21}). Moreover, although stellar evolution theory is successful in predicting numerous properties of stars, this is not the case for the evolution of angular momentum and its related physical processes (e.g., transport in stellar interiors, birth conditions, magnetic braking). Consequently, stellar models that incorporate it need to be calibrated with empirical observations (e.g., \citealt{gossage21}).

All of the above add to a compelling need for gyrochronology to be examined and tested.  For old ages ($\gtrsim$ few Gyr), these efforts have used field stars \citep{irwin11,angus15,newton16a,vansaders16,lorenzooliveira19} and wide separation binaries \citep{chaname12,janes17,godoyrivera18,otani21}. For young ages ($\lesssim 1$ Gyr), the calibrations have been based on the empirical rotational sequences (i.e., period versus mass or temperature diagrams) of star-forming regions, associations, and open clusters \citep{denissenkov10,gallet13,gallet15,angus19,defreitas21}.

With the advent of space-based missions, the field of stellar rotation is entering a new era. The unparalleled astrometry from the {\gaia} mission \citep{gaia18a} has already allowed new gyrochronology inspections to be carried out (e.g., \citealt{curtis19b,angus20}). In addition to this, results obtained from the observations by the {\kepler} and {\ktwo} missions \citep{borucki10,howell14} have showed striking trends at various ages. These include strongly mass-dependent rotation rates in populations younger than 10 Myr \citep{somers17b,rebull18,rebull20}, unusual spin-down behavior in $\sim 1\textendash 3$ Gyr-old clusters (\citealt{curtis19a,curtis20,gruner20}; see also \citealt{agueros18}), and anomalously rapid rotation in $\gtrsim 5$ Gyr-old field stars \citep{vansaders16,hall21}. 

The strengths of space data, namely unparalleled time coverage and photometric precision, have indeed provided exquisite data in selected regimes. However, for many interesting clusters, a combination of crowded fields, faint sources, and long rotation periods, make obtaining data from space surveys impractical. By contrast, ground-based surveys can measure precise and accurate rotation periods for even faint sources, because star spot modulation induces relatively large photometric signals. The native seeing even from average sites is also far better than the 4{\arcsec} pixels of {\kepler} and {\ktwo}, or the 21{\arcsec} pixels of the Transiting Exoplanet Survey Satellite ({\tess}; \citealt{ricker15}). As a result, ground-based surveys are still fundamental for studies of stellar rotation.

In this context, although star clusters have played a crucial role in our understanding of angular momentum evolution, a thorough decontamination of their rotational sequences is currently lacking. For the systems that have been observed from the ground, the field contamination in their sequences is expected to go from $\sim 20\%$ (e.g., \citealt{hartman09a}) and reach up to $\sim 40\textendash 60\%$ (e.g., \citealt{irwin07b,irwin08a,irwin09}). \editfinal{For the systems observed from space, a lower contamination rate may be expected due to more careful target selections, but given their importance as rotational benchmarks, detailed astrometric analysis are certainly needed.} \editfinal{Furthermore, it has been common in the literature to assume that rotational outliers of cluster sequences correspond to field contaminants. We are now in a position to test this assumption.} 

\editfinal{Ultimately, the unknown extent to which contamination is a problem, and what effect this could have on the patterns of the rotational sequences, make the revision of their memberships an imperative task.} This is something the unprecedented space-based data from {\gaia} is particularly well-suited for. With this, the goal of this paper is to use the high-precision {\gaia} astrometry to clean the rotational sequences of a sample of open clusters \editfinal{(observed from the ground and space)}, with the prospects of providing revised gyrochronology calibrators, and an updated view of stellar rotation as a function of mass and age. 

This paper is structured as follows. In \S\ref{sec:sample_selection_and_period_data} we present the sample of star clusters with rotation period measurements we study. \editfinal{In \S\ref{sec:method} we present our method for identifying likely cluster members using the {\gaia} astrometry. In \S\ref{sec:properties_clusters_and_stellar} we revise the properties of the clusters we study, and calculate stellar masses and temperatures for their members. In \S\ref{sec:results}} we show the effects that the membership analysis has on the rotational sequences. In \S\ref{sec:discussion} we discuss our findings and present an updated view of stellar rotation. We conclude in \S\ref{sec:conclusions}.
\section{Sample Selection and Rotation Period Data} 
\label{sec:sample_selection_and_period_data}

We compile a list of open clusters that satisfy four conditions: 1) they hold high scientific interest for constraining the processes that govern stellar rotation; 2) they have rotation period information, derived uniformly within each cluster, for $\sim$ hundreds of candidate members; 3) they have reliable {\gaia} data that allows us to accurately model them astrometrically; 4) they potentially have enough non member contamination that a careful membership analysis to clean their rotational sequences would have significant impact. The rotation periods we use come from both ground- and space-based optical photometric monitoring. \editfinal{Due to the target selections of the original rotation period references,} the membership contamination is expected to be higher for the clusters observed from the ground than for those observed from space \editfinal{(see the clusters' descriptions below)}. 

We limit our selection to open clusters but do not include star-forming regions and associations, which require a more complex treatment of membership (e.g., \citealt{gagne18,galli18,kuhn19}). We also exclude systems that are either too close or too distant for a reliable membership analysis, or that have poor {\gaia} data, or systems with smaller or heterogeneous rotation samples. \editfinal{We discuss the systems with available rotational data sets that we did not analyze in Appendix \ref{appendix:app_data_period_expansions}.}

Table \ref{tab:list_of_clusters} lists the seven open clusters we study\footnote{Alternative names for some of our systems are: Pleiades$=$M45$=$Melotte 22; M50$=$NGC 2323; M37$=$NGC 2099; Praesepe$=$M44$=$Melotte 88$=$NGC 2632; NGC 6811$=$Melotte 222. All throughout this paper we refer to them with the names listed in Table \ref{tab:list_of_clusters}.}, together with the number of stars in each of them with period measurements. Table \ref{tab:list_of_clusters} also reports their literature ages \editfinal{as compiled by} \citet{gallet15} \editfinal{(see their Table 1 for the specific references)}, but we highlight that for our analysis we perform an independent derivation of revised cluster properties using {\gaia} data and recent spectroscopic results (see \S\ref{sec:properties_clusters_and_stellar}). \editfinal{We now describe the seven clusters in our sample.}

\begin{table}
\scriptsize
\caption{List of the seven open clusters we study. \label{tab:list_of_clusters}}
\begin{tabular}{lcc}
\hline
\hline
Name &  Literature Age ({\it pre} {\gaia}) & N$_{\star \text{, Period}}$ \\
- & [Myr] & - \\
\hline
{\clusterone}    & 35  & 176\\
{\clustertwo}    & 125 & 759\\
{\clusterthree}  & 130 & 812\\
{\clusterfour}   & 150 & 362\\
{\clusterfive}    & 550 &  367\\
{\clustersix}     & 580 & 809\\
{\clusterseven} & 1000 & 235\\
\hline
\\
\end{tabular}
\tablecomments{The clusters are sorted by increasing age. The quoted ages come from \editfinal{the compilation by} \citet{gallet15} and are the \editfinal{representative} {\it pre} {\gaia} values. The period measurements are taken from the references cited throughout \S\ref{sec:sample_selection_and_period_data}.}
\end{table}
\subsection{\editfinal{Three Monitor Project Clusters}}
\label{subsec:data_period_monitor}

The Monitor Project\footnote{\url{https://www.ast.cam.ac.uk/ioa/research/monitor/}} performed an optical photometric survey of several young open clusters using 2- and 4-m class telescopes with wide field cameras \citep{irwin07b}. The derived lightcurves allowed for searches of transiting planets \citep{miller08} and eclipsing binaries \citep{irwin07c}, as well as measurements of rotation periods. 

Of the clusters observed by the Monitor Project, we include {\clusterone}, {\clusterthree}, and {\clusterfour} in our sample. All three of these clusters offer interesting constraints on stellar rotation evolution. With an age of $\sim$ 35 Myr, {\clusterone} is a well-populated system where solar analogs have just arrived on the main sequence, and data in this age range (younger than the {\clustertwo} and older than star-forming regions) is a sensitive test of theory. With ages of $\sim$ 130 and 150 Myr, {\clusterthree} and {\clusterfour} offer natural comparison points for the benchmark {\clustertwo} cluster. Additionally, the contamination rates quoted for these clusters are in the $\sim$ 40\textendash60\% range \citep{irwin07b,irwin08a,irwin09}, which hints that a revision of their rotational sequences could be significant.

{\clusterone}, {\clusterthree}, and {\clusterfour} were studied in a similar fashion by \citet{irwin08a}, \citet{irwin09}, and \citet{irwin07b}, respectively. These authors surveyed regions out to $\sim$ 50{\arcmin}, $\sim$ 30{\arcmin}, and $\sim$ 60{\arcmin} from the clusters' center, identified candidate members on the basis of $V$ versus $V-I$ color-magnitude diagram (CMD) selections, and reported rotation periods for 176, 812, and 362 stars, respectively.
\subsection{\editfinal{One MMT Cluster}}
\label{subsec:data_period_hartman}

{\clusterfive} (age of $\sim$ 550 Myr) is one of the very few open clusters with a large rotational sample that are older than 500 Myr. This, in addition to its age being close to the classical $\sim$ 580 Myr age for {\clustersix}, and a quoted $\sim$ 20\%  contamination rate \citep{hartman09a}, make it an interesting system to include in our sample.

\citet{hartman08a} observed the {\clusterfive} cluster with the 6.5-m MMT telescope in order to constrain its parameters, study variable stars \citep{hartman08b}, measure rotation periods \citep{hartman09a}, and study the occurrence rate of transiting planets \citep{hartman09b}. Given the exquisite optical observations, thorough lightcurve analysis, and the aforementioned expected contamination rate, we include {\clusterfive} in our sample.

\citet{hartman09a} surveyed regions out to $\sim$ 20{\arcmin} from the cluster center, identified candidate members on the basis of two CMD selections ($r$ versus $g-r$ and $g-i$), and reported rotation periods for 575 stars. Of these, however, only 367 are considered by \citet{hartman09a} to be the {\it clean} periodic sample, where the different algorithms for determining periods showed good agreement with each other, and their results did not differ by more than 10\%. We take this {\it clean} sample as our nominal {\clusterfive} data set.
\subsection{\editfinal{Three {\kepler} Clusters}}
\label{subsec:data_period_keplerk2}

The {\kepler} spacecraft in both its original 4-year mission and subsequent {\ktwo} mission, observed several open clusters of a range of ages. These observations produced lightcurves with exquisite photometric precision, and have been used to construct rotation period catalogs for a number of clusters. 

Of the clusters observed by {\kepler} and {\ktwo}, we include the {\clustertwo}, {\clustersix}, and {\clusterseven} in our sample. Given their proximity, the {\clustertwo} (distance of $\approx$ 136 pc; age of $\sim$ 125 Myr) and {\clustersix} (distance of $\approx$ 186 pc; age of $\sim$ 580 Myr) clusters have been used for rotation studies for several decades (e.g., \citealt{anderson66,kraft67,dickens68,stauffer84,stauffer87,queloz98,terndrup99,terndrup00,scholz07,delorme11,agueros11,douglas14,rebull16a,rebull16b,rebull17,stauffer16}), and the periods derived from the {\ktwo} observations comprise their state-of-the-art rotational samples. While the number of {\clusterseven} stars with {\kepler} periods is lower compared to the aforementioned clusters, its old age ($\sim$ \editfinal{1 Gyr}) makes it a remarkably interesting system for stellar rotation studies (e.g., \citealt{meibom11a,curtis19a,rodriguez20,velloso20}).

\begin{itemize}

\item {\clustertwo}: \citet{rebull16a} performed a comprehensive lightcurve analysis of the stars observed by {\ktwo} in regions out to $\sim$ 350{\arcmin} from the cluster center. \citet{rebull16b} expanded on this by meticulously classifying and identifying the different structures present in the periodograms and lightcurves, and \citet{stauffer16} then used this to study the Pleiades in the context of angular momentum evolution. The sample of candidate members used by these studies comprises 759 stars with measured periods that were classified as {\it Best+OK} members by \citet{rebull16a} on the basis of proper motions and position in the CMD selections. We take this sample as our nominal {\clustertwo} data set. We note that although the {\ktwo} observations missed part of the northern region of the cluster, this is unlikely to bias the period distribution \citep{rebull16a}.

\item {\clustersix}: a similar study to that of the {\clustertwo} was carried out for {\clustersix} by \citet{rebull17}. In this case, the candidate members extend out to $\sim$ 400{\arcmin} from the cluster center, and they were selected on the basis of proper motions and CMD position. The final \citet{rebull17} sample comprises 809 candidate members with measured periods.

\item {\clusterseven}: \citet{meibom11a} studied candidates that were previously vetted using radial velocity (RV) data in a region out to 30{\arcmin} from the cluster center, and reported periods for 71 stars. \citet{curtis19a} surveyed a region of 60{\arcmin} radius, selected candidate members on the basis of {\gaia} CMD position and astrometry, and reported periods for 171 stars. We complement these with the \citet{santos19} and \citet{santos21} {\kepler} field catalogs, which reported periods for 194 {\clusterseven} candidate members. To maximize the number of stars with measured periods, we combine all of these catalogs and end up with 235 stars after accounting for repetitions. For stars in common among the references, we prioritize the \citet{santos19} and \citet{santos21} periods first, \citet{curtis19a} second, and \citet{meibom11a} third (although a comparison of these values showed excellent agreement). While the periods of our joint {\clusterseven} catalog are derived from slightly different techniques, the studies carried out by these four references are all based on the same {\kepler} data.
\end{itemize}
\section{\editfinal{Membership Method}}
\label{sec:method}

In this section, we describe the method we use to analyze the clusters in our sample. In \S\ref{subsec:method_gaia}, \editfinal{taking {\clusterone} as a working example}, we use the {\gaia} astrometry to fit the cluster and the field in phase-space, and calculate membership probabilities. Since the periodic samples are deep and include faint stars with large formal astrometric errors, we classify stars into four different groups: highly likely cluster members, highly likely non members, an intermediate category of possible members (typically faint objects with large astrometric uncertainties), and a category of stars without enough information to be classified. We are interested in both single and binary stars, and while we attempt to avoid biasing our results against binaries, there are {\gaia} selection effects related to the excess astrometric noise from them \editfinal{(see \S\ref{sec:discussion})}. We do not apply a Hertzsprung–Russell (HR) diagram selection, but our final samples are strikingly clear in that plane. In \S\ref{subsec:method_crossmatch} we crossmatch the \editfinal{{\clusterone}} {\gaia} data with its periodic sample. \editfinal{In \S\ref{subsec:method_astrometric_classification}} we show the results of applying our \editfinal{membership} method to all our clusters.
\subsection{{\gaia} Data, Cluster Fitting, and Membership Probabilities}
\label{subsec:method_gaia}

In \S\ref{subsubsec:method_gaia_data} we describe the {\gaia} data, in \S\ref{subsubsec:method_gaia_fitting} we use them to characterize the cluster and the field in parallax and proper motion space by fitting a model that allows for intrinsic dispersions, and in \S\ref{subsubsec:method_gaia_membership} we classify the stars in different categories and calculate memberships probabilities, identifying highly likely cluster members.
\subsubsection{{\gaia} DR2 Data}
\label{subsubsec:method_gaia_data}

The astrometric data used in this work are the data release 2 (DR2) of the {\gaia} mission\footnote{We note that, although nearing the completion of this work the early data release 3 (EDR3) from {\gaia} became available \citep{gaia21a}, we do not anticipate that using these newer data would substantially change the results here presented. This arises from the fact we are studying predominantly nearby clusters, where the DR2 parallax and proper motion errors are already small. Additionally, EDR3 does not contain new RV data with respect to DR2.} \citep{gaia18a}. {\gaia} DR2 provides positions, proper motions, and parallaxes, as well as photometry in the $G$, $G_{\text{BP}}$, and $G_{\text{RP}}$ bands for $\sim$ 80\% of its $\sim$ 1.7 billion stars. Although {\gaia} also provides RV measurements for some of these stars, this subset only corresponds to $\lesssim 0.5\%$ of the {\gaia} sample, and we therefore do not use this parameter in our membership study.

We download the {\gaia} DR2 data for all the sources contained within the region where the {\clusterone} rotation period catalog reports candidate cluster members (out to $\sim$ 50{\arcmin} from the cluster center, see \S\ref{subsec:data_period_monitor}). 

Regarding the {\gaia} DR2 parallaxes, as reported by \citet{lindegren18} and confirmed by several other works \citep{chan20,riess18,schonrich19,stassun18,zinn19}, there is a zero-point offset that needs to be considered. For the remainder of this paper, we adopt the global mean value of 29 $\mu$as for all the clusters (in the sense that the {\gaia} parallaxes are too small) reported by \citet{lindegren18}, with the exception of {\clusterseven}, for which we adopt the {\kepler} field 53 $\mu$as value from \citet{zinn19}. Spatial variations in this zero-point offset are real but modest for the systems that we are interested in. Furthermore, since we are interested in studying stars at the same true distance, the exact zero-point value does not strongly impact our membership results (although it could impact our cluster ages, see \S\ref{sec:properties_clusters_and_stellar}).
\subsubsection{Cluster Fitting}
\label{subsubsec:method_gaia_fitting}

Our method is similar to membership probability studies found in the literature, and its goal is to separate the cluster population from the field population in phase-space. For instance, \citet{vasilevskis58} and \citet{sanders71} used a 2D-version of this method using proper motions to compute memberships for a number of clusters. Other studies that have used similar versions of this method are \citet{francic89}, \citet{jones88}, \citet{jones91}, and \citet{jones96}. A 3D-version including parallaxes as well as individual star measurement errors and correlations, in addition to allowing for intrinsic dispersions, has been recently used by \citet{franciosini18}, \citet{roccatagliata18}, and \citet{roccatagliata20}.

In our method, a given star $i$ has a probability of belonging to either population $P$ (field $F$ or cluster $C$), with the total likelihood being:
\begin{equation}
\mathcal{L}_{i}= f_F \mathcal{L}_{F,i} + f_C \mathcal{L}_{C,i} \text{ ,}  
\label{eq:total_likelihood}
\end{equation}
where $f_{F}$ and $f_{C}$ represent the fraction of stars that belong to the field and cluster (normalized such that $f_{C}=1-f_{F}$), and $\mathcal{L}_{F,i}$ and $\mathcal{L}_{C,i}$ are the likelihoods of each population, respectively. We assume that the likelihood of each population $P$ can be described by a 3D multivariate-Gaussian:
\begin{equation}
\mathcal{L}_{P,i}= \frac{1}{(2\pi)^{3/2} |\Sigma_{i}|^{1/2}}\exp{\left[-\frac{1}{2} (\vec{x}_i - \vec{x}_P)^{T} \Sigma_i^{-1} (\vec{x}_i - \vec{x}_P) \right]} \text{ ,}
\end{equation}
where
\begin{equation}
\vec{x}_i-\vec{x}_{P}= \left( \begin{array}{c} \varpi_i - \varpi_{P}\\ \mu_{\alpha_{i}}-\mu_{\alpha_{P}}  \\ \mu_{\delta_{i}}-\mu_{\delta_{P}} \\ \end{array} \right)
\end{equation}
and $\Sigma_i= C_i + \Sigma_P$, where $C_i$ is the individual covariance matrix, and $\Sigma_P$ is the matrix of intrinsic dispersions:
\begin{equation}
C_i =   \left( \begin{array}{ccc} 
\sigma_{\varpi_{i}}^2 & \rho_{\varpi \mu_{\alpha},i} \sigma_{\varpi_{i}} \sigma_{\mu_{\alpha_{i}}} & \rho_{\varpi \mu_{\delta},i} \sigma_{\varpi_{i}} \sigma_{\mu_{\delta_{i}}} \\
 \rho_{\varpi \mu_{\alpha},i} \sigma_{\varpi_{i}} \sigma_{\mu_{\alpha_{i}}} & \sigma_{\mu_{\alpha_{i}}}^2 & \rho_{\mu_{\alpha} \mu_{\delta},i} \sigma_{\mu_{\alpha_{i}}} \sigma_{\mu_{\delta_{i}}}  \\ 
\rho_{\varpi \mu_{\delta},i} \sigma_{\varpi_{i}} \sigma_{\mu_{\delta_{i}}} & \rho_{\mu_{\alpha} \mu_{\delta},i} \sigma_{\mu_{\alpha_{i}}} \sigma_{\mu_{\delta_{i}}} & \sigma_{\mu_{\delta_{i}}}^2 \\ \end{array} \right) \text{ ,}
\end{equation}
\begin{equation}
\Sigma_P= \left( \begin{array}{ccc} \sigma_{\varpi_{P}}^2 & 0 & 0 \\ 0 & \sigma_{\mu_{\alpha_{P}}}^2 & 0  \\ 0 & 0 & \sigma_{\mu_{\delta_{P}}}^2 \\ \end{array} \right).
\end{equation}
By using these equations on the {\gaia} data, we can derive the cluster and field parameters using a maximum likelihood fit.

To accurately derive these parameters, we apply a set of quality and geometric cuts. We highlight that these cuts are applied only when defining the \editfinal{subset of stars} we use to derive the cluster and field astrometric parameters, but in \S\ref{subsubsec:method_gaia_membership} we apply our membership classification to the full {\gaia} sample \editfinal{(i.e., the stars excluded during this exercise are still classified later on)}. We run our maximum likelihood calculation using the subset of stars with \texttt{astrometric\_excess\_noise}$=0$ and apparent $G<$18 mag. In this way, we only fit the stars with well-defined astrometric solutions, and derive representative cluster parameters that are not affected by uncertain measurements of individual faint stars. We additionally note that the photometric monitoring survey that provides rotation period information searched for candidate members far in the outskirts of \editfinal{the cluster}. In our cluster parameter calculation, however, we perform our fit using a subset of stars located closer to the cluster center (but not so close such that the intrinsic dispersions we derive are affected by this decision).

We use Python's \texttt{minimize} function (method$=$\texttt{SLSQP}) to run our maximum likelihood calculation and derive the cluster and field parameters (i.e., $\varpi_{C}$, $\mu_{\alpha_{C}}$, $\mu_{\delta_{C}}$, $\sigma_{\varpi_{C}}$, $\sigma_{\mu_{\alpha_{C}}}$, $\sigma_{\mu_{\delta_{C}}}$ for the cluster, and the analogous set of parameters for the field population). We report the calculated cluster parameters in Appendix \ref{appendix:app_comparison_parameters}. \editfinal{All in all, although our algorithm to derive cluster parameters could have a higher degree of complexity (e.g., by calculating the parameters in a magnitude-dependent way), the membership classification presented in what follows is robust.}
\subsubsection{Membership Probability Calculation}
\label{subsubsec:method_gaia_membership}

Now that we have calculated the cluster and field parameters, we \editfinal{seek to} evaluate Equation (\ref{eq:total_likelihood}) for individual stars and calculate membership probabilities. First, however, \editfinal{the availability and quality of the {\gaia} astrometry need to be considered}, as the data are not of equal \editfinal{precision} for all stars, and they strongly depend on the target brightness. For instance, typical parallax (proper motion) uncertainties are of order $\sim$ 0.05 mas (0.05 mas yr$^{-1}$) for a $G=16$ mag star, and of order $\sim$ 0.5 mas (1.5 mas yr$^{-1}$) for a $G=20$ mag star.

Additionally, the presence of binary stars needs to be considered. Binary stars with measured periods are interesting targets for rotation studies (e.g., \citealt{stauffer18,tokovinin18}), and can provide important clues when investigating rotation in stellar populations (e.g., \citealt{simonian19,simonian20}). Because of this, and to not bias our rotational sequences, we purposely avoid discarding binary stars with our astrometric quality cuts.

\editfinal{We proceed as follows. From the full {\gaia} sample}, we select \editfinal{the stars} with \editfinal{available} positions, proper motions and parallax \editfinal{values}. Additionally, following \citet{gaia18b}, we select stars with \texttt{visibility\_periods\_used}$>$8 and with $\sqrt{\chi^2/(\nu'-5)}$$<$$1.2 \text{max}(1,\text{exp}(-0.2(G-19.5)))$, where $\chi^2$$\equiv$\texttt{astrometric\_chi2\_al} and $\nu'$$\equiv$\texttt{astrometric\_n\_good\_obs\_al}. The latter of these quality cuts removes most of the artifacts while retaining genuine binaries (where naturally the single-star solution does not provide a perfect fit; e.g., see \citealt{belokurov20}). 

The stars excluded by these quality cuts do not have enough information for a reliable membership probability to be calculated. Nonetheless, we do not remove them from our sample, as some of them could correspond to stars with measured rotation periods. We assign them the classification flag of \textbf{“no info”} stars and include them in the {\clusterone} rotational sequence shown in \S\ref{sec:results}.

The stars that survive these astrometric quality cuts have enough information such that we can perform a thorough membership classification. At this point we separate stars that {\it could be} cluster members from those that are most likely not members. For a given star $i$, we calculate the quantity:
\begin{equation}
\Delta_i = \frac{1}{\sqrt{3}} \sqrt{\frac{(\varpi_C - \varpi_i)^2}{(\sigma_{\varpi_{C}}^2 + \sigma_{\varpi_{i}}^2)}+\frac{(\mu_{\alpha_C} - \mu_{\alpha_i})^2}{(\sigma_{\mu_{\alpha_C}}^2 + \sigma_{\mu_{\alpha_i}}^2)}+\frac{(\mu_{\delta_C} - \mu_{\delta_i})^2}{(\sigma_{\mu_{\delta_C}}^2 + \sigma_{\mu_{\delta_i}}^2)}} \text{ ,}
\label{eq:definition_Delta}
\end{equation}
where values with the $C$ index represent the cluster parameters derived in \S\ref{subsubsec:method_gaia_fitting}. We classify the stars with values of $\Delta_i > 3$ as \textbf{“non members”}, as they are located outside the cluster's 3$\sigma$ ellipsoid in phase-space.

\begin{figure}[h]
\epsscale{1.1}  
\plotone{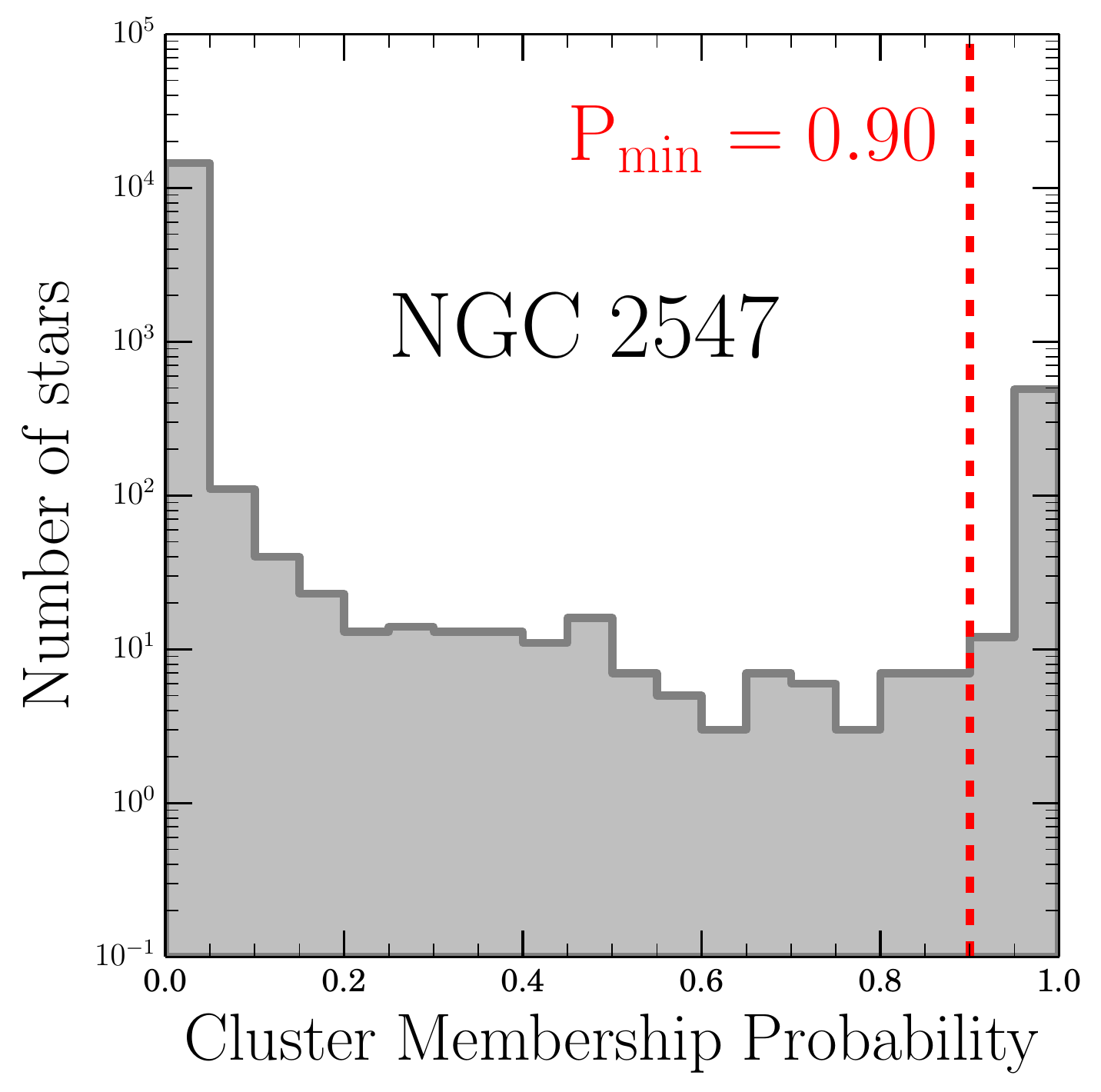}
\caption{{\clusterone}: membership probability distribution for the stars with $\Delta_i \leq 3$ (i.e., stars that {\it could be} cluster members, as they are inside the cluster's 3$\sigma$ ellipsoid in phase-space). The peak at probabilities close to 1 represents the population of likely cluster members. The vertical red dashed line, $P_{\text{min}}=0.90$, shows the probability threshold we use to separate the {\it probable}  members ($P_{C}\geq P_{\text{min}}$) from the {\it possible} members ($P_{C}<P_{\text{min}}$).}
\label{fig:Figure_working_example_cluster_membership_probability_cluster}
\end{figure}

All of the remaining unclassified stars have $\Delta_i \leq 3$ and are technically consistent with the cluster phase-space parameters (at the 3$\sigma$ level). This population, however, has a contribution from objects that have large astrometric uncertainties (particularly in parallax), for which a reliable classification is hard to determine with the existing data. Therefore, another selection criterion is required in order to separate the likely cluster members from this ambiguous population. For this purpose we again use the method described in \S\ref{subsubsec:method_gaia_fitting}, and calculate membership probabilities for all objects with $\Delta_i \leq 3$. Following Equation (\ref{eq:total_likelihood}), for a given star $i$, the cluster membership probability is calculated as:
\begin{equation}
P_{C,i}= \frac{ f_C \mathcal{L}_{C,i}}{\mathcal{L}_{i}}.
\label{eq:membership_probability}
\end{equation}
Conversely, this corresponds to $1-P_{F,i}$, where $P_{F,i}$ is the field membership probability for the star $i$. The distribution of cluster membership probabilities for all stars with $\Delta_i \leq 3$ in the {\clusterone} field is shown in Figure \ref{fig:Figure_working_example_cluster_membership_probability_cluster}. 

For {\clusterone}, we find the distribution of cluster membership probabilities to be strongly bimodal, with most of the sample having probabilities close to either 0 or 1. Although not  explicitly shown, the membership probability distribution has a strong dependence on the stars' apparent $G$ band magnitude, with the distribution being mostly bimodal for bright stars, and becoming {\it blurrier} for fainter stars (with increasingly more stars having intermediate probability values; e.g., see Figure 5 of \citealt{jones96}). This magnitude-dependent outcome is typical, as bright stars tend to have more precise astrometric measurements, and the method provides a {\it yes or no} answer regarding their memberships. On the other hand, faint stars typically have larger astrometric uncertainties, which only allow the method to provide an intermediate answer.

\begin{figure}[h]
\epsscale{0.9}  
\plotone{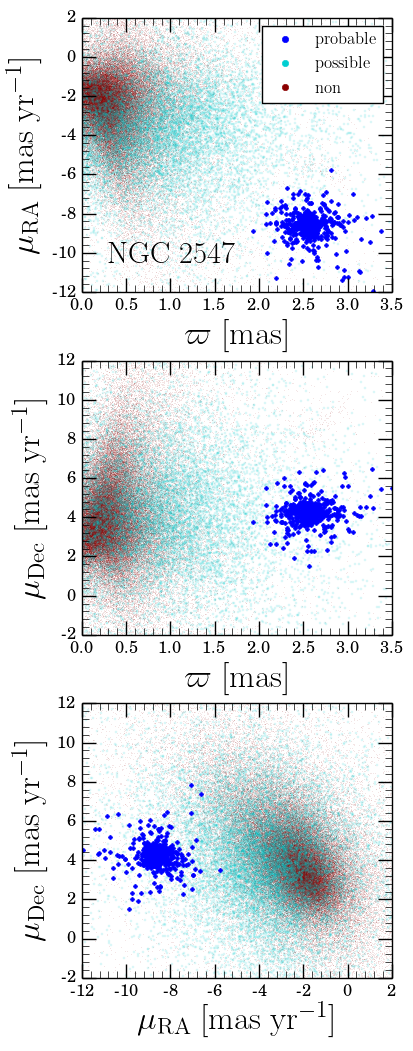}
\caption{{\clusterone}: phase-space projections of the {\it probable} (blue), {\it possible} (cyan), and {\it non} (red) members defined in \S\ref{subsubsec:method_gaia_membership}. The {\it probable} members are clustered in phase-space, while the {\it non} members (i.e., field stars) populate a much larger region at smaller parallax and proper motion values. The {\it possible} members correspond to an intermediate population, dominated by faint stars with large astrometric uncertainties.}
\label{fig:Figure_working_example_cluster_phase_space}
\end{figure}

\begin{figure}[h]
\epsscale{1.1}  
\plotone{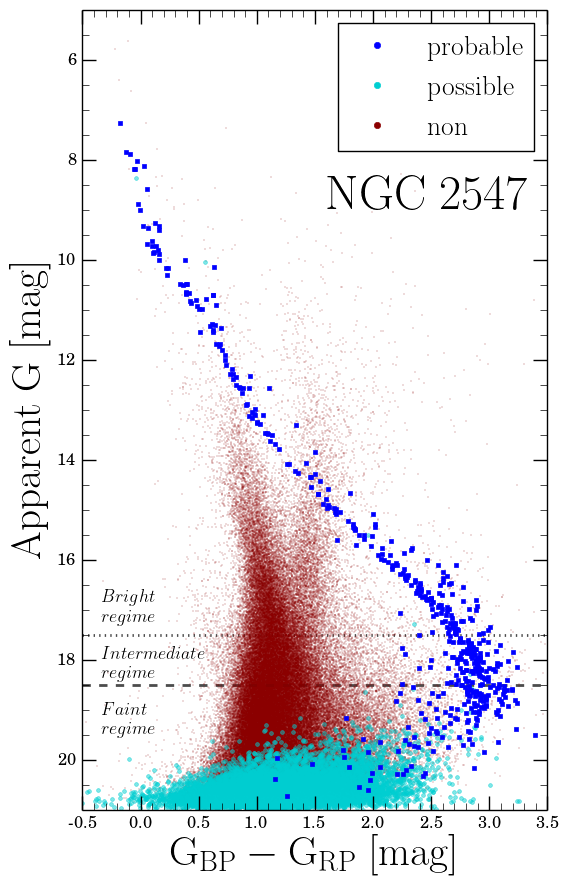}
\caption{{\clusterone}: apparent $G$ band versus $G_{\text{BP}}-G_{\text{RP}}$ CMD of {\it probable} (blue), {\it possible} (cyan), and {\it non} (red) members defined in \S\ref{subsubsec:method_gaia_membership}. Although no photometric selection criteria was used to define the three populations, the {\it probable} members form an isochrone-like sequence, confirming that our selection method is identifying likely cluster members. The {\it non} members show the typical behavior of the background field, while the {\it possible} members mainly consist of faint stars ($G \gtrsim$ 19.5 mag). The “turnover” to bluer colors for apparent $G \gtrsim$ 18.5 mag for the {\it probable} members sequence is discussed in \S\ref{subsubsec:method_gaia_membership}. The dotted and dashed lines show the $G=$ 17.5 and 18.5 mag limits, which we use to define different regimes in our mass and temperature calculation (see \S\ref{subsec:properties_stellar}).}
\label{fig:Figure_working_example_cluster_apparent_CMD}
\end{figure}

With this in mind, we apply a final selection criterion and classify the stars with $P_{C} \geq$ 0.90 as \textbf{“probable members”}, and those with $P_{C}<$ 0.90 as \textbf{“possible members”}. We show the resulting phase-space projections and apparent CMD of these populations, in addition to the \textbf{“non member”} population, in Figures \ref{fig:Figure_working_example_cluster_phase_space} and \ref{fig:Figure_working_example_cluster_apparent_CMD}. (The \textbf{“no info”} population is absent from these figures, as no astrometric information is available for those stars.)

We choose the probability threshold of $P_{\text{min}}=0.90$ to separate between {\it probable} and {\it possible} members as a compromise between completeness in our sample of probable cluster members on the one hand, and the presence of contamination on the other. Several tests with different $P_{\text{min}}$ values showed that the phase-space and CMD of the {\it probable} members seemed to lose bona fide cluster stars when using higher $P_{\text{min}}$ values, while lower values included stars with incoherent kinematics and photometry. We nonetheless note that the overall number of {\it probable} and {\it possible} members do not considerably change unless very high ($P_{\text{min}} \gtrsim $0.95) or very low ($P_{\text{min}} \lesssim $0.05) threshold values are used (see Figure \ref{fig:Figure_working_example_cluster_membership_probability_cluster}). Furthermore, Appendix \ref{appendix:app_tables_of_membership_probabilities} reports the {\gaia} data for the {\it probable} and {\it possible} members and their cluster membership probabilities, and the reader is free to experiment with customized threshold values. 

Figure \ref{fig:Figure_working_example_cluster_phase_space} shows that the {\it probable} members (blue points) tend to be clustered in phase-space with a small number of stars located in the outskirts. On the other hand, the {\it non} members (i.e., the field; red points) occupy a much larger region in the phase-space projections (with stars populating virtually all corners of the diagrams), but tend to be concentrated at smaller parallax and proper motion values. The {\it possible} members (cyan points) appear as an intermediate population of stars, located in between the cluster and the field.

Figure \ref{fig:Figure_working_example_cluster_apparent_CMD} complements what is seen in Figure \ref{fig:Figure_working_example_cluster_phase_space}. The {\it probable} members form a tight isochrone-like sequence in the apparent CMD (with the exception of the color turnover for apparent $G \gtrsim$ 18.5 mag; see discussion below), with even a parallel sequence being visible indicating the presence of photometric binaries (e.g., \citealt{hurley98}). The {\it non} members behave as expected from the field, with the two main branches (main-sequence and giant branch) being clearly visible in the data. The {\it possible} members mainly populate the bottom part of the plot, demonstrating that this population is dominated by faint stars that, consequently, have large astrometric uncertainties, which in turn causes them to have $\Delta_i \leq 3$ values (differentiating them from the {\it non} members). While some of these {\it possible} members could very well be real cluster members, the current precision of their astrometric information does not allow for a more stringent classification.

We highlight that even though our classification method is completely agnostic to the photometry of the stars, and it is solely based on kinematics, it seems to appropriately select stars that form a main-sequence in the CMD. We take this as a confirmation that our method is properly identifying likely cluster members and distinguishing them from the field.

A peculiar feature can be seen in Figure \ref{fig:Figure_working_example_cluster_apparent_CMD} for apparent $G \gtrsim$ 18.5 mag. At this apparent magnitude, instead of continuing to the bottom-right part of the diagram, the {\it probable} members start to turn to bluer colors for fainter magnitudes. This feature has already been found by other works when using the {\gaia} DR2 photometry (e.g., \citealt{arenou18,lodieu19b,lodieu19a,smart19}), and it arises from an overestimation of the flux in the $G_{\text{BP}}$ passband for faint, red sources \citep{riello18,riello21}.
\subsection{Crossmatch}
\label{subsec:method_crossmatch}

Now that we have kinematically classified every {\gaia} DR2 star in the field of {\clusterone} as either a {\it probable}, {\it possible}, or {\it non} member, or a {\it no info} star, we proceed to connect this with the rotation sample of \citet{irwin08a}. 

We crossmatch both catalogs using the CDS X-Match Service on VizieR\footnote{\url{http://vizier.u-strasbg.fr/}}. First, we download all the {\gaia} DR2 matches to a given {\clusterone} star within 3{\arcsec}. This produces matches for all stars, with some of them ($\sim$ 10\%) being matched to two {\gaia} sources. The distribution of angular separations is strongly peaked at small values, with $\sim$ 80\% of the stars having a match within 0.2{\arcsec}, a tail extending out to 0.5{\arcsec}, and some matches extending past this value out to 3{\arcsec}.

To select the best possible matches for stars matched to two {\gaia} sources, we compare the $V$ magnitudes from \citet{irwin08a} with the $G$ magnitudes from {\gaia}. This exercise helps us break the ties and decide which star is the correct match. In all cases, the {\gaia} star with the more similar brightness corresponds to the closer match in angular separation. The second matched star is typically several magnitudes fainter, with angular separations greater than 1{\arcsec} (the outliers of the separation distribution). We only keep one {\gaia} match per star, and are left with an angular separation distribution where $\sim$ 95\% of the stars have a match within 0.3{\arcsec}, and all of them have a match within 0.9{\arcsec}.

In order to test the reliability of our crossmatching approach, we use the Pleiades cluster as a benchmark for which a completely independent crossmatching is available. We first replicate the same crossmatching procedure used in {\clusterone} with our {\clustertwo} data. We then take the 2MASS IDs for the Pleiades stars from \citet{rebull16a}, and look them up in the precomputed 2MASS-{\gaia} DR2 crossmatch (\texttt{gaiadr2.tmass\_best\_neighbor}) that can be found in the ESA's webpage of the {\gaia} Archive\footnote{\url{http://gea.esac.esa.int/archive/}}. We find an excellent agreement ($>$ 99.7\% of coincidence) between ESA's crossmatch and ours for the stars found by both methods (and moreover, VizieR produces matches for 10 {\clustertwo} stars not found by ESA). We take this as a confirmation that our crossmatching is properly combining the periodic samples with the {\gaia} sources. 
\subsection{Applying our Method to all the Clusters}
\label{subsec:method_astrometric_classification}

We follow the method described in \S\ref{subsec:method_gaia} for {\clusterone}, and apply it to the other six clusters listed in Table \ref{tab:list_of_clusters}. 
We report the astrometric parameters we calculate for the clusters in Appendix \ref{appendix:app_comparison_parameters}, where we also compare them with the values reported in the literature. We find our astrometric parameters to be in good agreement with those reported by other studies, with fractional differences being typically at the $\sim$1\textendash2\% level or less.

\begin{figure}[h]
\epsscale{1.1}  
\plotone{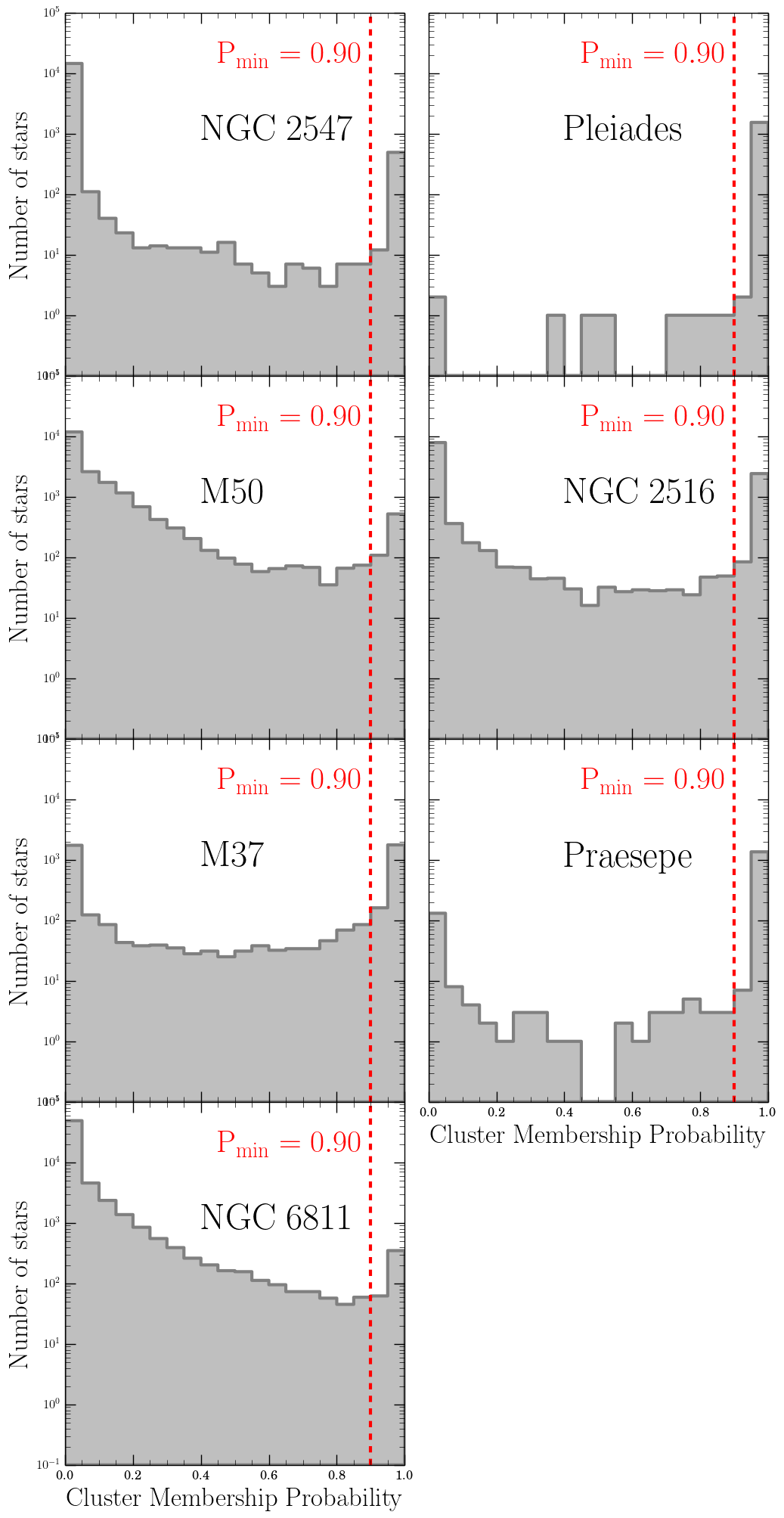}
\caption{Membership probability distribution for all the clusters we study, analogous to Figure \ref{fig:Figure_working_example_cluster_membership_probability_cluster}.}
\label{fig:Figure_all_clusters_membership_probability_cluster}
\end{figure}

\begin{figure}[h]
\epsscale{0.85}  
\plotone{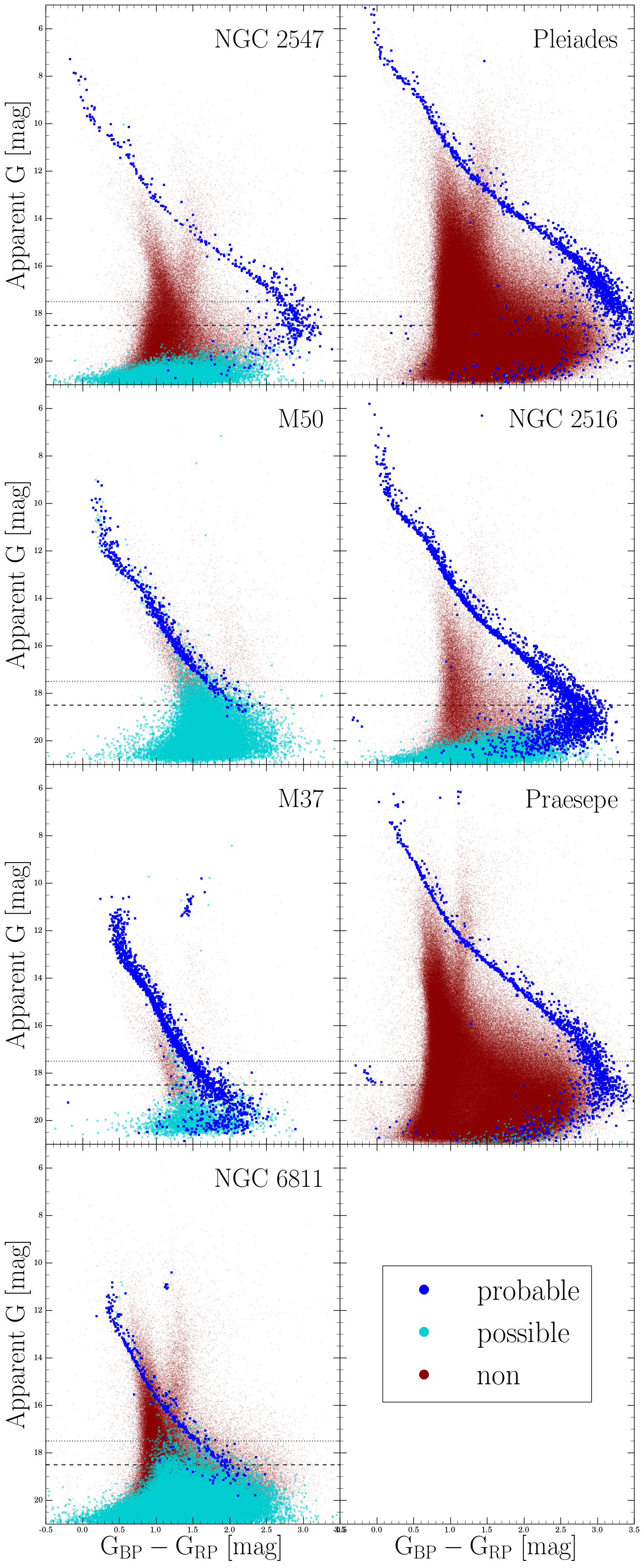}
\caption{Apparent $G$ band versus $G_{\text{BP}}-G_{\text{RP}}$ CMD of {\it probable} (blue), {\it possible} (cyan), and {\it non} (red) members for all the clusters we study, analogous to Figure \ref{fig:Figure_working_example_cluster_apparent_CMD}.}
\label{fig:Figure_all_clusters_apparent_CMD}
\end{figure}

For each cluster, we classify every {\gaia} star contained within the region where the corresponding rotation period catalog reports candidate members, into one of the four previously described categories ({\it no info}, {\it probable}, {\it possible}, or {\it non} member). We report tables with the corresponding list of {\it probable} and {\it possible} members for every cluster in Appendix \ref{appendix:app_tables_of_membership_probabilities}. Analogously to Figures \ref{fig:Figure_working_example_cluster_membership_probability_cluster} and \ref{fig:Figure_working_example_cluster_apparent_CMD} for {\clusterone}, Figures \ref{fig:Figure_all_clusters_membership_probability_cluster} and \ref{fig:Figure_all_clusters_apparent_CMD} show the membership probability distribution and apparent CMDs for all the clusters we study. The phase-space projections for all clusters, analogous to Figure \ref{fig:Figure_working_example_cluster_phase_space} for {\clusterone}, can be found in Appendix \ref{appendix:app_phase_space_projections_all_clusters}.

Figure \ref{fig:Figure_all_clusters_membership_probability_cluster} shows that, for every cluster, in the sample of {\gaia} stars that satisfy $\Delta_i \leq$ 3, there is a population of stars with membership probabilities close to 1. This population represents a set of likely cluster members, and we classify those with $P_{C} \geq$ 0.90 as {\it probable} members (and the rest as {\it possible} members). On the low-probability end, however, clear cluster-by-cluster differences can be observed. While clusters such as the {\clustertwo} and {\clustersix} show only a small number of {\it possible} members, {\clusterthree} and {\clusterseven} are in the opposite situation. These differences are explained by considering how similar or different the astrometric parameters of a given cluster are, compared with the field (which is typically a diffuse background population with small parallax and low proper motion). In other words, a cluster with a large parallax and proper motion (e.g., the {\clustertwo} or {\clustersix}) will be more distinct and separated from the field in phase-space, and therefore it is highly unlikely that unassociated field stars will have similar astrometry (i.e., $\Delta_i \leq$ 3), and hence a small number of {\it possible} members is expected. On the other hand, field stars are more likely to mimic the astrometry of a cluster that has a small parallax and a low proper motion (e.g., {\clusterthree} or {\clusterseven}), and therefore the number of stars that are kinematically consistent with the cluster increases.

The apparent CMDs of all our clusters, shown in Figure \ref{fig:Figure_all_clusters_apparent_CMD}, display an isochrone-like sequence of {\it probable} members, with clear binary sequences being identified in all of them. The older clusters {\clusterfive}, {\clustersix}, and {\clusterseven} show not only a main-sequence population, but also a few giant stars (with {\clustersix} even showing a clear white dwarf sequence). We highlight that, in order to obtain unbiased cluster memberships, our method is completely agnostic to the {\gaia} photometry, and these results are a consequence of our careful astrometric analysis. Additionally, as previously found for {\clusterone}, we observe the turn over to bluer $G_{\text{BP}}-G_{\text{RP}}$ colors at apparent magnitudes $G \gtrsim$ 18.5 mag in the {\it probable} member sequence of all the clusters. We take this as further evidence that this corresponds to an artifact of the {\gaia} photometry (see \S\ref{subsubsec:method_gaia_membership} for details).

As discussed in \S\ref{subsubsec:method_gaia_membership}, our ability to cleanly separate between {\it possible} and {\it probable} members is a strong function of the stars' apparent $G$ band magnitude, as the quality of their astrometry decreases with decreasing brightness. This can be clearly seen, for instance, in the CMD of {\clusterseven}, where all the stars fainter than apparent $G \simeq$ 19 mag are classified as either {\it possible} or {\it non} members, and none of them are classified as {\it probable} members. Real {\clusterseven} members with $G >$ 19 mag are simply too faint for {\gaia} DR2 to provide precise astrometry, preventing us from reliably differentiating them from the field stars.

The crossmatching between the periodic samples and the {\gaia} DR2 data is done in identical fashion for all of the clusters, following the approach of \S\ref{subsec:method_crossmatch}. Unlike the case of {\clusterone}, however, we do not find a {\gaia} DR2 counterpart for every periodic star in most of the other clusters. Our recovery rates (and number of stars not found in the crossmatching over the number of stars with measured periods) are: 100\% (0/176) for {\clusterone}, 99.9\% (1/759) for {\clustertwo}, 94.3\% (46/812) for {\clusterthree}, 99.4\% (2/364) for {\clusterfour}, 96.7\% (12/367) for {\clusterfive}, 99.8\% (2/809) for {\clustersix}, and 100\% (0/232) for {\clusterseven}. The only clusters with recovery rates below 99\% are {\clusterthree} and {\clusterfive}, which correspond to distant systems where the periodic samples extend to magnitudes fainter than the apparent $G \approx 21$ mag limit of {\gaia} DR2. In the following, we classify the stars not found in the crossmatching as {\it no info} stars, as their membership is uncertain and we do not have evidence to discard them.
\section{\editfinal{Cluster and Stellar Properties}} 
\label{sec:properties_clusters_and_stellar}
\subsection{\editfinal{Cluster Properties}}
\label{subsec:properties_clusters}

To perform meaningful inter-cluster comparisons of stellar rotation, reliable cluster properties are needed. In the {\it pre} {\gaia} era, classic works such as \citet{denissenkov10}, \citet{gallet13}, and \citet{gallet15} thoroughly compiled periodic samples as well as properties for a large number of clusters. Nevertheless, the heterogeneous nature of their parameter compilation (adopting values from varied techniques and data sets), can result in inhomogeneities in the adopted masses, ages, and metallicity scales. Nowadays, the systematic observations of star clusters by spectroscopic surveys (\citealt{netopil16}; {\gaia}-ESO, \citealt{spina17}, \citealt{magrini17}; APOGEE, \citealt{majewski17}, \citealt{jonsson20}), as well as the uniform all-sky {\gaia} data, provide an opportunity to study these systems in a more consistent and homogeneous fashion. In the following, we attempt, to the extent that it is possible, to compile and derive cluster properties that are in a consistent and uniform scale.

The mass and temperature values \editfinal{we calculate in \S\ref{subsec:properties_stellar}}, as well as the \editfinal{results and} discussion presented in \editfinal{\S\ref{sec:results}} and \S\ref{sec:discussion}, require four cluster properties to be known: distance modulus, reddening, age, and metallicity. The former two are needed to translate the observed photometry to absolute and dereddened CMD space, while the latter two are required to compare the data with the appropriate stellar evolutionary model. For distance modulus, we adopt the values derived from our astrometric analysis (considering the parallax zero-points described in \S\ref{subsubsec:method_gaia_data}), and for metallicity we compile values reported from recent high-resolution spectroscopic results. In particular, we adopt the \editfinal{state-of-the-art} values reported by \citet{netopil16} for {\clustertwo}, {\clusterfour}, {\clusterfive}, {\clustersix}, and {\clusterseven}, and the value reported by the {\gaia}-ESO survey \citep{spina17} for {\clusterone}. The only cluster without a spectroscopic metallicity measurement is {\clusterthree}, for which we adopt a solar value. \editfinal{We now describe our procedure to obtain cluster reddenings and ages.}

Instead of adopting reddening values from the literature, we calculate our own using a dedicated fitting routine. For a given cluster, we start by correcting the observed $G$ band magnitudes by distance modulus. Once the spectroscopic metallicity is known, we download a set of PARSEC models\footnote{We initially also considered using the MIST \citep{dotter16,choi16} and BHAC15 \citep{baraffe15} models, but the PARSEC models showed better agreement when compared with the data. This is not surprising, as the PARSEC models include empirical corrections to fit the mass-radius relation of dwarf stars, which in turn produces a better agreement with the CMDs of clusters.} \citep{bressan12,chen14} of that composition with a range of plausible ages (e.g., 100, 200, and 300 Myr for {\clusterfour}). Then, we define a region in the absolute CMD where the different models agree with each other, which excludes the phases near and more evolved than the turnoff. We then iterate over a range of possible reddenings and find the value that produces the best agreement between the observed data with the series of models. 

\editfinal{For a given $E(B-V)$ value}, we assume $A_{V}= 3.1 \times E(B-V)$ \citep{cardelli89} \editfinal{and deredden the photometry as follows.} For stars with measured photometry in the three {\gaia} bands, we deredden the observed magnitudes following the approach of \citet{gaia18b}, where the extinction coefficients depend on $G_{\text{BP}}-G_{\text{RP}}$ color and extinction itself. For stars that lack a measured $G_{\text{BP}}-G_{\text{RP}}$ color, we adopt the PARSEC coefficients\footnote{\url{http://stev.oapd.inaf.it/cgi-bin/cmd_3.3}} for a G2V star ($A_{G}/A_{V}=0.85926$, $A_{G_{\text{BP}}}/A_{V}=1.06794$, and $A_{G_{\text{RP}}}/A_{V}=0.65199$).

\editfinal{For every $E(B-V)$ value we are iterating over}, we deredden the {\it probable} cluster members' photometry, define a set of bins in the $(G_{\text{BP}}-G_{\text{RP}})_0$ color coordinate, and calculate the mean colors and 75th magnitude percentiles (which accounts for the photometric binaries). We then find the $E(B-V)$ value that minimizes the sum of the squared difference of the data minus the models given the bins, and adopt this as our global cluster reddening. Finally, we compare the CMD location of the full sample of {\it probable} members with models of varying ages, and estimate the age following the envelope of hot stars below and near the turn-off \editfinal{(typically stars hotter than $\sim$ 7,000 K)}. We estimate that this procedure leads to typical errors of $\sim 20\%$ for reddenings and \editfinal{$\sim$ 10\textendash20\% for ages.}

We summarize the revised cluster properties we adopt in Table \ref{tab:revised_cluster_properties}. Note that for the nearby and young clusters {\clusterone} and the {\clustertwo}, we adopt the lithium-depletion ages from \citet{jeffries05} (see also \citealt{naylor06}) and \citet{stauffer98}, respectively. This technique yields reliable and almost completely model-independent ages in the $10 \textendash 200$ Myr range. These adopted ages are in good agreement with the values we derive from our CMD analysis.

\begin{table}
\scriptsize
\begin{center}
\caption{Revised \editfinal{and updated} cluster properties. \label{tab:revised_cluster_properties}}
\begin{tabular}{lcrrrr}
\hline
\hline
Name & Age & [Fe/H] & DM & \editfinal{Distance} & $E(B-V)$  \\
- & [Myr] & [dex] & [mag] & \editfinal{[pc]} & [mag]   \\
\hline
{\clusterone}  & 35 & $-$0.01 & 7.925 & \editfinal{384.6} & 0.044 \\
{\clustertwo}  & 125 & $-$0.01 & 5.670 & \editfinal{136.2} & 0.051 \\
{\clusterthree}  & 150 & 0.00 & 9.936 & \editfinal{970.9} & 0.210 \\
{\clusterfour}  & 150 & $+$0.05 & 8.058 & \editfinal{408.8} & 0.103 \\
{\clusterfive}  & 500 & $+$0.02 & 10.787 & \editfinal{1436.8} & 0.246 \\
{\clustersix}  & 700 & $+$0.16 & 6.345 & \editfinal{185.8} & 0.014 \\
{\clusterseven}  & 950 & $+$0.03 & 10.167 & \editfinal{1080.1} & 0.047 \\
\hline
\\
\end{tabular}
\end{center}
\tablecomments{Summary of the revised \editfinal{and updated} cluster properties (age, metallicity, distance modulus, and reddening) that we adopt to calculate masses and temperatures, as well as in our discussion throughout \S\ref{sec:discussion}. \editfinal{The cluster distances are also shown for reference.} We calculate distance moduli from our astrometric analysis, survey the literature for \editfinal{updated} high-resolution spectroscopic metallicity measurements, and estimate reddening and age from a CMD analysis of the {\it probable} cluster members.}
\end{table}
\subsubsection{Comparing our Cluster Properties with the Literature}
\label{subsubsec:cluster_properties_comparewithliterature}

We now compare the cluster properties we have derived, with the values that can be found in the literature from similar approaches. The main references we use for this are \citet{gaia18b} and \citet{bossini19}, who also studied open clusters based on the {\gaia} DR2 data, and reported parameters for most of the clusters in our sample based on comparisons with PARSEC models. We also use as reference the works by \citet{cummings16}, \citet{cummings18a}, and \citet{cummings18b}. Although these did not use {\gaia} data, they reported parameters for many of the clusters in our sample based on CMD fits to $UBV$ photometry and PARSEC models.

For distance modulus, our values are calculated from the global cluster parallaxes, accounting for a zero-point value of 53 $\mu$as in {\clusterseven} and of 29 $\mu$as for the other clusters. While \citet{gaia18b} and \citet{bossini19} (whose membership and astrometry come from \citealt{cantatgaudin18a}) do not consider these zero-points, in Appendix \ref{appendix:app_comparison_parameters} we add these offsets to their parallax values for an appropriate comparison, and find differences to be within $\approx \pm 1$\%. For metallicity, we share references and use high-resolution spectroscopic values from \citet{netopil16} or {\gaia}-ESO when available. 

For reddening, if we exclude {\clusterthree} from the comparison, we find good agreement with \citet{gaia18b} and \citet{bossini19}, and the differences between our values and theirs are contained within $\pm 0.032$ mag in $E(B-V)$. On the other hand, {\clusterthree} is the only cluster for which the \citet{gaia18b} and \citet{bossini19} reddenings differ between each other. The former reports $E(B-V) = 0.105$ mag, the latter reports 0.153 mag, and we obtain a best-fit value of 0.210 mag. We note, however, that \citet{cummings16} and \citet{cummings18a} report $E(B-V)=0.230$ for {\clusterthree}, in better agreement with our estimate. Ultimately, the above differences in reddening are perhaps unsurprising considering the different CMD analysis approaches and parallax zero-point considerations.

Regarding stellar ages, we find our revised values to be in good agreement with the {\it pre} {\gaia} values \editfinal{compiled by} \citet{gallet15} listed in Table \ref{tab:list_of_clusters} (see also \citealt{denissenkov10, gallet13}), especially considering the associated uncertainties. The largest difference is seen for {\clustersix}, for which we derive an age of 700 Myr, while \citet{gallet15} adopt 580 Myr. Given the crucial importance that clusters' ages play in models of angular momentum evolution, the overall good agreement between our values and the {\it pre} {\gaia} ones is a meaningful result of our work.

Comparing our ages with those by \citet{gaia18b}, we find our results to be in good agreement. A similar comparison with \citet{bossini19} shows that their ages seem to be systematically underestimated by factors of $30\textendash60\%$ for the younger clusters. The exception to the above is {\clusterfour}, for which \citet{gaia18b} and \citet{bossini19} report ages of $\approx$ 300 and 250 Myr, respectively. If confirmed, this would position {\clusterfour} as an important anchor for stellar rotation at intermediate ages, as few systems older than the {\clustertwo} and younger than {\clusterfive} have such comprehensive periodic data sets. Nevertheless, our analysis favors the classical $\sim$ 150 Myr age, and this value is similar to the results by \citet{cummings16}, \citet{cummings18a}, and \citet{cummings18b} (see also \citealt{fritzewski20}, \citealt{healy20}, and \citealt{bouma21}).
\subsection{Stellar Properties}
\label{subsec:properties_stellar}

In order to perform comprehensive comparisons of the rotational sequences of the open clusters, we need to place them on a common scale. For this, we derive mass and temperature values for the cluster members by comparing the photometric data with state-of-the-art evolutionary models in absolute magnitude space. Although the papers that reported periods also reported photometry in at least two bands for most of our clusters (e.g., $V$ and $I$ for {\clusterone}; see \S\ref{sec:sample_selection_and_period_data}), the specific filters vary on a cluster by cluster basis. Instead of using these, we take advantage of the uniform $G$, $G_{\text{BP}}$, and $G_{\text{RP}}$ photometry provided by {\gaia} DR2. This also allows us to derive masses and temperatures for all the {\gaia} cluster members, not just the subset with measured periods.

\begin{figure}[h]
\epsscale{1.1}  
\plotone{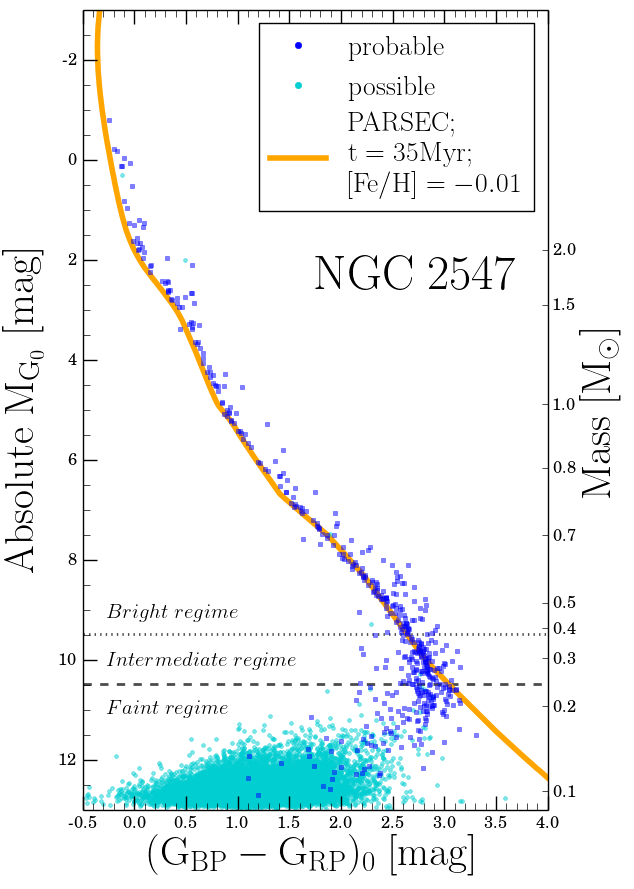}
\caption{{\clusterone}: absolute and dereddened CMD of the {\it probable} (blue) and {\it possible} (cyan) members. The points from Figure \ref{fig:Figure_working_example_cluster_apparent_CMD} have been corrected by extinction and distance modulus. The best-fit {\clusterone} PARSEC isochrone is shown as the orange line. The dotted and dashed lines correspond to the apparent $G=$ 17.5 and 18.5 mag limits from Figure \ref{fig:Figure_working_example_cluster_apparent_CMD} translated to absolute CMD space, which define different regimes in our calculation. We use this figure to interpolate the model and derive mass and temperature values (see \S\ref{subsec:properties_stellar}).}
\label{fig:Figure_working_example_cluster_absolute_and_dereddened_CMD}
\end{figure}

\editfinal{We again illustrate our procedure using {\clusterone} as a working example.} We use its $E(B-V)$ and distance modulus values (see \S\ref{subsec:properties_clusters}) and calculate absolute and dereddened photometry. We show the extinction-corrected absolute CMD of the {\it probable} and {\it possible} {\clusterone} members in Figure \ref{fig:Figure_working_example_cluster_absolute_and_dereddened_CMD}. The data can now be directly compared with stellar models to infer mass and temperature values. We use a PARSEC isochrone with the corresponding {\clusterone} age and metallicity, and show it as the orange line in Figure \ref{fig:Figure_working_example_cluster_absolute_and_dereddened_CMD}.

The bright {\it probable} cluster members show an excellent agreement with the model in Figure \ref{fig:Figure_working_example_cluster_absolute_and_dereddened_CMD}. As discussed in Figure \ref{fig:Figure_working_example_cluster_apparent_CMD}, the $G_{\text{BP}}-G_{\text{RP}}$ color misbehaves for apparent $G \gtrsim 18.5$ mag, and the {\it probable} member sequence becomes bluer for fainter magnitudes. This limit is translated to the absolute and dereddened CMD and shown as the horizontal dashed line in Figure \ref{fig:Figure_working_example_cluster_absolute_and_dereddened_CMD}, with the dotted line showing a similar apparent $G=17.5$ mag limit. We use these two apparent magnitude limits, here translated to absolute magnitude space, to define different regimes in our mass and temperature calculation.

Additionally, whenever possible, the effects of binarity on the CMD need to be considered. Photometric binaries appear as brighter/redder stars compared to the main-sequence locus, and to account for them we follow the procedures of \citet{stauffer16} and \citet{somers17b}. In this approach, when the {\gaia} $(G_{\text{BP}}-G_{\text{RP}})_0$ color is reliable, stars are projected down (or up) onto the main-sequence locus at fixed color, effectively removing the contribution of close companions (i.e., our mass and temperature estimates for these stars correspond to those of the primary star). We designate the projected absolute magnitudes as $M_{G_{0,proj}}$, and describe its calculation in detail  below.

To acknowledge the misbehavior of the $G_{\text{BP}}-G_{\text{RP}}$ color and account for the contribution of photometric binaries when possible, we define three regimes to calculate projected magnitudes $M_{G_{0,proj}}$ in the absolute CMD of Figure \ref{fig:Figure_working_example_cluster_absolute_and_dereddened_CMD}:
\begin{itemize}
\item {\it Bright regime:} For stars brighter than the dotted line (apparent $G=17.5$ mag translated to absolute magnitude), we calculate $M_{G_{0,proj}}$ by projecting the stars $M_{G_{0}}$ value down (or up) onto the model at the measured $(G_{\text{BP}}-G_{\text{RP}})_0$ color.
\item {\it Faint regime:} For stars fainter than the dashed line (apparent $G=18.5$ mag translated to absolute magnitude), or stars that lack a $G_{\text{BP}}$ or $G_{\text{RP}}$ magnitude, we cannot rely on the $(G_{\text{BP}}-G_{\text{RP}})_0$ color to do a de-projection. Instead, we simply keep the calculated $M_{G_{0}}$ value as is, and define $M_{G_{0,proj}}$ to be equal to it. This is therefore a regime where we do not account for unresolved companions.
\item {\it Intermediate regime:} For stars in between the dotted and dashed lines ($17.5 \leq G \leq 18.5$ mag range translated to absolute magnitude), we do an intermediate, {\it ramp-like} calculation. We calculate the values we would have obtained following the {\it bright} and {\it faint} regimes separately, and combine them in a progressive way \editfinal{with linear weighting. In this way, the} extremes match the corresponding methods, effectively avoiding a sharp transition between the different regimes.
\end{itemize}

For every star, this procedure collapses the observed photometry to the single quantity $M_{G_{0,proj}}$. We use this value to interpolate the corresponding {\clusterone} PARSEC model in the magnitude-mass and magnitude-temperature space. We only interpolate in the regime of stars fainter than the main-sequence turnoff (i.e., we do not report masses for evolved stars) and down to the faintest $M_{G_{0}}$ magnitude allowed by the model. We report our derived masses and temperatures for the {\clusterone} {\it probable} and {\it possible} members in Appendix \ref{appendix:app_tables_of_membership_probabilities}.

\begin{figure}[h]
\epsscale{1.20}  
\plotone{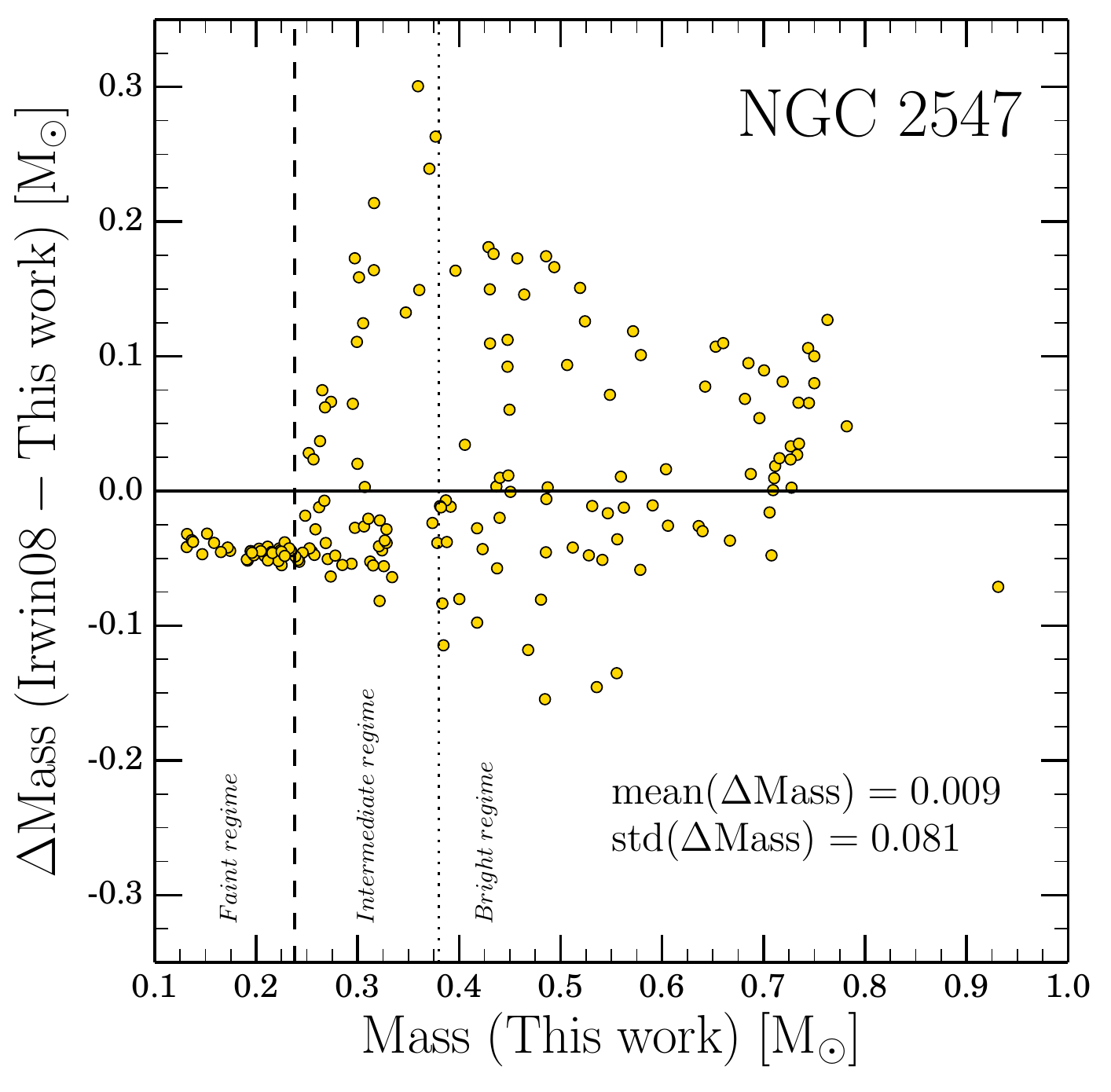}
\caption{{\clusterone}: Comparison of our mass estimates with those of \citet{irwin08a} for the stars in the periodic sample. We show mass difference (in the sense of \citet{irwin08a} minus this work) versus mass from this work. Both estimates come from entirely independent photometric data sets and interpolated models. The dotted and dashed lines correspond to the apparent $G=$ 17.5 and 18.5 mag limits from Figure \ref{fig:Figure_working_example_cluster_apparent_CMD} translated to mass coordinates, and these values define the three regimes used for our mass calculation. We find a good overall agreement, and the comparison suggests a possible systematic uncertainty of $\sim$ 0.05 $M_{\odot}$ in the masses derived by this method. In the {\it intermediate} and {\it bright} regimes, the \citet{irwin08a} masses tend to be higher due to their method not accounting for photometric binaries. }
\label{fig:Figure_working_example_cluster_mass_comparison}
\end{figure}

For the subset of {\clusterone} stars with period measurements, \citet{irwin08a} calculated masses using their $I$ band magnitudes and the NextGen models \citep{baraffe98}. These values offer a completely independent reference point to validate our own masses, and we compare our estimates with theirs in Figure \ref{fig:Figure_working_example_cluster_mass_comparison}. It is important to highlight one key difference between both methods: we account for the presence of photometric binaries (when possible), while \citet{irwin08a} do not. Accordingly, we find the mass difference to be a strong function of mass itself. For stars in the {\it faint regime}, where we do not account for binaries, we see a good agreement but with an approximately constant offset of $\Delta$Mass $\approx -0.05 M_{\odot}$ that is likely due to the different underlying models used in the interpolation. For higher masses, the scatter increases and many stars have positive mass differences, indicating cases where \citet{irwin08a} did not account for the contribution of unresolved companions. Overall, we find a good agreement, with mean and standard deviation values of $\Delta$Mass being $\approx$ 0.01 and 0.08 $M_{\odot}$, respectively.

\begin{figure}[h]
\epsscale{0.90}  
\plotone{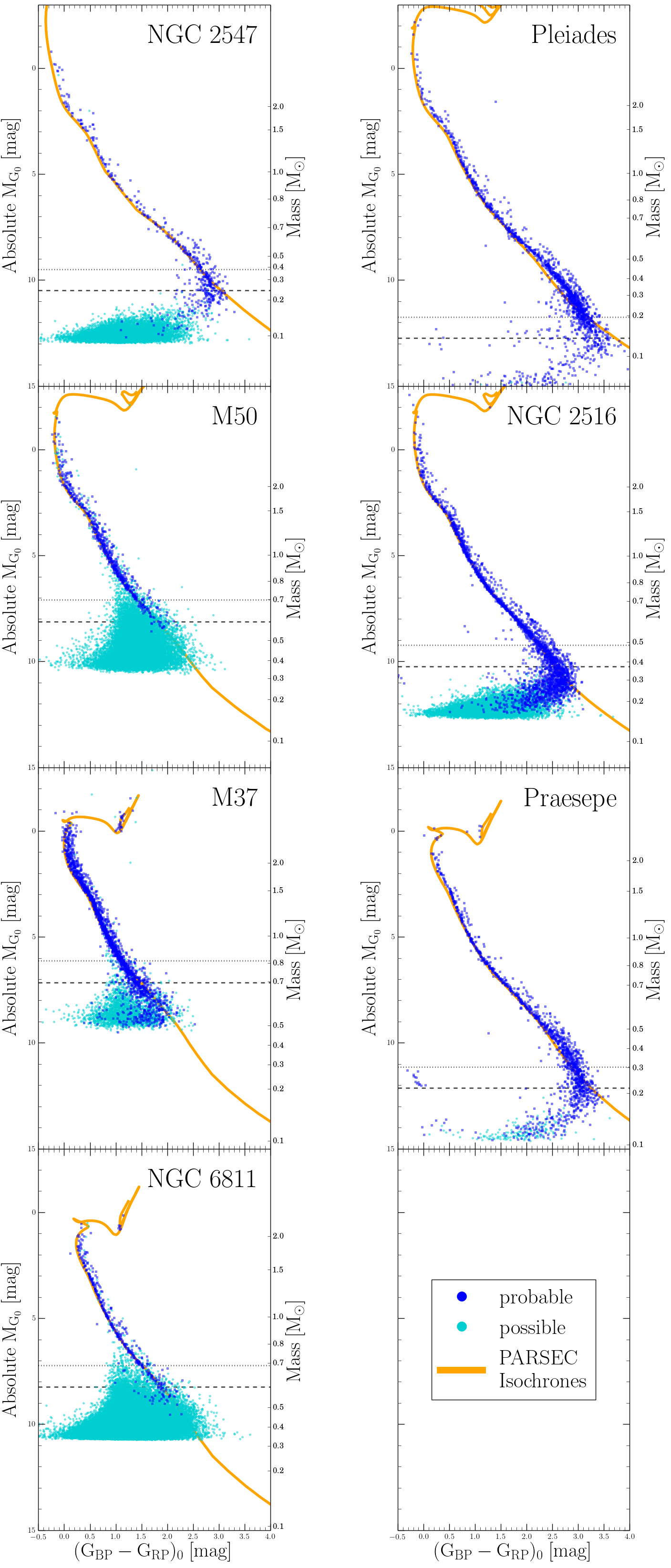}
\caption{Absolute and dereddened CMD of the {\it probable} (blue) and {\it possible} (cyan) members for all the clusters we study, analogous to Figure \ref{fig:Figure_working_example_cluster_absolute_and_dereddened_CMD}. The orange lines are the respective best-fit PARSEC isochrones.}
\label{fig:Figure_all_clusters_absolute_and_dereddened_CMD}
\end{figure} 

The above comparison validates our method, and suggests that masses derived in this way are subject to a systematic uncertainty of order $\sim$ 0.05 $M_{\odot}$. \editfinal{Statistical uncertainties are of order $\sim$ 0.01\textendash0.02 $M_{\odot}$, considerably smaller than the systematic ones.} More importantly, by using the {\gaia} photometry, we have defined a procedure to calculate the masses and temperatures that can be uniformly applied to all our clusters.

Analogously to Figure \ref{fig:Figure_working_example_cluster_absolute_and_dereddened_CMD} for {\clusterone}, Figure \ref{fig:Figure_all_clusters_absolute_and_dereddened_CMD} shows the extinction-corrected absolute CMD for the {\it probable} and {\it possible} members of all the clusters. Some discrepancies arise for low mass stars, where the {\it probable} members appear brighter/redder than the model (e.g., see the $\sim 0.4 M_{\odot}$ stars in {\clustertwo} and {\clustersix}). As other studies suggest, these differences could arise due to the known radius inflation problem in cool dwarfs (e.g., \citealt{torres02,clausen09,torres10,kraus11,somers17a,jackson18,jackson19}). Accounting for the underlying physical process that causes this in the models is beyond the scope of this paper, but we refer interested readers to the works by \citet{somers15b} and \citet{somers20}. Regardless, while evidently not perfect, Figure \ref{fig:Figure_all_clusters_absolute_and_dereddened_CMD} shows a good overall agreement of the data with models across the HR diagram for all the clusters.

We follow the procedure described \editfinal{for {\clusterone}}, and use the absolute CMDs to calculate masses and temperatures for the {\it probable} and {\it possible} cluster members. We report these values in Appendix \ref{appendix:app_tables_of_membership_probabilities}. For the subset of stars with independent mass and temperature estimates from other studies, we compare these values with ours in Appendix \ref{appendix:app_mass_and_Teff_comparison}. Globally, we find our masses and temperatures to be in good agreement with those reported by the period references. The comparisons suggest possible systematic uncertainties of $\sim 0.05 \textendash 0.1 M_{\odot}$ in mass and $\sim 150$ K in temperature, which are modest considering the different methods, photometric data, and underlying stellar models employed.
\section{Results} 
\label{sec:results}

\editfinal{For the remainder of this paper we focus} on the subset of main-sequence stars with period measurements, the periodic samples. \editfinal{Particularly, in this section} we proceed to investigate how \editfinal{the clusters' rotational sequences} change when we consider the astrometric classifications previously derived, \editfinal{to what extent field contamination is an issue, and whether rotational outliers are real cluster members}. For this, we remove from our sample all the {\clustertwo} and {\clustersix} stars that \citet{rebull16b} and \citet{rebull17} classify as pulsators, as well as the giant {\clusterseven} stars for which \citet{santos21} report rotation periods (which are longer than 100 days). Additionally, for the {\clustertwo} and {\clustersix} stars with multiple period measurements, we only consider their main periods as the adopted rotation period ($P1$ from \citealt{rebull16a,rebull16b} and \citealt{rebull17}). Tables with the membership classifications and periods, as well as other important parameters (e.g., {\gaia} DR2 IDs, derived masses and temperatures, membership probabilities) for the periodic samples are reported in Appendix \ref{appendix:app_tables_rotation_period_sample}.

\begin{figure*}
\gridline{\fig{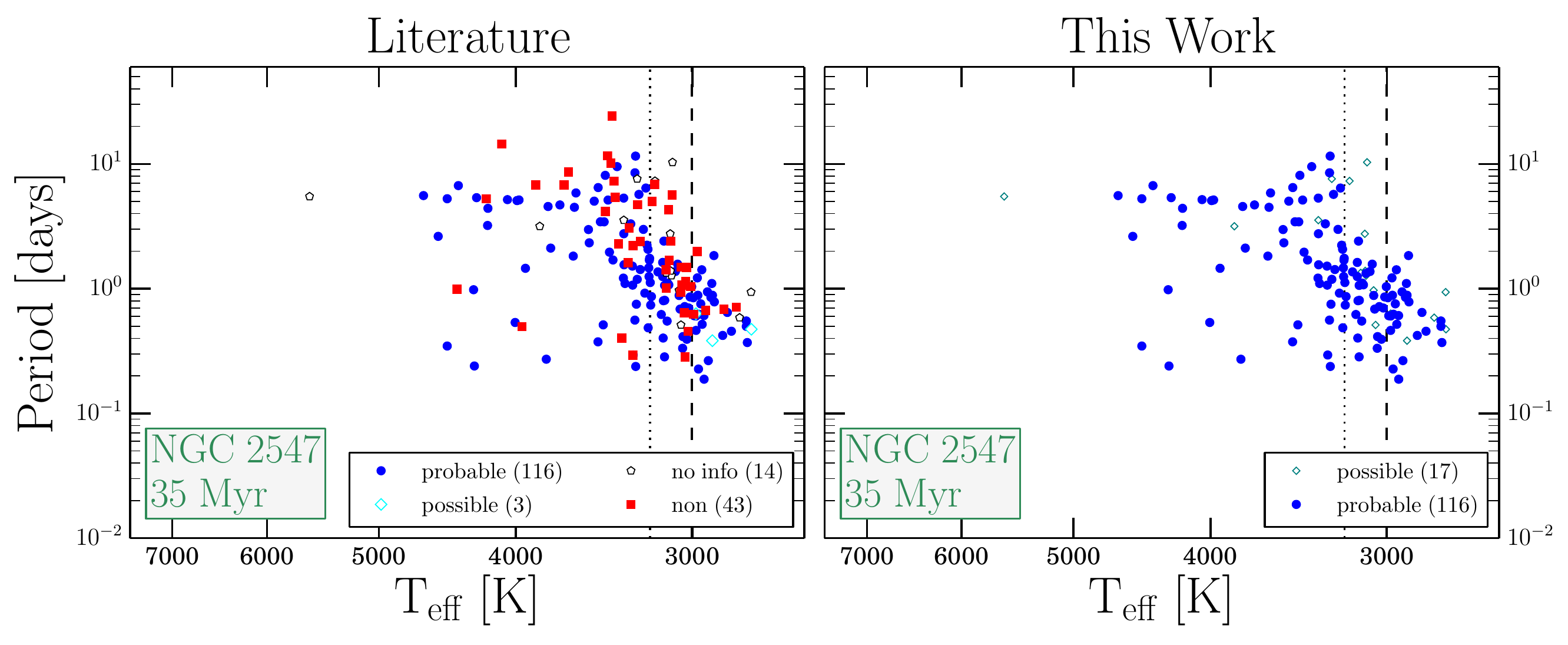}{0.50\textwidth}{a)}
          }
\gridline{
	    \fig{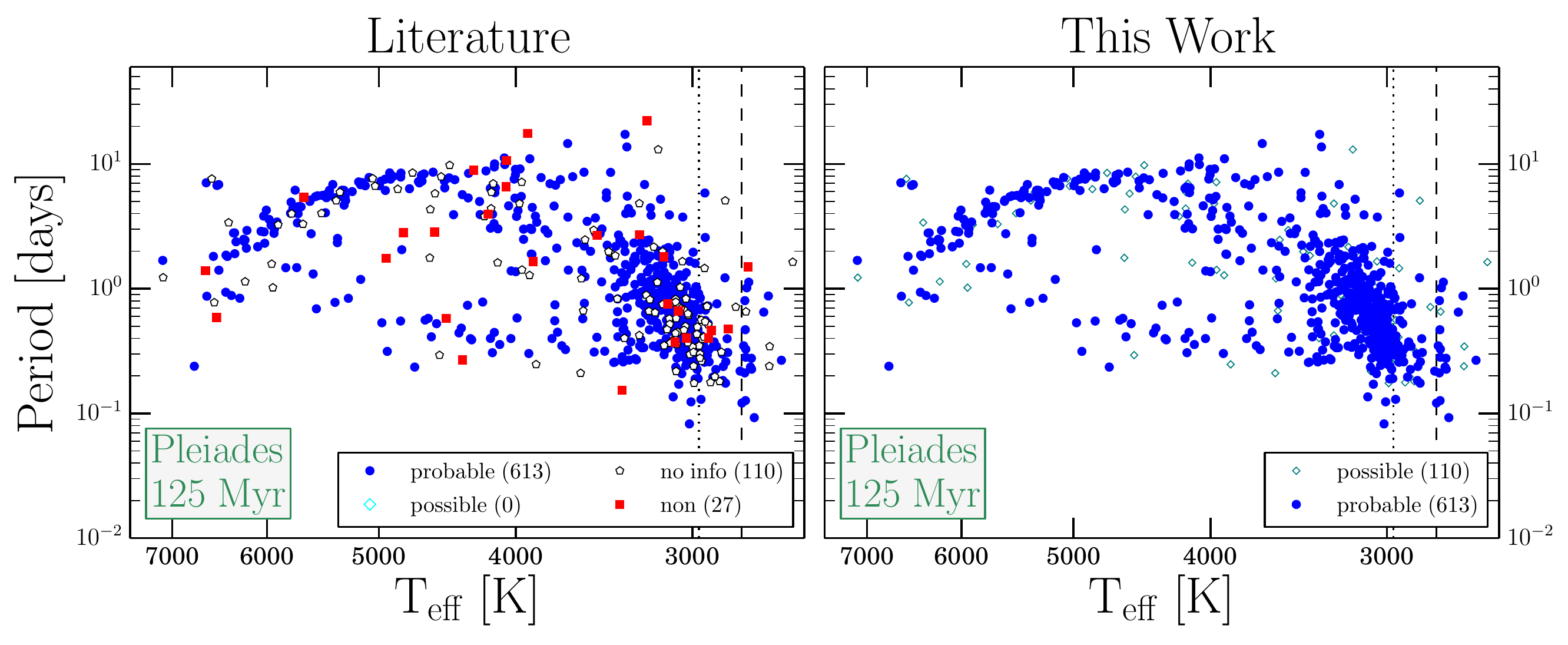}{0.50\textwidth}{b)}
	    \fig{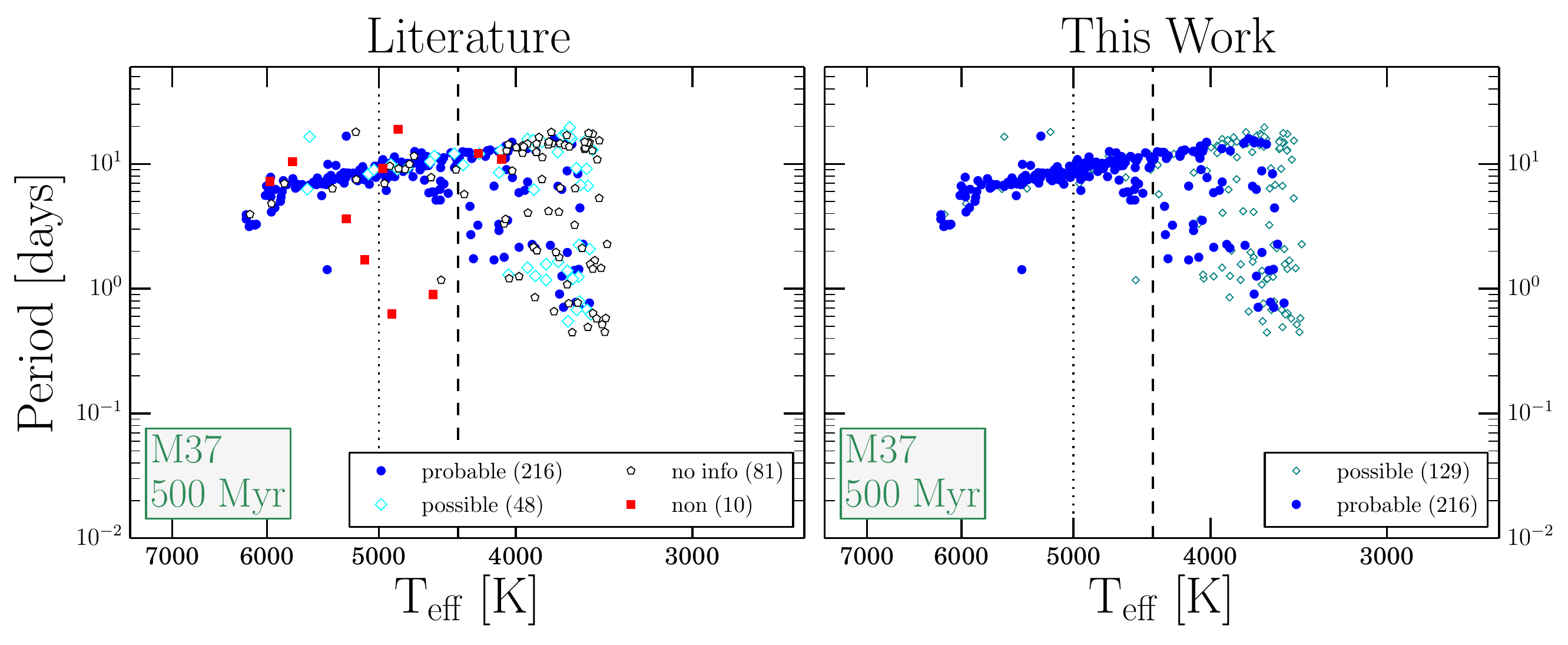}{0.50\textwidth}{e)}
          }
\gridline{
	    \fig{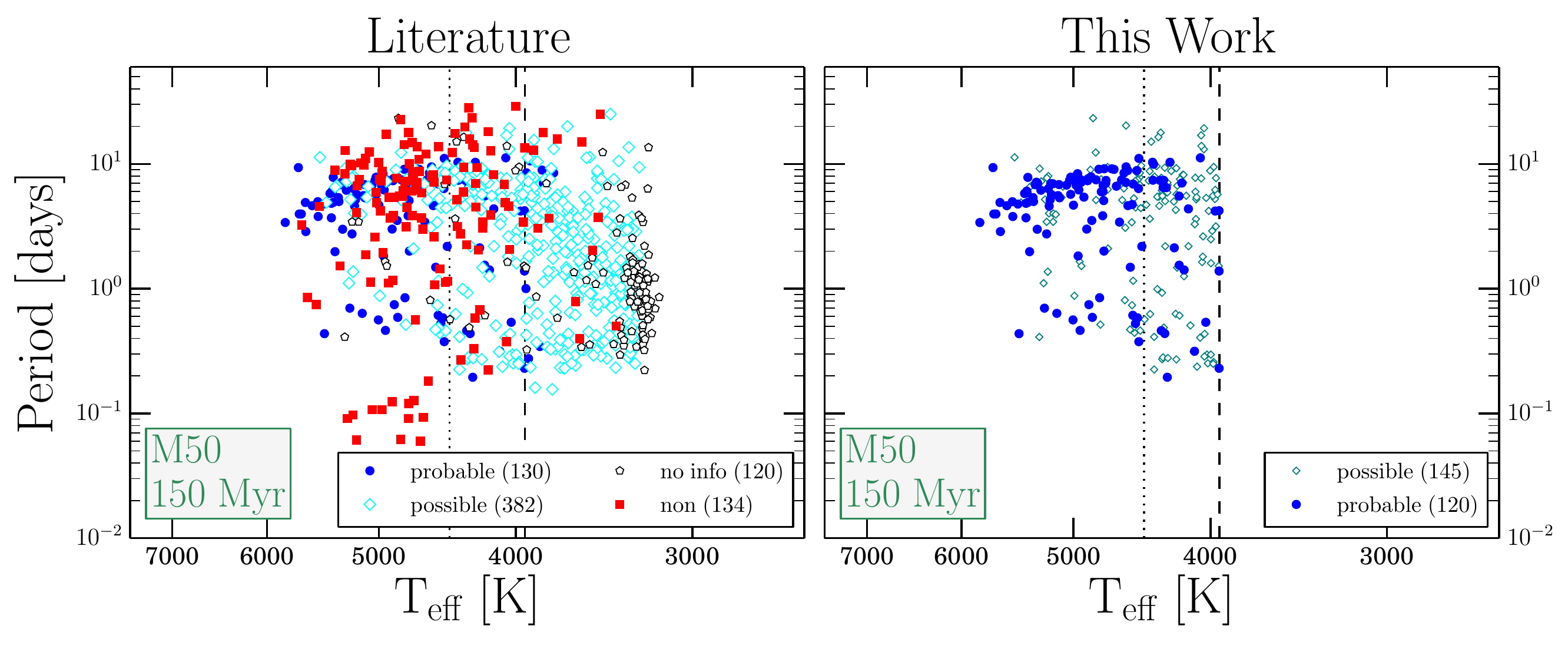}{0.50\textwidth}{c)}
	    \fig{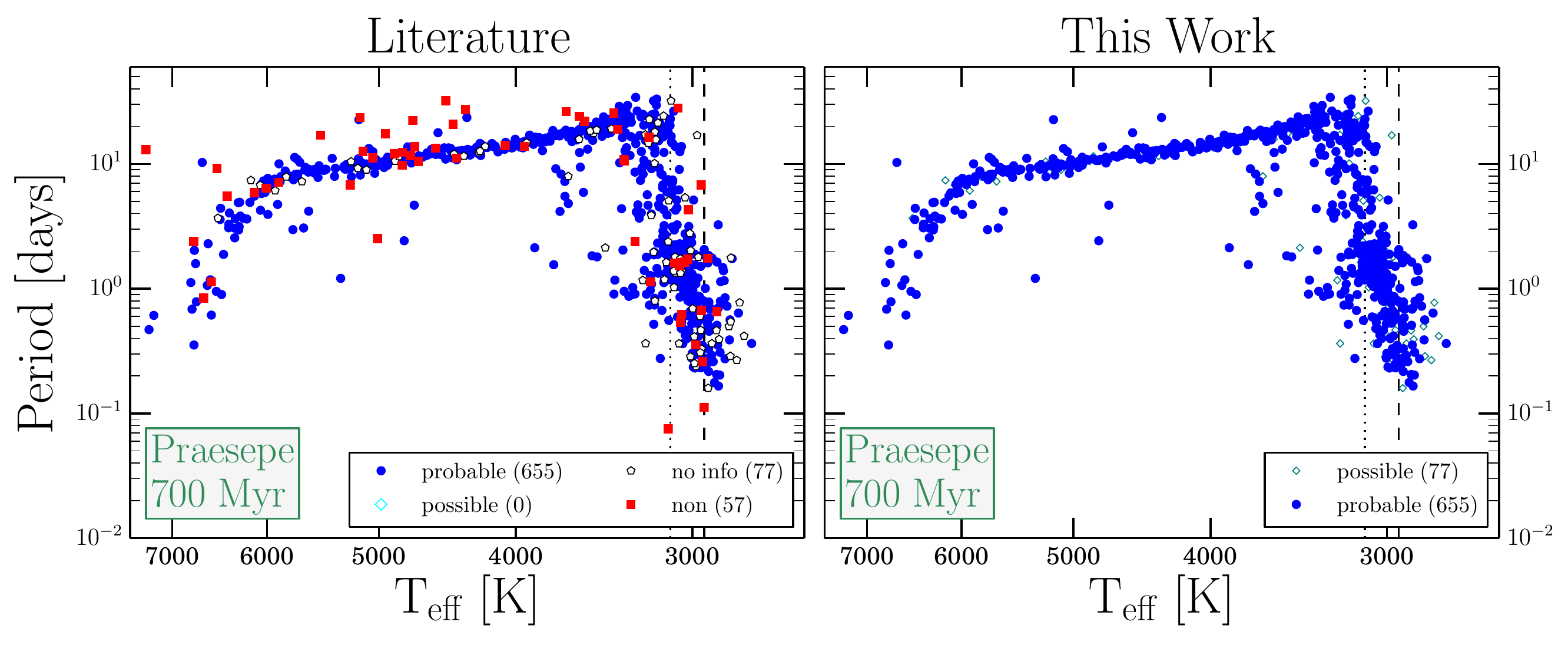}{0.50\textwidth}{f)}
          }
\gridline{
	    \fig{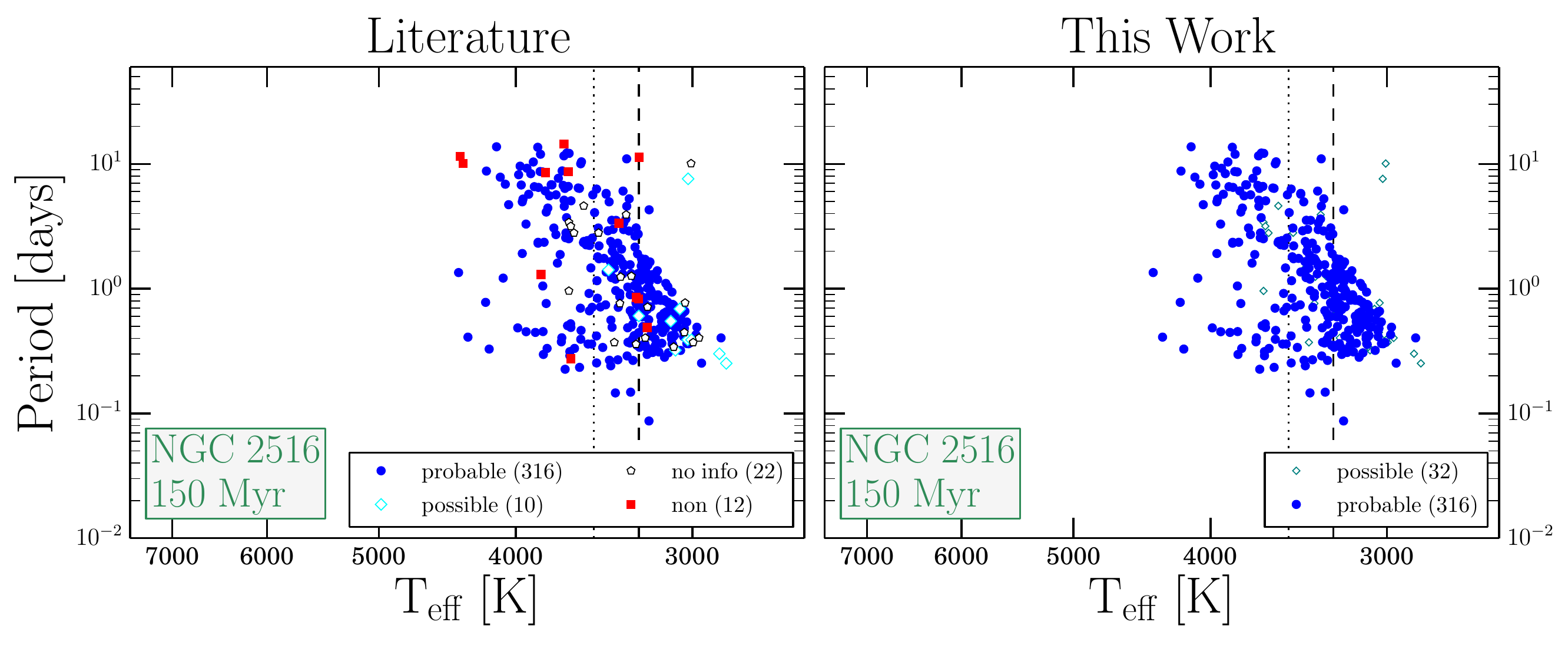}{0.50\textwidth}{d)}
	    \fig{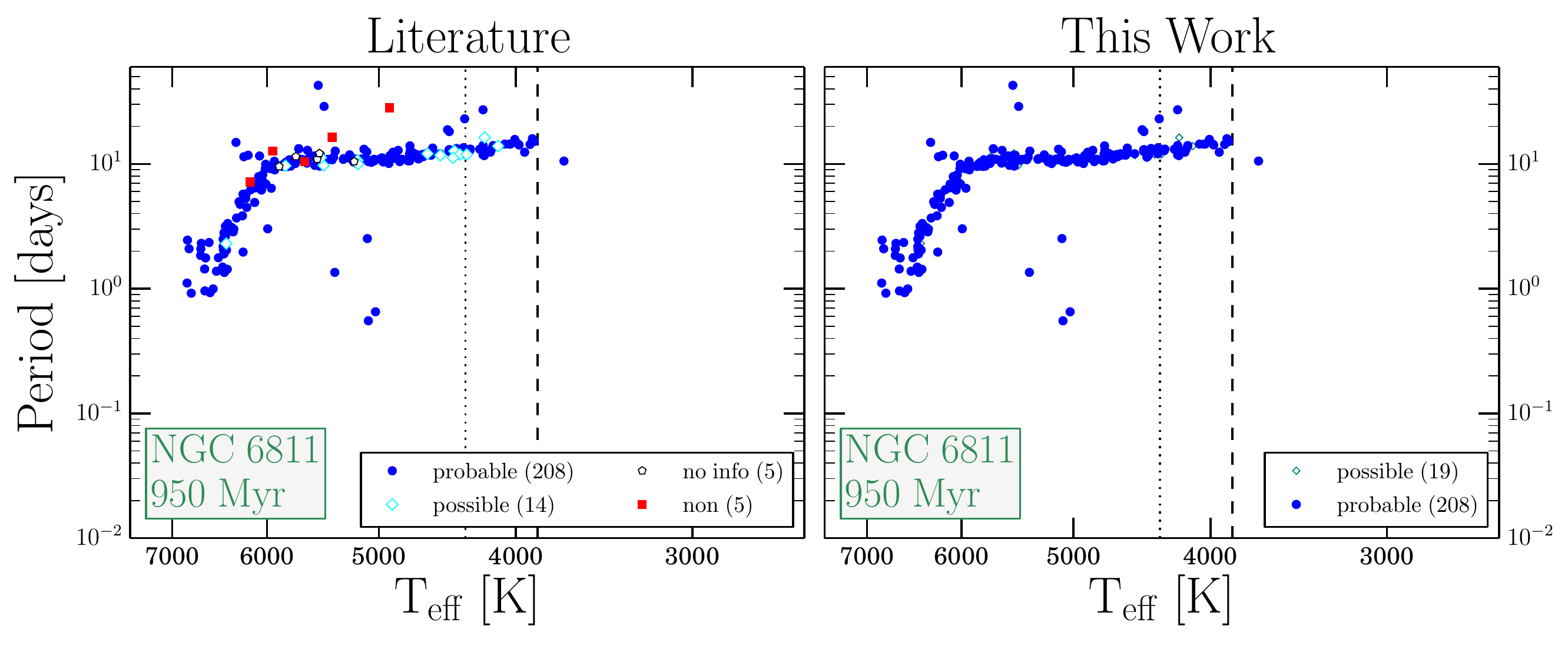}{0.50\textwidth}{g)}
          }
\caption{Period versus temperature diagrams for all the clusters we study. For each cluster we show two panels: the literature sequence (i.e., {\it pre} {\gaia} DR2 membership analysis; left panel) and its revised version reported in this work (i.e., {\it post} {\gaia} DR2; right panel). The literature sequences show all the stars reported in the original period references, and we color code them according to our astrometric classification: {\it no info} star (open black pentagons), {\it non} member (red squares), {\it possible} member (open cyan diamonds), or {\it probable} member (blue circles). The number of stars in each category is shown in the parenthesis accompanying the labels. In the revised sequences we do not show the {\it non} members to demonstrate the effects of removing the field contamination, and we combine the {\it no info} stars with the {\it possible} members (open green diamonds). The dotted and dashed lines correspond to the apparent $G=17.5$ and 18.5 mag limits from Figure \ref{fig:Figure_all_clusters_apparent_CMD} translated to {\Teff} coordinates. We observe varying degrees of {\it non} member contamination, with the clusters observed from the ground showing higher rates, and the more distant clusters having more uncertain revised sequences (see \S\ref{sec:results} and \S\ref{sec:discussion}).}
\label{fig:Figure_all_cluster_before_after_GDR2_rotseq}
\end{figure*}

Figure \ref{fig:Figure_all_cluster_before_after_GDR2_rotseq} shows the results of applying our astrometric classification to the rotational sequences of the periodic samples. For each cluster, we show both the literature sequence (i.e., {\it pre} {\gaia} DR2 membership analysis), and its revised version reported in this work (i.e., {\it post} {\gaia} DR2). The literature rotational sequences (left column) include all the candidate cluster members reported in the original period references, and we color code them according to our astrometric classifications. The revised sequences (right column), only show {\it probable} members, {\it possible} members, and {\it no info} stars, but do not show stars classified as {\it non} members. Since at this point we have no means to either confirm or reject the membership status of the {\it no info} stars, we simply add them to the {\it possible} members category in the revised sequences of Figure \ref{fig:Figure_all_cluster_before_after_GDR2_rotseq}. 

Before comparing the literature and revised rotational sequences in detail, we note that a strong artifact inherited from the {\gaia} astrometry is present in the membership classification of our most distant clusters. In {\clusterthree} (distance of $\sim$ 970 pc), for $T_{\text{eff}}\gtrsim 4500$ K, most of the stars are classified as either {\it probable} or {\it non} members. On the other hand, stars cooler than this limit are mostly classified as {\it possible} members (and furthermore, for $T_{\text{eff}}\lesssim 3500$ K the predominant category is {\it no info} stars). This is a direct consequence of the quality of the {\gaia} astrometry decreasing monotonically for fainter, cooler targets, where our classification method can no longer reliably identify {\it probable} or {\it non} members, and we instead classify most stars as {\it possible} members. A similar effect is observed in {\clusterfive} (distance of $\sim$ 1440 pc), with the limit being around $T_{\text{eff}}\simeq 4000$ K. This effect is absent in the rotational sequence of {\clusterseven} (distance of $\sim$ 1080 pc), but only because the periodic sample does not extend to such faint, cool stars.

We now examine Figure \ref{fig:Figure_all_cluster_before_after_GDR2_rotseq} in detail and discuss how the inclusion of the astrometric information affects the rotational sequence of each cluster independently:
\begin{itemize}

\item {\clusterone}: for this cluster we classify $\sim$ 25\% of the \citet{irwin08a} candidates as {\it non} members. By removing this contamination, the rotational sequence changes considerably. At a given {\Teff}, most of the {\it non} members actually correspond to stars with the longest periods, and after including our membership analysis, only 2 stars with period $>$ 10 days remain (one of them a {\it probable} member, while the other a {\it no info} star). Interestingly, the revised {\clusterone} sequence exhibits strong {\Teff}- and mass-dependent trends in the period distribution, with the mean period decreasing with decreasing {\Teff} for $T_{\text{eff}}<3500$ K. Given its young age ($\sim$ 35 Myr), we expect the period versus {\Teff} distribution of {\clusterone} to be strongly affected by processes that regulate stellar rotation at birth.

\item {\clustertwo}: we find the {\it non} member contamination to be small ($\lesssim$ 4\%), and that removing it, for the most part, does not empty specific regions in the rotational sequence. This finding is not surprising, as the {\clustertwo} stars studied with {\ktwo} by \citet{rebull16a}, \citet{rebull16b} and \citet{stauffer16} had been previously vetted by proper motion surveys. Further examination of the stars classified as {\it non} members reveals that $\approx$ 70\% of them have $\Delta_i$ values between 3 and 5 (i.e., are outside the cluster's $3\sigma$ ellipsoid in phase-space, but within $5\sigma$), suggesting that a less stringent astrometric selection could have classified them as possible cluster members. Additionally, for the {\clustertwo} we observe an interesting phenomenon (also observed in {\clustersix}): there is a large number of {\it no info} stars across the entire {\Teff} range. This is noticeably different from the cases of {\clusterthree} and {\clusterfive}, where the {\it no info} stars are heavily concentrated at the faint, cool end of the distributions. We further discuss this in \S\ref{sec:discussion}, but in short, these {\it no info} stars mainly correspond to photometric binaries, which points to biases in the {\gaia} DR2 astrometry.

\item {\clusterthree}: this cluster shows the largest differences of all when we compare the literature and revised rotational sequences. For $T_{\text{eff}} > 4500$ K, we find the literature sequence to have substantial {\it non} member contamination ($\approx 36\%$), and we observe that the different astrometric populations occupy particular regions of the rotational sequence. Firstly, we note that all of the stars with periods $\lesssim$ 0.1 day are cleanly identified as {\it non} members. This is \editfinal{perhaps} not surprising, as these periods most certainly do not correspond to rotational signals but rather pulsation signals from field stars. Secondly, we note that most of the long period stars ($\gtrsim$ 10 days) are actually classified as {\it non} members. Thirdly, after removing the field contamination, the {\it probable} members do form a rather clean sequence of rotation period as a function of {\Teff}, and both branches of slowly and rapidly rotating stars observed in the {\clustertwo} can now be seen in the revised {\clusterthree} sequence. For $T_{\text{eff}} \lesssim 4000$ K, the quality of the {\gaia} DR2 astrometry does not allow us to classify stars reliably, but we expect the contamination fraction to be of similar or higher significance than that of stars with $T_{\text{eff}} > 4500$. Given how uncertain this part of the diagram is, in the analysis that follows we simply discard the {\clusterthree} stars cooler than $\sim$ 4000 K (vertical dashed line in Figure \ref{fig:Figure_all_cluster_before_after_GDR2_rotseq}, i.e., the apparent $G=18.5$ mag limit translated to {\Teff} coordinate), and only consider the stars hotter than this limit.

\item {\clusterfour}: this cluster shows a very low contamination rate, with only $\approx 3\%$ of the stars being classified as {\it non} members, and $\approx 88\%$ of the \citet{irwin07b} stars being classified as {\it probable} members. We suspect this arises from the fact that this cluster is very rich and therefore dominates (in terms of number counts; see Figure \ref{fig:Figure_all_clusters_apparent_CMD}) over the field population in the CMD selection done by \citet{irwin07b}. Accordingly, both literature and revised rotational sequences are almost identical, with the period distribution showing strong mass dependent trends. We note, however, that the periodic {\clusterfour} stars \editfinal{span} a narrow range in temperature (and mass), and the sample is limited to $T_{\text{eff}} < 4400$ K ($M < 0.7 M_{\odot}$).

\item {\clusterfive}: for $T_{\text{eff}} \gtrsim 4000$ K, we find the {\it non} member contamination to be $\approx$ 4\%, which is small compared to the $\sim 20\%$ contamination rate expected by \citet{hartman08a}. Similarly to {\clusterfour}, we think this arises from the richness of the cluster compared to the field, in addition to the double CMD selection done by \citet{hartman08a}. Interestingly, however, the locations of many of the {\it non} members do not seem to be arbitrary. Most of them are stars that lie well off of the converged period sequence (at both longer and shorter periods), and are clearly separated from the {\it probable} members in the rotational sequence. This result implies that the converged sequence at this age ($\sim$ 500 Myr; \S\ref{sec:properties_clusters_and_stellar}) is actually stronger than previously thought. We note, nonetheless, that a few rotational outliers still survive the astrometric selection at both longer and shorter periods than the converged sequence. Additionally, and similarly to {\clusterthree}, {\clusterfive} also exhibits a clear break in the astrometric classification, in this case around 4000 K. For stars cooler than this limit, our method cannot confidently separate {\it probable} members from {\it non} members, and we instead classify most of them as {\it possible} members or {\it no info} stars. Nevertheless, and in contrast to {\clusterthree}, given how clean the \citet{hartman08a} membership was in the $T_{\text{eff}} \gtrsim 4000$ K regime, we expect most of the stars cooler than this value to be real {\clusterfive} members, and we do include them in the analysis that follows.

\item {\clustersix}: the {\it non} members correspond to $\approx$ 7\% of the sample, and this small contamination rate is not surprising given the previous CMD and proper motion vetting in the {\ktwo} stars studied by \citet{rebull17}. Further examination of these {\it non} member stars reveals that $\approx$ 40\% of them have $\Delta_i$ values between 3 and 5, and therefore a less stringent astrometric selection could have classified them as possible cluster members. Similarly to the {\clustertwo}, the {\it non} members do not seem to empty specific regions in the rotational sequence, with the exception of the stars located in the $ 4300 < T_{\text{eff}} < 5500$ K range. Many of these \editfinal{{\it non} members} have longer periods than the converged sequence, and the revised rotational sequence appears narrower than previously thought for this {\Teff} range at this age ($\sim$ 700 Myr; \S\ref{sec:properties_clusters_and_stellar}). We find this result to be similar to that of {\clusterfive}, \editfinal{hinting to a convergence of rotation rates that is considerably stronger than previously thought (see \S\ref{sec:discussion}).} Nonetheless, we also note that a few rotational outliers survive the astrometric analysis and are still present in the revised sequence. Finally, and similarly to the {\clustertwo}, we find that the {\it no info} {\clustersix} stars populate the entire {\Teff} range of the sequence. These stars mostly appear as photometric binaries in the CMD, and we further discuss them in \S\ref{sec:discussion}.

\item {\clusterseven}: this cluster shows a very low contamination rate, with only $\approx$ 2\% of the stars being classified as {\it non} members. This is consistent with expectations, given the samples that the {\clusterseven} rotation period catalogs were based on. \citet{meibom11a} vetted stars on the basis of RV data, \citet{curtis19a} combined CMD selections with a set of astrometric cuts to ensure consistency with the cluster's phase-space projections (although they did not take the stars' uncertainties or intrinsic cluster dispersion into account), and \citet{santos21} performed their lightcurve analysis based on our sample of {\it probable} members. While all of the above work together to produce a low contamination rate by construction, three of the five stars classified as {\it non} members are clear outliers when compared with the {\it probable} members sequence. Similar to what we see in the other old clusters ({\clusterfive} and {\clustersix}), however, the revised sequence does show a number of long period outliers, hinting to a rare channel that produces slow rotation for a small fraction of members in $\gtrsim 0.5$ Gyr-old clusters.
\end{itemize}

Finally, we point out a feature that is ubiquitous in all the clusters we study: the revised rotational sequences show a number of rapid rotators that survive all the astrometric cuts. These are stars that likely correspond to synchronized binaries, and they stand out more clearly in the older {\clusterfive}, {\clustersix}, and {\clusterseven} clusters with periods of $0.5\textendash 2$ days and $5000 < T_{\text{eff}} < 6000$ K.

After having analyzed how the clusters' rotational sequences change when the {\it non} member contamination is excluded, we emphasize that properly accounting for cluster membership remains a fundamental piece of empirical studies of stellar rotation. Incorrect or biased conclusions could be derived from rotational sequences that lack the appropriate vetting. Replicating our analysis with improved astrometry should further refine the astrometric classifications, and expand the mass and temperature range where the {\it probable} and {\it non} members can be reliably distinguished from each other.
\section{Discussion} 
\label{sec:discussion}

In this section, we use the periodic samples to study stellar rotation as a function of mass and age. 
\subsection{Rotation Data in the {\gaia} Era}
\label{subsec:discussion_rotation_in_gaia_era}

At this point, for every cluster in our sample, we have classified the periodic stars using our astrometric analysis (\S\ref{sec:method}), presented a set of revised properties that are needed to perform meaningful inter-cluster comparisons (\S\ref{sec:properties_clusters_and_stellar}), and discussed the effects that removing the field contamination has on the individual rotational sequences (\S\ref{sec:results}). Now, we combine all of these to construct an updated portrait of the evolution of stellar rotation. 

\begin{figure}[h]
\epsscale{1.20} 
\plotone{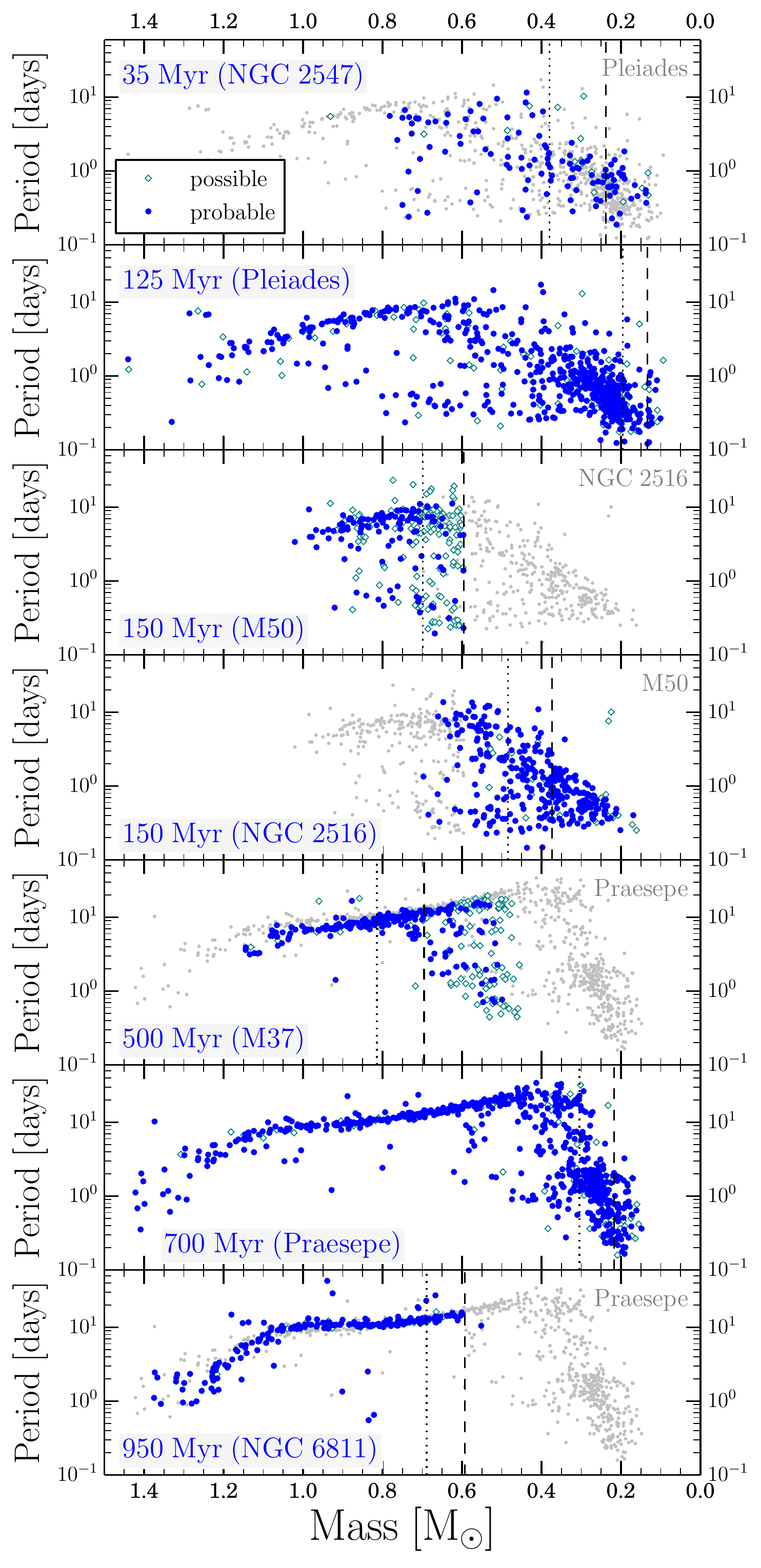}
\caption{Revised period versus mass diagrams for all the clusters we study (i.e., after having removed the {\it non} member contamination). The clusters are sorted by age, and the ages and masses are the values calculated in \S\ref{subsec:properties_stellar}. The points are color-coded in the same fashion as the revised sequences of Figure \ref{fig:Figure_all_cluster_before_after_GDR2_rotseq}. The dotted and dashed lines correspond to the apparent $G=17.5$ and 18.5 mag limits from Figure \ref{fig:Figure_all_clusters_apparent_CMD} translated to mass coordinates. To aid inter-cluster comparisons, in some panels we add an additional cluster in grey in the background. This figure represents the state-of-the-art data for studies of stellar rotation (see \S\ref{subsec:discussion_rotation_in_gaia_era}).}
\label{fig:period_mass_evolution_aftergaiadr2}
\end{figure}

We illustrate this in Figure \ref{fig:period_mass_evolution_aftergaiadr2}, where we show the revised period versus mass diagrams as a function of age. The color-coding is the same as the revised sequences of Figure \ref{fig:Figure_all_cluster_before_after_GDR2_rotseq}, and the {\it non} member contaminants have been removed. This represents the state-of-the-art for rotation studies in terms of clean samples with stellar masses, temperatures, and ages derived in a consistent scale. We make these data publicly available in Appendix \ref{appendix:app_tables_rotation_period_sample}.

One of the major results of our study can be seen in the rotational sequence of {\clusterthree} in Figure \ref{fig:period_mass_evolution_aftergaiadr2}. When the {\it non} member contamination is removed, the revised {\clusterthree} sequence clearly resembles that of the {\clustertwo}, as we would expect given their similar ages (150 and 125 Myr, respectively). Incidentally, the low mass end of the {\clusterthree} sequence coincides with the high-mass end of the co-eval {\clusterfour} sequence ($\approx$ 0.65 $M_{\odot}$). This is illustrated in their respective panels, where we show the {\clusterthree} sequence in blue and green, and the {\clusterfour} sequence in grey, and vice versa. Given their indistinguishable ages, the merged {\clusterthree} and {\clusterfour} data can provide a valuable comparison point for the benchmark {\clustertwo} cluster. Likewise, a similar comparison can be made for the older {\clusterfive} and {\clustersix} clusters. Ultimately, the above demonstrates a decisive finding: when careful membership analysis are performed, the rotational sequences that have been constructed from ground-based observations (e.g., {\clusterthree} by \citealt{irwin09}), can be as informative for stellar rotation studies as those constructed from space-based observations (e.g., the {\clustertwo} by \citealt{rebull16a}). \editfinal{In this context, future dedicated ground-based monitoring combined with careful astrometric selections using {\gaia} could provide unprecedented constraints for angular momentum evolution (e.g., \citealt{curtis20}).} This conclusion is particularly relevant in the {\it post} {\kepler} and {\ktwo} era, and considering that only a fraction of the {\tess} targets are being observed with long baselines. 

Regarding the field contaminants, our membership analysis yields varying degrees of contamination rates. As mentioned in \S\ref{sec:sample_selection_and_period_data}, we expected the clusters observed from the ground to show higher contamination rates compared to those observed from space. This turns out to be the case for {\clusterone} and {\clusterthree}, where we find the contamination to be $\approx$ 25\% and 36\%, respectively. While high, these values are lower than the $\sim 40\textendash 60\%$ rates anticipated by \citet{irwin08a} and \citet{irwin09}, which were based on predictions from Galactic models. For the rest of the sample we find the contamination rates to be lower, of order $\lesssim 5\%$ (see \S\ref{sec:results}).

Interestingly, although our membership study has predominantly removed rotational outliers at both long and short periods for several clusters, many outliers are nonetheless classified as {\it probable} members and remain in the revised sequences. An interesting example of this are the slowly rotating stars (periods $\approx 7\textendash 10$ days) in the young {\clusterone} (age of $\sim$ 35 Myr), which hints to strongly mass-dependent initial conditions for stellar rotation (see also \citealt{somers17b,rebull18,rebull20}). Similarly, all three $\gtrsim 0.5$ Gyr-old clusters show confirmed members that are rotating faster ($\sim$ 1 day) and slower ($\sim$ 20 days) than their slowly rotating branches. Now that the membership status of these outliers have been confirmed, their presence can no longer be ignored or attributed to field contaminants. 

\editfinal{The rapid rotators in systems with a converged sequence cannot be explained by single star evolution in current models. The most plausible explanation is that they have experienced tidal synchronization or are merger products. Regarding the slowly rotating outliers, a quick inspection reveals that most appear as typical cluster members in astrometric and photometric regards, but a fraction of them are photometric binaries in the CMDs \footnote{\editfinal{Particularly EPIC 211898294 in {\clustersix}, and KIC 9349106 and KIC 9656987 in {\clusterseven}.}}. We suspect that they could correspond to the long-period end of the distribution of tidally synchronized binaries (e.g., see \citealt{lurie17}). Another explanation could be that they were born with unusually low angular momentum, but given that the young {\clustertwo} sequence does not show many of these stars, we find this hypothesis less likely. Ultimately,} unless explained by modern theories of angular momentum evolution, \editfinal{these outliers} could potentially weaken the applicability of gyrochronology in field stars.
\subsection{A Revised Picture of Angular Momentum Evolution}
\label{subsec:discussion_trends}

\begin{figure}[h]
\epsscale{1.20} 
\plotone{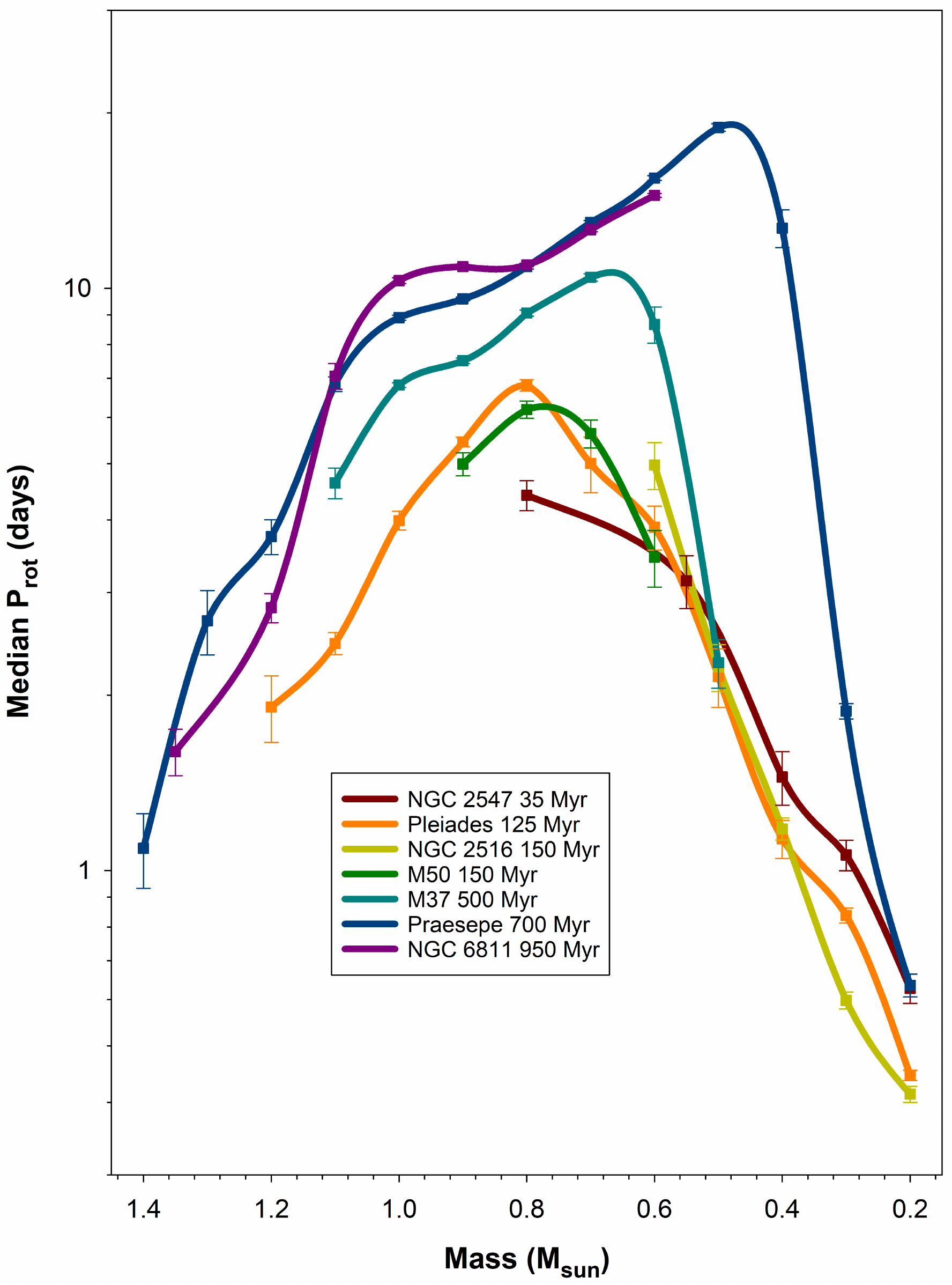}
\caption{Period versus mass diagram for the median rotators. The squares show the median periods and their standard errors, computed in $0.1 M_{\odot}$ bins. The solid lines show a spline fit to the data. Each cluster is shown as its own line, with the names, revised ages, and colors indicated in the legend. For masses $< 0.5 M_{\odot}$, the median rotators have a non-monotonic behavior, showing a spin up from ages of $\sim$ 35 Myr to $125\textendash150$ Myr, and a spin down afterwards at $\sim$ 700 Myr. For masses $> 0.5 M_{\odot}$, they show a monotonic spin down as a function of time until $\sim$ 700 Myr, after which the behavior becomes complicated (see \S\ref{subsec:discussion_trends}).}
\label{fig:period_mass_median_rotators}
\end{figure}

\begin{figure*}
\gridline{\fig{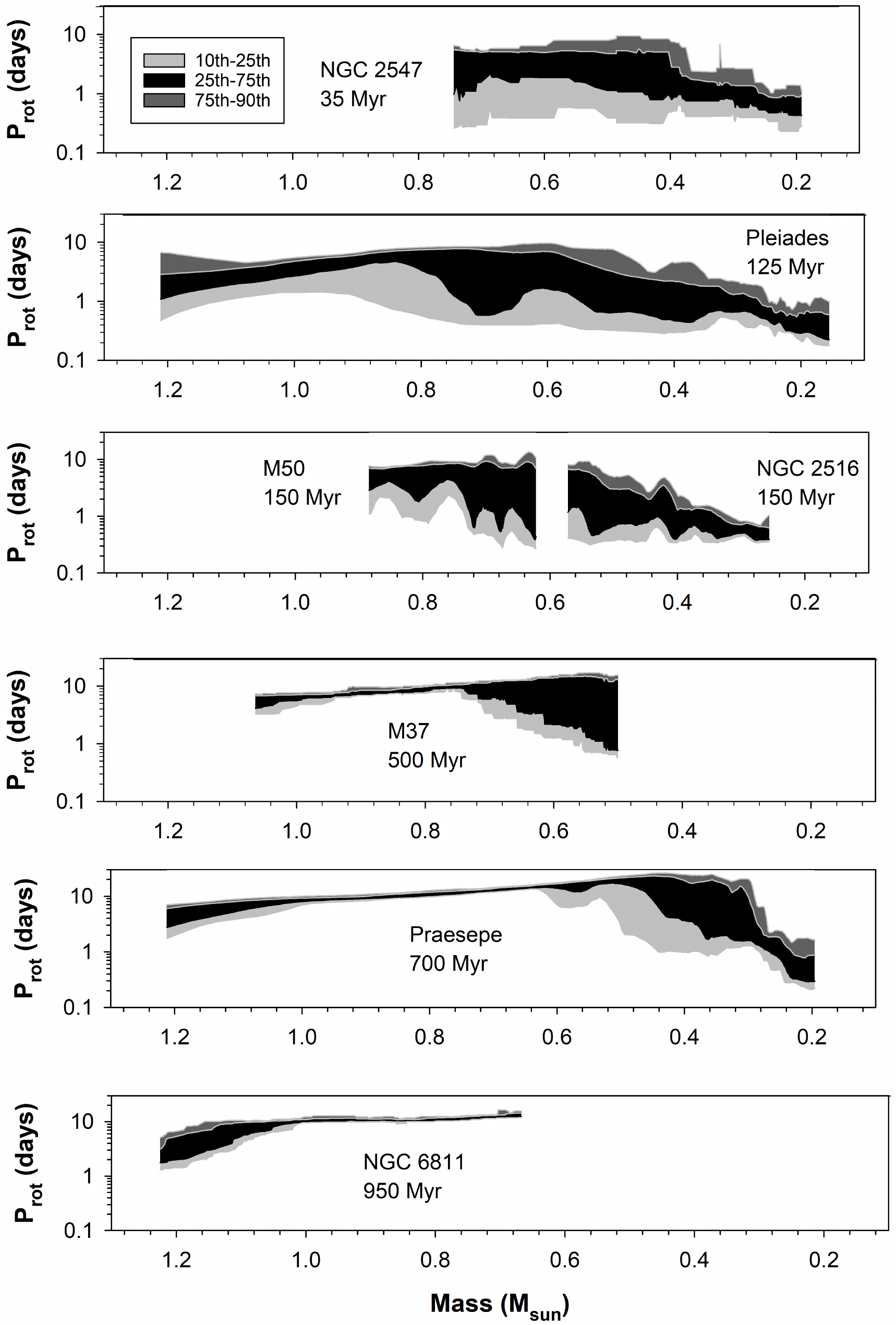}{0.70\textwidth}{}
          }
\caption{Distribution of rotation rates as a function of mass for the clusters considered in our sample. The 10th, 25th, 75th and 90th percentiles of the period distributions are computed for 40 star bins in data rank ordered by mass, and then boxcar smoothed. The region of rapid rotators is shown in light grey, that of intermediate rotators is shown in black, and that of slow rotators is shown in dark grey. {\clusterthree} and {\clusterfour} are shown in a combined panel, given their virtually identical ages and complementary mass ranges. We see a clear convergence of rotation rates into a narrow sequence in the older clusters, with the limit (i.e., lowest mass) of this converged branch progressively extending to lower masses for older ages (e.g., $\sim 0.85 M_{\odot}$ in the {\clustertwo} to $\sim 0.65 M_{\odot}$ in {\clustersix}). For masses below this, the interquartile range is actually broader in the older clusters than in the younger ones. This suggests that, at fixed mass, the rapid rotators lose less angular momentum (or experience a lower torque) than the slow rotators. Current models of angular momentum evolution do not reproduce this feature.}
\label{fig:period_mass_bands}
\end{figure*}

\begin{figure}[h]
\epsscale{1.20} 
\plotone{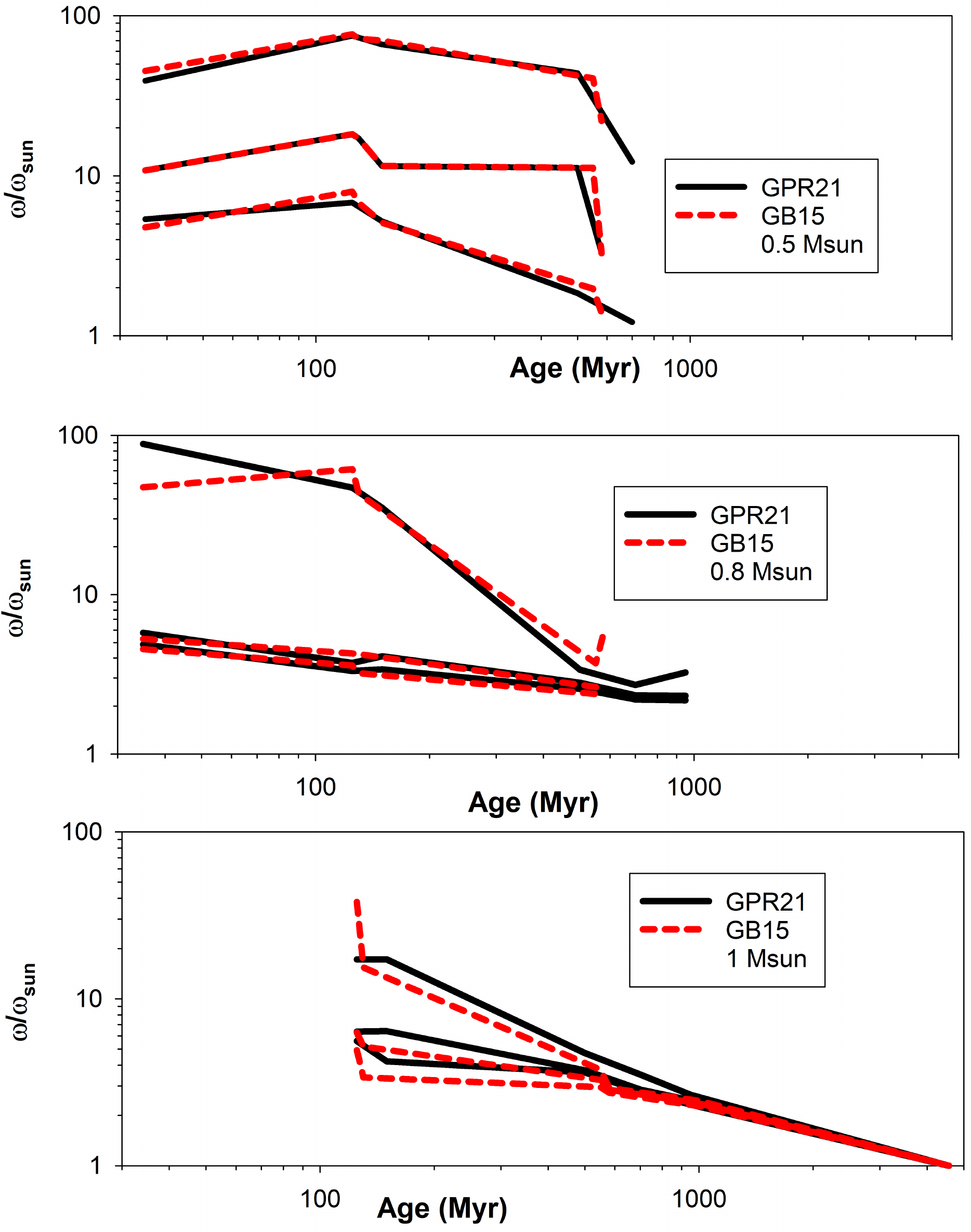}
\caption{Temporal evolution of the 25th, 50th, and 90th percentiles of the angular velocity distributions (normalized to the solar value). We show this for different masses, with 0.5 $M_{\odot}$ stars shown in the top, 0.8 $M_{\odot}$ stars shown in the middle, and 1 $M_{\odot}$ stars shown in the bottom. Our values are shown as the black solid line (GPR21), and the literature values from \citet{gallet15} are shown as the red dashed line (GB15).  Our analyses of the clusters' memberships and properties play a role in shifting our values with respect to the literature ones in both axes (see \S\ref{subsec:discussion_trends}).}
\label{fig:percentile_evolution_vs_GB15}
\end{figure}

We now quantify the trends that can be obtained from the revised sequences of Figure \ref{fig:period_mass_evolution_aftergaiadr2}. We display this in three different approaches, and these are shown in Figures \ref{fig:period_mass_median_rotators}, \ref{fig:period_mass_bands}, and \ref{fig:percentile_evolution_vs_GB15}. Additionally, we calculate the percentiles of the rotational distributions of our clusters in different mass bins, and we make them publicly available in Appendix \ref{appendix:app_table_percentiles}.

Figure \ref{fig:period_mass_median_rotators} shows, for each cluster in our sample, the period versus mass diagram of the median rotators in $0.1 M_{\odot}$ bins. This allows us to simultaneously analyze the mass and age dependence of stellar rotation. For stars with masses below $0.5 M_{\odot}$, the median rotation periods for the young {\clusterone} (35 Myr) are longer than those seen in the older {\clustertwo} and {\clusterfour} clusters  (125 and 150 Myr, respectively). This is consistent with expectations, considering that stars of these masses take a few hundred Myr to reach the ZAMS, and are therefore still contracting by the age of {\clusterone}. By the age of {\clustersix} (700 Myr), these stars have spun down in a strongly mass-dependent fashion, with lowest masses still showing periods below 1 day, and $\sim 0.4 M_{\odot}$ stars rotating at $\gtrsim$ 10 days. 

The stars more massive than $0.5 M_{\odot}$, on the other hand, show a monotonic behavior with time in most of the age range probed by our clusters. They consistently spin down as they age from 35 to 700 Myr, and their median rotator trends in Figure \ref{fig:period_mass_median_rotators} do not intersect each other, in agreement with expectations from a Skumanich-type spin-down (i.e., period $\propto \text{age}^{1/2}$). The notable exception to this, however, is the comparison of {\clustersix} and {\clusterseven}. For these two clusters, their rotational sequences do not seem to be simple translations of each other to longer or shorter periods. Although their age difference is non-negligible ($\sim$ 250 Myr), their median rotators are virtually overlapping at $\sim$ 1.1 $M_{\odot}$ and in the $0.8\textendash0.6 M_{\odot}$ range. The latter of these has already been noted by \citet{curtis19a} (see also \citealt{meibom11a}), and a similar overlapping was reported by \citet{agueros18} for a smaller sample in the 1.4 Gyr-old cluster NGC 752. In terms of the interpretation of this feature, \citet{agueros18} and \citet{curtis19a} have formulated it as a temporary epoch of stalling in the spin-down of K-dwarfs, and \citet{spada20} have proposed that it arises from the competing effect between magnetic braking and a strongly mass-dependent internal redistribution of angular momentum. We leave a detailed examination of this as future work, but given the considerable difference in metallicity between both clusters ([Fe/H]$=+0.16$ dex for {\clustersix} and $+0.03$ dex for {\clusterseven}; \citealt{netopil16}), we highlight the importance of incorporating chemical composition in comprehensive models of angular momentum evolution (e.g., \citealt{amard20a,amard20b,claytor20}).

Figure \ref{fig:period_mass_median_rotators} can also be used to remark upon global properties of stellar rotation. First, regardless of the age, no sharp transition is seen in the rotation periods near the boundary between partially convective and fully convective stars ($\sim 0.35 M_{\odot}$). Second, as illustrated by {\clusterone} in our sample, and by the young Upper Scorpius and Taurus associations in \citet{somers17b}, \citet{rebull18}, and \citet{rebull20}, the initial conditions of stellar rotation are strongly mass-dependent, and need to be accounted for in models of angular momentum evolution. Third, when considering the \editfinal{young ages probed by our sample ($\sim$ 35 to 200 Myr)}, we note that the stars in the $1.0\textendash0.6 M_{\odot}$ range populate a global maximum in terms of rotation periods. In other words, the median rotators in this mass interval never rotate more rapidly than $\sim 3\textendash4$ days. Given the intimate connection between stellar rotation and activity (e.g., \citealt{wright11,wright18a,lehtinen20}), where more rapidly rotating stars tend to expose their planets to higher levels of potentially harmful radiation, this feature could provide an optimal window in the search for habitable worlds. 

In Figure \ref{fig:period_mass_bands}, we show the rotational sequences for the clusters we study, but in this case separating the data into different percentiles of rotation. For each cluster, across the mass range where they have period information, we show a band of rapid rotators (star within the 10th to 25th percentiles of the period distribution; light grey region), intermediate rotators (25th to 75th percentiles; black region), and slow rotators (75th to 90th percentiles; dark grey region). At young ages, we observe a well-defined upper limit in the rotation periods, with virtually no stars showing periods longer than $\sim$ 10 days. At late ages, we clearly see the convergence of rotation periods to a tight sequence in a heavily mass-dependent fashion, in agreement with expectations. The convergence in our revised sequences is so strong that the upper and lower edges of the distribution in {\clusterfive}, {\clustersix}, and {\clusterseven} in the $1.0\textendash0.8M_{\odot}$ range are barely visible. Additionally, in both Figures \ref{fig:period_mass_median_rotators} and \ref{fig:period_mass_bands}, there is another important clue about the torques from magnetized winds. For all clusters, there is: \editfinal{1)} a characteristic mass where the distribution has collapsed down into a narrow range \editfinal{(the un-saturated domain, where the torque is $\mathrm{d}J/\mathrm{d}t \propto \omega^3$)}; and \editfinal{2)} a mass range just below it where stars still retain a range of surface rotation rates \editfinal{(with the rapid rotators still being in the saturated domain, where the torque is $\mathrm{d}J/\mathrm{d}t \propto \omega$)}. For this latter mass range, the width of the interquartile range is broader in the older systems ($\gtrsim 500$ Myr) than it is in the young ones ($\lesssim 150$ Myr). This indicates a relative divergence in the surface rotation rates, in contrast to the convergence that is predicted by canonical models of angular momentum loss (e.g., \citealt{vansaders13,matt15}).

In other words, the data in Figure \ref{fig:period_mass_bands} appear to require lower torques for rapid rotators than for intermediate rotators, which acts to create the bimodal distribution that has been noted in the literature (and which is now clearly seen in our sample). One possible explanation for this, as proposed by \citet{garraffo18}, is that there is a change in field topology, such that rapid rotators have more complex magnetic field configurations than intermediate rotators, and are therefore less efficient at losing angular momentum. Although this concept is interesting, the data could also be fit with a wind model that incorporates a super-saturation mechanism, where the overall field strength declines. Alternatively, it could indicate a dependence of the convective overturn timescale on the rotation rate at fixed mass, as loss laws are typically parametrized by the ratio of the rotation period to this timescale (i.e., the Rossby number). Ultimately, a number of additional physical properties that are involved in the loss of angular momentum could be influencing the empirical sequences (e.g., mass loss rate, dipole field strength, Alfv\'en radius). It will be informative to compare different proposed mechanisms, but an adjustment to canonical models is clearly required. We note that the late time behavior of the distribution is not sensitive to the treatment of this feature, as older stars have lost memory of their initial conditions, and typical semi-empirical approaches are tuned to reproduce the spin-down of the slow and rapid rotator branches of the data. However, the properties of intermediate rotators in systems that retain rapid rotators will not be correctly modeled by classical methods.

Finally, in Figure \ref{fig:percentile_evolution_vs_GB15}, we show the temporal evolution of the rotational percentiles for three different masses ($0.5, 0.8$, and $1 M_{\odot}$). We frame this as a comparison with \citet{gallet15}\footnote{Note that no \editfinal{rotational} data had been published for the $0.8M_{\odot}$ stars in {\clusterseven} by the time \citet{gallet15} performed their analysis.} in terms of the 25th, 50th, and 90th percentiles of the distribution of angular velocities ($\omega=2\pi/P_{\text{rot}}$), normalized to the solar value ($\omega_{\odot}=2.87 \times 10^{-6} \text{s}^{-1}$). Our membership analysis plays a role in shifting our values with respect to the literature ones in both axes, as the revised cluster properties change the ages (\S\ref{sec:properties_clusters_and_stellar}), and the revised rotational sequences change the angular velocity distributions (\S\ref{sec:results}). In general, our percentile evolution appear less noisy and shows smoother patterns. In the $0.5$ and $0.8 M_{\odot}$ stars, our data capture the same broad trends. Importantly, the overall slow rotation in young stars remains, and this provides further evidence for the need of core-envelope decoupling in angular momentum models (e.g., \citealt{denissenkov10}). Before our work, the classification of all of the slow rotators in {\clusterone} as field contaminants was a real possibility, and while our analysis did remove a number of them, some still remain. For the $1 M_{\odot}$ stars, our data suggests a faster rotation in all percentiles (as our analysis has predominantly removed slow rotators), and a globally steeper spin-down slope that remains approximately constant (and similar to the Skumanich value) in the entire age range.
\subsection{{\gaia} Biases against Photometric Binaries}
\label{subsec:discussion_photometric_binaries}

\begin{figure*}
\gridline{\fig{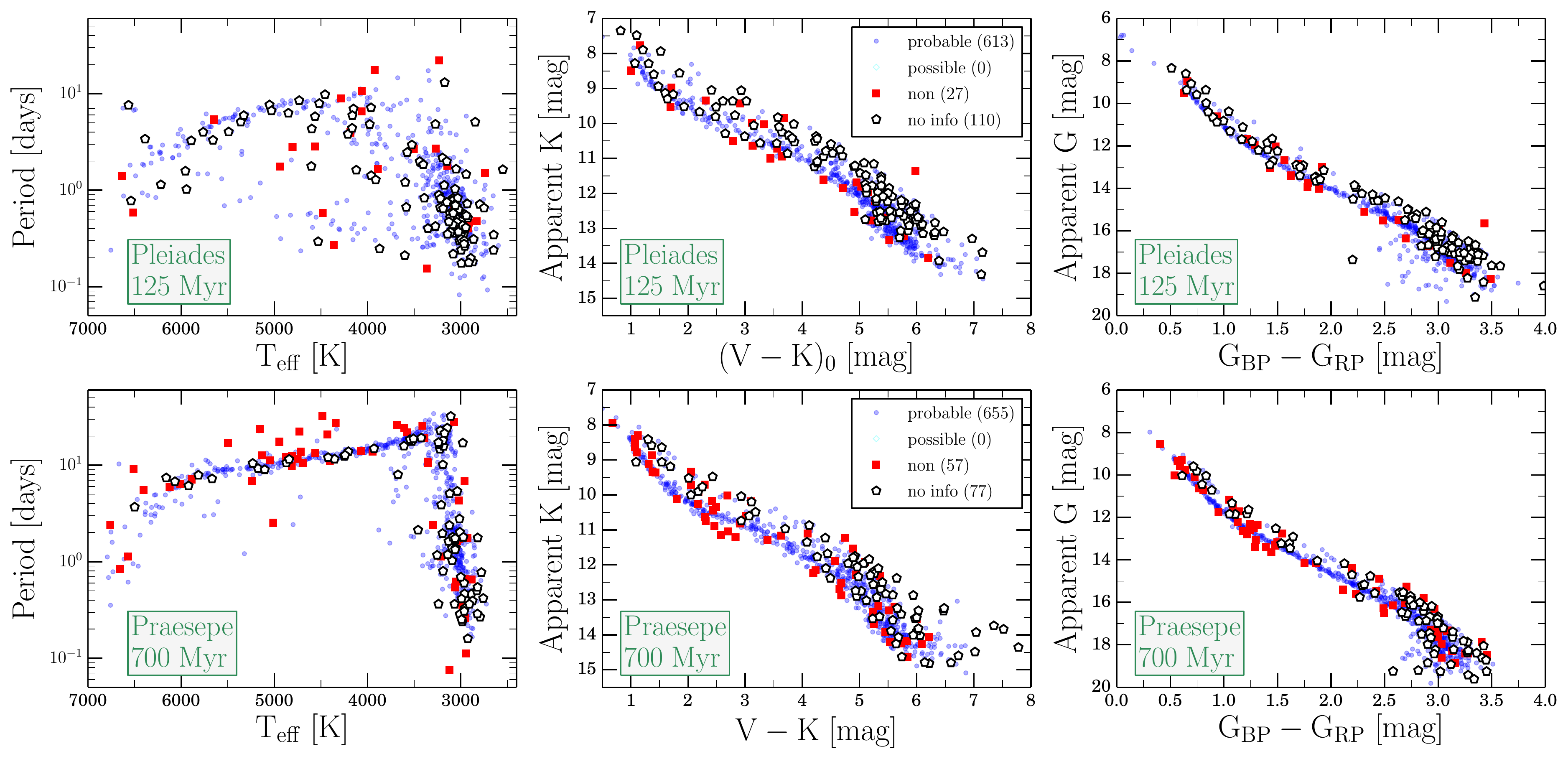}{1.00\textwidth}{}
          }
\caption{Period versus temperature (left column), apparent $V$ versus $V-K$ CMD (middle column), and apparent $G$ versus $G_{\text{BP}}-G_{\text{RP}}$ CMD (right column) for the nearby {\clustertwo} (top row) and {\clustersix} (bottom row) clusters. We take the $V$ and $K$ photometry from \citet{rebull16a} and \citet{rebull17}. The color and symbol coding are the same from the literature sequences of Figure \ref{fig:Figure_all_cluster_before_after_GDR2_rotseq}. For the purposes of this figure, we highlight the {\it no info} stars and {\it non} members over the {\it probable} members. For both clusters, the stars classified as {\it no info} stars appear predominantly as photometric binaries on both CMDs, hinting to biases in the {\gaia} astrometry for these systems. Notably, while in the $\sim$ 125 Myr {\clustertwo} sequence many {\it no info} stars correspond to rapid rotators, almost all of them have converged onto the slowly rotating branch in the $\sim$ 700 Myr {\clustersix} sequence. This indicates that, although unresolved, these likely binary systems are spinning down as single stars do (see \S\ref{subsec:discussion_photometric_binaries}).}
\label{fig:rotational_sequence_and_CMD_highligh_noinfo}
\end{figure*}

Investigating the behavior of binary stars in empirical rotation studies is important to better understand stellar populations and angular momentum evolution. In the regime of wide separations, binaries can be used to place novel gyrochronology constraints in unexplored age and metallicity domains \citep{chaname12,janes17,godoyrivera18}. In the regime of close separations, recent studies have found a correlation between rapid rotation and field binaries \citep{simonian19,simonian20}. Specifically in the M-dwarf domain, \citet{stauffer18} found that in the young Upper Scorpius association ($\sim$ 8 Myr) and the {\clustertwo} cluster, the stars with multiple period measurements from {\ktwo} are predominantly photometric binaries (see also \citealt{rebull16a}, \citealt{rebull17}, and \citealt{tokovinin18}). Furthermore, \citet{stauffer18} showed that the rotation periods of these young, unresolved binaries, are much shorter as well as similar to each other than otherwise expected for single M-dwarfs. This hints at the crucial role that binarity plays in star formation and the rotation rates imprinted at birth (see also \citealt{messina19}).

In this context, in Figure \ref{fig:rotational_sequence_and_CMD_highligh_noinfo} we examine the rotational sequences and CMDs of the {\it no info} stars in our most nearby clusters (i.e., the targets that did not pass our astrometric quality cuts; see \S\ref{subsubsec:method_gaia_membership}). In the revised rotational sequences of {\clusterthree} and {\clusterfive}, the stars classified as {\it no info} predominantly occupy their very lowest {\Teff} (and therefore apparent magnitude) values, which indicates that their classification is purely due to low quality astrometry. In contrast to this, for the {\clustertwo} and {\clustersix}, the {\it no info} stars span the entire {\Teff} range (see \S\ref{sec:results}), and we would have expected them to have high-quality {\gaia} astrometry. Considering the previous selections made by \citet{rebull16a} in the {\clustertwo} and by \citet{rebull17} in {\clustersix}, and given the low contamination rates found for these systems in \S\ref{sec:results}, we expect most of these {\it no info} stars to be real members of these nearby clusters.

The CMDs of Figure \ref{fig:rotational_sequence_and_CMD_highligh_noinfo} show that the {\it no info} stars appear, predominantly, as photometric binaries with respect to the sequences of {\it probable} members in both the {\clustertwo} and {\clustersix}. This result hints to biases in the {\gaia} DR2 data against these unresolved systems, even with conservative quality cuts, where their separation is seemingly too close for the current astrometry to properly distinguish their orbital and systemic motions. Interestingly, many of these {\it no info} stars appear as rapid rotators in the {\clustertwo} rotational sequence. By the age of {\clustersix}, however, they are overlapping with the bulk of the {\it probable} members on the slowly rotating branch. This indicates that, while their separations are close enough to be unresolved in {\gaia}, these binaries are not synchronized and they are actually spinning down in a similar fashion as the single stars do. This finding is similar to that of \citet{stauffer18}, who reported that the rapid rotation of stars in binaries with respect to single stars is gone by the age of {\clustersix}, but in this case we extend it to higher mass stars, beyond the M-dwarf regime. 

Figure \ref{fig:rotational_sequence_and_CMD_highligh_noinfo} also provides an interesting distinction in our membership classification. While presumably some of the stars classified in the {\it no info} category could turn out to be field contaminants, the stars classified as {\it non} members appear to occupy different locations in the rotational sequences and CMDs. Most of these {\it non} members appear below the track of {\it probable} members on the CMD, and correspond to rotational outliers in the converged branch of {\clustersix}. In addition to their astrometric distinction, we take this as further confirmation that the {\it non} members and {\it no info} stars correspond to markedly different populations. 

Finally, although our astrometric selections were designed to avoid biases against binary stars, we conclude that newer {\gaia} data will be needed to fully resolve these photometric binaries. In the meanwhile, while not being classified as {\it probable} members by our membership analysis, we advise future rotation studies of binary stars to include these {\it no info} stars in their samples for completeness purposes. 
\section{Conclusions} 
\label{sec:conclusions}

Open clusters have played a key role in stellar rotation studies, but their memberships are often uncertain, and the expected contamination rates can reach up to $\sim$ 60\%. In this paper, we use the {\gaia} DR2 astrometry to revise the memberships of several open clusters with rotational data (Table \ref{tab:list_of_clusters}), and provide an updated portrait of stellar rotation as a function of mass, temperature, and age. 

For a given cluster, we carry out an unrestricted search for cluster members in the {\gaia} data by performing astrometric modeling in parallax and proper motion space. From this, we calculate membership probabilities and identify {\it probable} cluster members, {\it non} members (i.e., field stars), and an intermediate category of {\it possible} members (typically faint stars with large astrometric uncertainties). An additional category denominated {\it no info} stars is assigned to stars that do not have enough astrometric information to be classified. Although our membership selection only includes astrometric (but not photometric) data, the CMD projections of the {\it probable} members show clean isochrone-like sequences for all the clusters (Figure \ref{fig:Figure_all_clusters_apparent_CMD}). 

We use the homogeneous {\gaia} photometry to derive revised cluster properties (age, distance modulus, reddening; see Table \ref{tab:revised_cluster_properties}) by comparing with state-of-the-art evolutionary models. Importantly, we find our revised ages to be in good agreement with {\it pre} {\gaia} literature estimates, and in some cases, in disagreement with more recent analyses (see \S\ref{sec:properties_clusters_and_stellar}). We further exploit the {\gaia} photometry and compare it with models to calculate stellar masses and temperatures for our {\it probable} and {\it possible} cluster members. All of this allows us to perform an analysis of the clusters' rotational sequences in a common mass, temperature, and age scale, something that has been missing in previous studies.

We crossmatch the {\gaia} stars with the rotation period catalogs, and use our membership analysis to remove the {\it non} member contaminants from the rotational sequences of the clusters (Figure \ref{fig:Figure_all_cluster_before_after_GDR2_rotseq}). We find varying degrees of {\it non} member contamination, with some of the clusters observed from the ground showing high rates (e.g., $\approx 25\%$ and $36\%$ for {\clusterone} and {\clusterthree}), and the clusters observed from space showing lower rates (e.g., $\approx 4\%$ for the {\clustertwo}). {\clusterthree} is an exceptional case, where our revised sequence is considerably different from the previous literature one, and its revised sequence now clearly resembles that of the benchmark and similar-age {\clustertwo} cluster.

We present an updated and self-consistent portrait of the evolution of stellar rotation, and quantify important trends that can be obtained from it. Our most important findings in these regards are:
\begin{itemize}
\item Once the {\it non} member contamination has been removed, the rotational sequences of clusters observed from the ground can be as constraining as those observed from space (Figure \ref{fig:period_mass_evolution_aftergaiadr2}). This is particularly relevant for the {\it post} {\kepler} and {\ktwo} era.
\item Although our membership analysis has predominantly removed rotational outliers in all the clusters, many of these still survive at both short and long periods (Figure \ref{fig:period_mass_evolution_aftergaiadr2}). It is likely that \editfinal{some of these correspond to synchronized binaries (especially the rapid rotators), but most slow rotators appear as typical single stars in astrometric and photometric regards.} \editfinal{An alternate explanation for slow rotators would be that there is a small population with very low birth angular momentum.} \editfinal{In any case, these stars} can no longer be attributed to field contamination and they need to be explained by future theories of angular momentum evolution.
\item \editfinal{At young ages,} stars in the $1.0\textendash0.6 M_{\odot}$ range populate a global maximum in terms of rotation periods. Given the strong rotation-activity connection in stars, this potentially provides their planets with an optimal window for habitability (Figure \ref{fig:period_mass_median_rotators}).
\item At the age of the young {\clusterone} cluster ($\sim$ 35 Myr), there is a clear-cut set of maximum periods ($\sim 10$ days), \editfinal{which provides clues about the initial conditions of rotation}. At old ages ($\gtrsim 500$ Myr), stars in the $1.1\textendash0.6M_{\odot}$ rage converge to a sequence that is even narrower than previously observed (Figure \ref{fig:period_mass_bands}).
\item In the saturated domain, where stars have not yet converged into a uniform sequence at fixed mass, the separation (in terms of rotation periods) between the intermediate and rapid rotators increases with age. This feature is not predicted by classical models, and we interpret it as the latter group experiencing lower torques compared to the former one (Figure \ref{fig:period_mass_bands}).
\item We find biases in the {\gaia} data against photometric binaries, which predominantly appear as rapid rotators at young ages, but nonetheless spin-down as single stars do (Figure \ref{fig:rotational_sequence_and_CMD_highligh_noinfo}).
\end{itemize}

The data products of this paper will be a useful reference for the community. For the clusters in our sample, we make publicly available both the catalog of {\it probable} and {\it possible} members (regardless of whether they have period measurements; Appendix \ref{appendix:app_tables_of_membership_probabilities}), as well as the catalog of the periodic stars (regardless of their membership classification; Appendix \ref{appendix:app_tables_rotation_period_sample}). Additionally, we report the percentiles of the rotational distributions as a function of mass and age, which can be used to calibrate future models of angular momentum evolution (Appendix \ref{appendix:app_table_percentiles}).

\editfinal{In a broader sense}, the field of stellar rotation is entering a new era thanks to the wealth of photometric and astrometric data provided by space-based missions (e.g., {\kepler}, {\ktwo}, {\tess}, and {\gaia}). As shown in this work, these can be adequately complemented with ground-based observations. Still, although the rotational sequences of numerous star-forming regions, stellar associations, and open clusters exist in the literature, properly accounting for their memberships remains a crucial part of exploiting them in full. Extensions of our work would increase the sample size and take advantage of improved astrometry and RV measurements (e.g., {\gaia} DR3) to perform full 6D kinematic analyses.

Ultimately, future rotation studies should attempt to answer the several questions that remain open. \editfinal{Besides the ones mentioned in Section \S\ref{sec:intro},} some of these include: the importance of shifts in magnetic field topology on the magnetic braking \citep{metcalfe19b,metcalfe19a}; the role that bright faculae play in stellar variability (e.g., \citealt{montet17,reinhold20}); and the deep connection between rotation and the various aspects of stellar activity (e.g., \citealt{stelzer13,zhang20,dixon20,godoyrivera21,ilin21}).
\acknowledgments

We thank \^Angela Santos, Rafael Garc\'ia, and Savita Mathur for providing period measurements for the {\clusterseven} stars. We also thank John Stauffer for useful discussions.

DGR and MHP acknowledge support from NASA grant 80NSSC19K0597.

This work has made use of data from the European Space Agency (ESA) mission {\it Gaia} (\url{https://www.cosmos.esa.int/gaia}), processed by the {\it Gaia} Data Processing and Analysis Consortium (DPAC, \url{https://www.cosmos.esa.int/web/gaia/dpac/consortium}). Funding for the DPAC has been provided by national institutions, in particular the institutions participating in the {\it Gaia} Multilateral Agreement. 

This publication makes use of data products from the Two Micron All Sky Survey, which is a joint project of the University of Massachusetts and the Infrared Processing and Analysis Center/California Institute of Technology, funded by the National Aeronautics and Space Administration and the National Science Foundation.

\appendix
\restartappendixnumbering
\editfinal{In Appendix \ref{appendix:app_data_period_expansions}, we discuss the systems that were discarded from this analysis but that could be studied in future works}. In Appendix \ref{appendix:app_comparison_parameters}, we report the astrometric cluster parameters we derive, and compare them with the literature values. In Appendix \ref{appendix:app_tables_of_membership_probabilities}, we report a table with the {\gaia} DR2 information for the {\it probable} and {\it possible} cluster members (independently of whether they belong to the periodic sample or not). In Appendix \ref{appendix:app_phase_space_projections_all_clusters}, we show the phase-space projections of the {\it probable}, {\it possible}, and {\it non} members for all the clusters, and provide a brief astrometric analysis for each of them. In Appendix \ref{appendix:app_mass_and_Teff_comparison}, we show the comparison of our mass and temperature values with independent estimates from the literature. In Appendix \ref{appendix:app_tables_rotation_period_sample}, we report a table with the main information and properties for the stars in the periodic samples. In Appendix \ref{appendix:app_table_percentiles}, we report the percentiles of the rotation period distributions as a function of mass and age.

\section{\editfinal{Potential Expansions to our Cluster Sample}} 
\label{appendix:app_data_period_expansions}
\restartappendixnumbering

There are a number of systems that we initially considered including in our sample given their high scientific interest, but that we discarded for specific reasons, which we discuss below. We leave an equivalent investigation of them as future work to be done with a more sophisticated underlying model, and when higher precision astrometry becomes available.

\begin{itemize}

\item $\rho$ Ophiuchus, Taurus, Upper Scorpius, and h Per: \citet{rebull18}, \citet{rebull20}, and \citet{moraux13} reported periods for these young associations (ages of $\sim$ 1, 3, 8, and 13 Myr, respectively). Given their young ages, these systems offer interesting pre main-sequence stellar rotation constraints. \editfinal{For the first three of these,} however, their more complex kinematics and reddening distributions would likely require somewhat different analysis techniques. \editfinal{For h Per, on the other hand, its small parallax ($\approx$ 0.399 mas; \citealt{cantatgaudin18a}) and the current precision of the faint star {\gaia} data precludes a thorough examination}.  Therefore, we defer the consideration of these systems to a subsequent paper.

\item IC 2391/2602 and $\alpha$ Per: with ages of $\sim$ 50 and $\sim$ 80 Myr respectively, these clusters offer snapshots of the evolution of stellar rotation at young ages that could be compared with slightly younger and older systems (e.g., {\clusterone} and the {\clustertwo}). Unfortunately, the rotation samples for these systems are still small ($<$ 40 stars each) and heterogeneous \citep{denissenkov10,gallet15}.

\item M34: rotation periods for stars in this $\sim$ 220 Myr-old cluster have been reported by \citet{irwin06} (as one of the Monitor Project targets), \citet{james10}, and \citet{meibom11b}. Although we initially attempted to include M34 in our sample, the {\gaia} data of this system showed odd gaps in spatial and magnitude space, possibly related to problems with the scanning pattern of this region of the sky. This prevented us from carrying out a similar study for this cluster at this time.

\item Hyades: rotation periods for stars in this $\sim$ 625 Myr-old cluster have been reported by a number of authors (e.g., \citealt{delorme11}, \citealt{douglas16}, \citealt{douglas19}), and the Hyades is sufficiently close ($\approx$ 47 pc; \citealt{gaia18b}) and well studied that we do not anticipate substantial membership issues.

\item NGC 6819: rotation periods for 30 stars in this cluster were reported by \citet{meibom15}. Given its age of $\sim$ 2 Gyr \citep{bossini19}, studying this cluster is of high scientific interest. However, given its small parallax ($\approx$ 0.356 mas; \citealt{cantatgaudin18a}), and the current precision of the {\gaia} data, this cluster would naturally be harder to fit with our current method.

\end{itemize}
\section{Astrometric Cluster Parameters and Comparison with the Literature}
\label{appendix:app_comparison_parameters}
\restartappendixnumbering

\begin{figure}[h]
\epsscale{1.05}  
\plotone{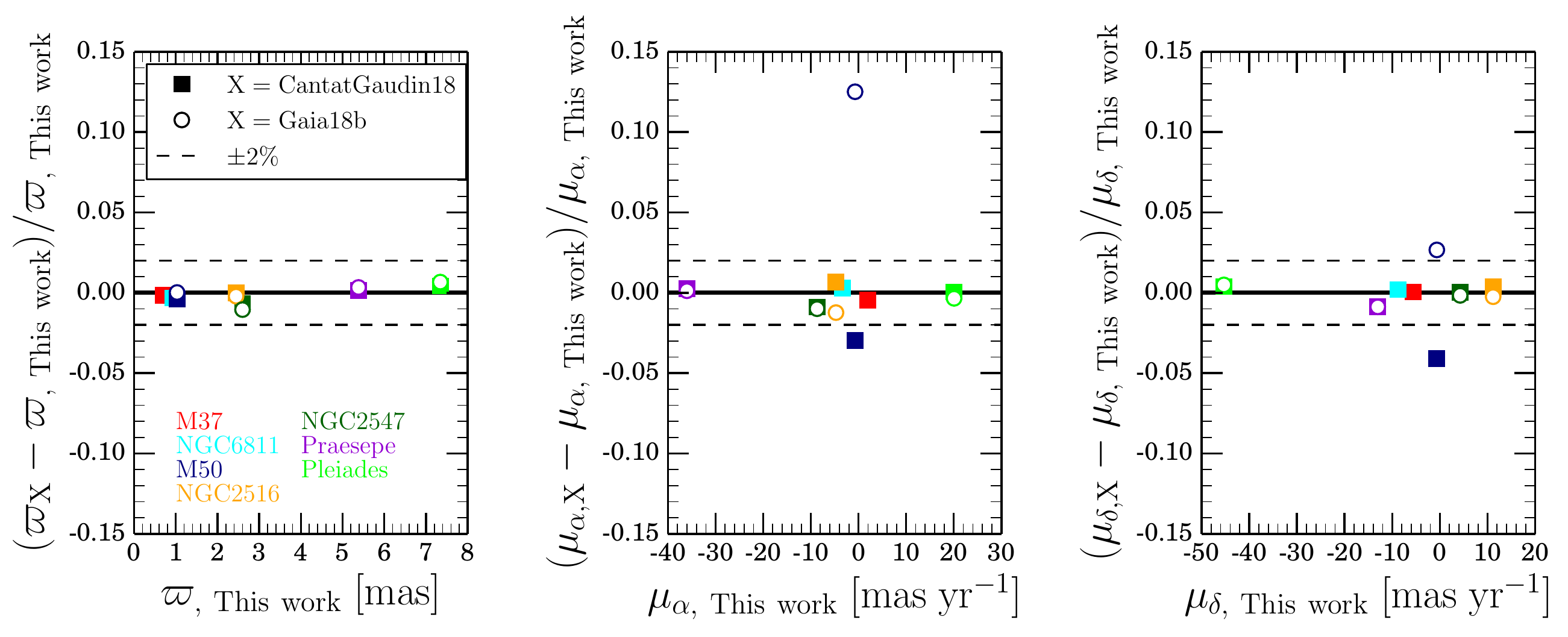}
\caption{Comparison of our cluster parameters with literature reference values from \citet{cantatgaudin18a} and \citet{gaia18b}. Left Panel: Parallax difference divided by our parallax value, as a function of our parallax value. The comparison with \citet{cantatgaudin18a} is shown as the filled squares, while the comparison with \citet{gaia18b} is shown as the open circles. Each of the seven clusters we study is plotted with a different color, and this is indicated by the appropriate text near the bottom. The black dashed line shows a $\pm$ 2\% fractional difference for reference. To make an accurate parallax comparison, we are adding the same zero-point value to the parallax of all three sources. Middle and Right Panels: same as the left panel for $\mu_{\alpha}$ and $\mu_{\delta}$. Note that \citet{gaia18b} do not report cluster parameters for {\clusterfive} and {\clusterseven}.}
\label{fig:Figure_comparison_astrometric_parameters_literature}
\end{figure}

\begin{table}
\scriptsize
\centering
\caption{Astrometric cluster parameters. \label{tab:cluster_astrometric_parameters}}
\begin{tabular}{lcccccc}
\hline
\hline
Cluster Name & $\varpi$ & $\sigma_{\varpi}$ & $\mu_{\alpha}$ & $\sigma_{\mu_{\alpha}}$ & $\mu_{\delta}$ & $\sigma_{\mu_{\delta}}$ \\
- & [mas] & [mas] & [mas yr$^{-1}$] & [mas yr$^{-1}$] & [mas yr$^{-1}$] &  [mas yr$^{-1}$] \\
\hline
{\clusterone}   & 2.600 & 0.072 & -8.686   & 0.550 & 4.261     & 0.457 \\
{\clustertwo}   & 7.344 & 0.269 & 20.066    & 1.505 & -45.325 & 2.312 \\
{\clusterthree} & 1.030 & 0.023 & -0.709   & 0.257 & -0.637   & 0.170 \\
{\clusterfour}  & 2.446 & 0.065 & -4.716   & 0.592 & 11.179   & 0.484 \\
{\clusterfive}   & 0.696 & 0.031 & 1.933     & 0.184 & -5.645   & 0.171 \\
{\clustersix}    & 5.382 & 0.120 & -36.001 & 1.171 & -13.031 & 0.967 \\
{\clusterseven}& 0.926 & 0.018 & -3.389   & 0.120 & -8.794   & 0.122 \\
\hline
\\
\end{tabular}
\tablecomments{List of cluster parameters (parallax, proper motion in $\alpha$ and $\delta$, and their respective dispersions) for the seven clusters we study. These are the values we use in Equation (\ref{eq:membership_probability}) to calculate membership probabilities. Note that the reported parallax values include the global 29 $\mu$as zero-point offset described in \S\ref{subsubsec:method_gaia_data}, with the exception of {\clusterseven}, for which we use the {\kepler} field 53 $\mu$as zero-point value from \citet{zinn19}.}
\end{table}

Using the method described in \S\ref{subsec:method_gaia}, we calculate the astrometric parameters for the seven clusters we study. These values are reported in Table \ref{tab:cluster_astrometric_parameters}. We compare our cluster parameters with those reported by \citet{cantatgaudin18a} and \citet{gaia18b} in Figure \ref{fig:Figure_comparison_astrometric_parameters_literature}. 

We find an excellent agreement with the literature, with most of the fractional differences being less than $\pm$ 2\%. The only exceptions to this are the proper motion parameters of {\clusterthree} (dark blue points in Figure \ref{fig:Figure_comparison_astrometric_parameters_literature}). This is perhaps not surprising, as the {\clusterthree} proper motion in both coordinates is quite small (and similar to that of the field). We would therefore naturally expect this system to show the largest differences. We nonetheless note that \citet{cantatgaudin18a} and \citet{gaia18b} also get the largest disagreements between their values for {\clusterthree}, and our parameters are in between their values.
\section{Data Table for the {\it Probable} and {\it Possible} Cluster Members}
\label{appendix:app_tables_of_membership_probabilities}
\restartappendixnumbering

Table \ref{tab:table_membership_information_probandposs_members} lists the {\it probable} and {\it possible} cluster members for all the clusters we study. We report the main {\gaia} DR2 astrometric and photometric information, as well as the membership parameters and classifications we derive.

We caution that, while Table \ref{tab:table_membership_information_probandposs_members} reports the same set of values for all stars regardless their membership classification, the $M_{G_{0,\text{proj}}}$, mass, and {\Teff} values are calculated assuming as if all the stars are true main-sequence cluster members (and are all at the same respective global cluster parallax). For stars that turn out not to be real members, this will naturally yield an incorrect mass and {\Teff} estimate. Therefore, we suggest that interested readers should only consider the mass and {\Teff} values for the {\it probable} members as being meaningful in future studies, until higher quality astrometry fully resolves the membership classification of the {\it possible} members here reported.

\begin{table}
\scriptsize
\caption{Main information and properties for the {\it probable} and {\it possible} cluster members. \label{tab:table_membership_information_probandposs_members}}
\begin{tabular}{lll}
\hline
\hline
Column & Source & Description\\
\hline 
Cluster & This paper & Cluster with respect to which this star's membership was calculated\\
{\gaia} DR2 & {\gaia} DR2 Catalog & {\gaia} DR2 Source ID\\
$\alpha$ & {\gaia} DR2 Catalog & RA coordinate\\
$\delta$ & {\gaia} DR2 Catalog & Dec coordinate\\
$\varpi$ & {\gaia} DR2 Catalog & Parallax (not including zero-point value)\\
$\sigma_{\varpi}$ & {\gaia} DR2 Catalog & Parallax error\\
$\mu_{\alpha}$ & {\gaia} DR2 Catalog & Proper motion in RA  \\
$\sigma_{\mu_{\alpha}}$ & {\gaia} DR2 Catalog & Proper motion in RA error\\
$\mu_{\delta}$ & {\gaia} DR2 Catalog & Proper motion in Dec  \\
$\sigma_{\mu_{\delta}}$ & {\gaia} DR2 Catalog & Proper motion in Dec error\\
$G$ & {\gaia} DR2 Catalog & $G$ band magnitude \\
$G_{\text{BP}}$ & {\gaia} DR2 Catalog & $G_{\text{BP}}$ band magnitude \\
$G_{\text{RP}}$ & {\gaia} DR2 Catalog & $G_{\text{RP}}$ band magnitude \\
$G_{\text{BP}}-G_{\text{RP}}$ & {\gaia} DR2 Catalog & $G_{\text{BP}}-G_{\text{RP}}$ color\\
$\Delta$ & This paper & $\Delta$ quantity (as defined in Equation (\ref{eq:definition_Delta}))\\
Probability & This paper & Membership probability (as defined in Equation (\ref{eq:membership_probability}))\\
Classification & This paper & Membership classification (i.e., {\it probable} or {\it possible} member)\\
 $M_{G_{0,\text{proj}}}$ & This paper & Projected absolute and dereddened $G$ band magnitude\\
Mass & This paper & Mass obtained by interpolating the best-fit cluster model\\
{\Teff} & This paper & Effective temperature obtained by interpolating the best-fit cluster model\\
Flag Mass \& {\Teff} & This paper & Regime in which the $M_{G_{0,\text{proj}}}$, mass, and {\Teff} values were calculated\\
\hline
\\
\end{tabular}
\tablecomments{(The full table is available online in machine-readable format.) We report the main {\gaia} DR2 astrometric and photometric information, as well as the membership parameters we calculate, for the stars we classify as {\it probable} and {\it possible} cluster members. We also report our derived projected absolute and dereddened $G$ band magnitudes ($M_{G_{0,\text{proj}}}$), as well as the respective mass and temperature values obtained by interpolating the models, and the regime in which they were calculated (see \S\ref{subsec:properties_stellar} for details). The parallaxes are reported as they appear in {\gaia} DR2 and do not include the zero-point values we have used elsewhere. For abbreviation purposes we only list the {\it possible} members with membership probabilities greater than 50\%. We are glad to provide extended tables (including the same information for all the {\it possible} members as well as for the {\it non} members and {\it no info} stars) to interested readers upon request.}
\end{table}
\section{Astrometric Analysis of Clusters and Phase-Space Projections}
\label{appendix:app_phase_space_projections_all_clusters}
\restartappendixnumbering

After applying our classification method to all the clusters, we obtain a sample of {\it probable} members, {\it possible} members, {\it non} members, and {\it no info} stars for each of them. Figure \ref{fig:Figures_all_clusters_phase_space} shows the phase-space projections of these populations (except for the {\it no info} stars, which are absent from the plot due to the lack of astrometric information; see \S\ref{subsubsec:method_gaia_membership}). We now provide an analysis of the astrometric classification in each of the clusters:

\begin{figure}[h]
\epsscale{0.55}  
\plotone{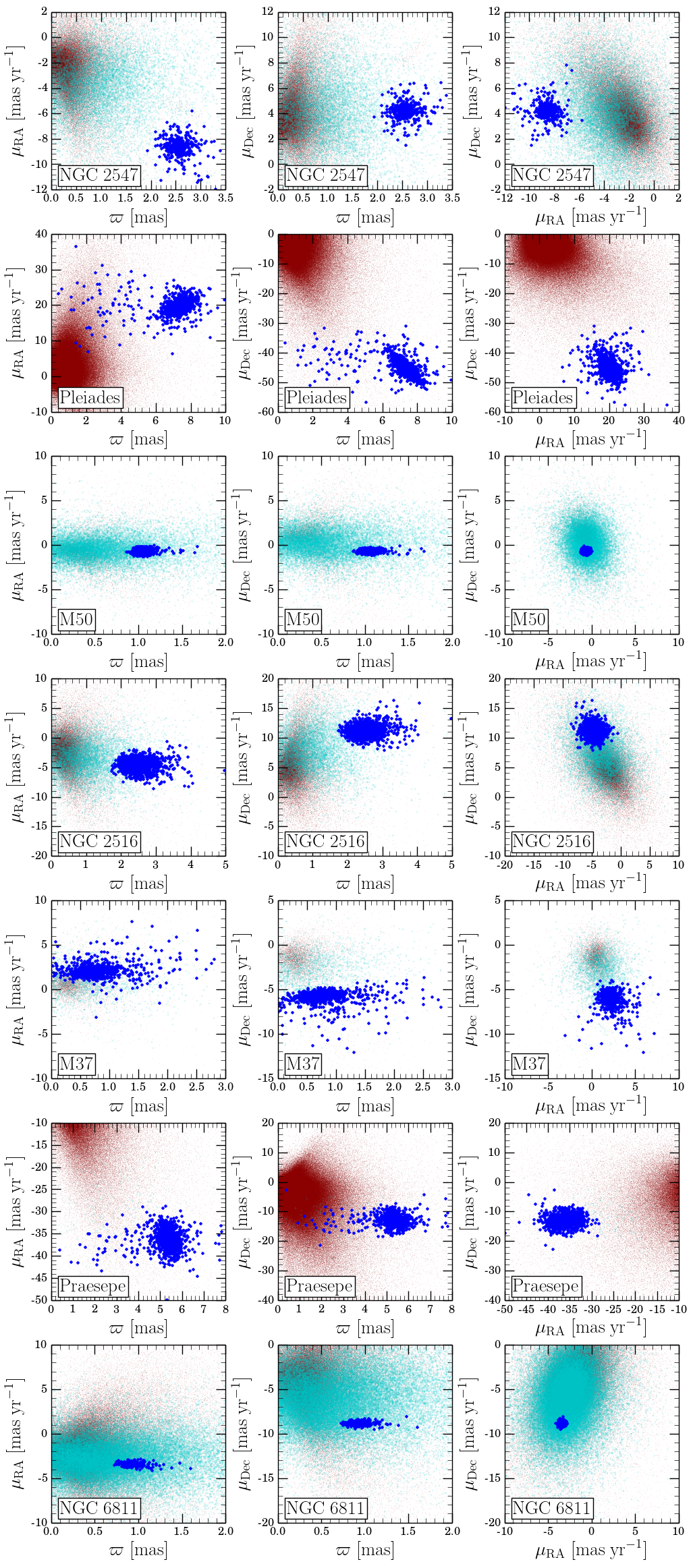}
\caption{Phase-space projections of the {\it probable} (blue), {\it possible} (cyan), and {\it non} (red) members for all the clusters we study (analogous to Figure \ref{fig:Figure_working_example_cluster_phase_space}). Each cluster corresponds to a row, and they are sorted as presented in Table \ref{tab:list_of_clusters}. Note that the range of the parallax and proper motions vary considerably on a cluster by cluster basis.}
\label{fig:Figures_all_clusters_phase_space}
\end{figure}

\begin{itemize}
\item {\clusterone}: for this cluster, the parallax and proper motion in RA are the variables that allow us to distinguish the cluster from the field population in phase-space. On the other hand, the proper motion in Dec of the cluster is similar to that of the field. The {\it possible} members are an intermediate population located in between the cluster and the field in the phase-space projections. 
\item {\clustertwo}: of all the clusters in our sample, this cluster is the one that is the most separated from the field in phase-space. Given its large parallax, as well as proper motion (particularly in Dec), this cluster can be readily differentiated from the background population. Some of the {\it probable} members appear as outliers in the projections including parallax, but they correspond to \editfinal{faint} stars \editfinal{(apparent $G \gtrsim$ 18.5 mag)} with uncertain parallax measurements. The small number of stars classified as {\it possible} members is a direct consequence of the large kinematic differences between the cluster and the field. In other words, it is unlikely that a star will have a value of $\Delta_i \leq 3$ (inside the cluster's 3$\sigma$ ellipsoid in phase-space) and will have a low cluster membership probability.
\item {\clusterthree}: in this case, the proper motion of the cluster is quite similar to that of the field, and therefore the parallax plays a crucial role in identifying the {\it probable} cluster members. Nonetheless, given the small cluster parallax ($\sim$ 1.0 mas) and the therefore larger astrometric uncertainties associated (as the stars in {\clusterthree} are naturally fainter than those of, for instance, the {\clustertwo}), together with the proper motion similarities with the field, the number of {\it possible} members for this cluster is much larger than for others. In other words, given the large astrometric uncertainties, there are a large number of stars with values of $\Delta_i \leq 3$ (inside the cluster's 3$\sigma$ ellipsoid) that have low membership probabilities ($P<P_{\text{min}}$). The CMD of this cluster (see Figure \ref{fig:Figure_all_clusters_apparent_CMD}) demonstrates this, as for magnitudes fainter than apparent $G \simeq 17$ mag, the {\it probable} members are surrounded by an envelope of {\it possible} members. Additionally, the \editfinal{proper motion projection} shows that the {\it possible} members are overlapping with the {\it non} members altogether.
\item {\clusterfour}: for this cluster, the parallax plays an important role when identifying the {\it probable} members. Although there are differences between the cluster and field proper motions (more in the Dec coordinate than in RA), the parallax is the dominant means to separate the populations. As in the case of {\clusterone}, the {\it possible} members appear as an intermediate population in the phase-space projections.
\item {\clusterfive}: this cluster has the smallest parallax of our entire sample, in addition to having a proper motion in RA rather similar to that of the field. Thus, the proper motion in Dec plays an important role in the astrometric classification. {\clusterfive} is also the richest cluster in our sample, and we identify a large number of {\it probable} members that densely populate the CMD. Moreover, a few giant stars can be seen in the CMD, highlighting that our classification method is agnostic to the stars photometry, and they naturally arise from a careful astrometric selection. \editfinal{Similarly as for the {\clustertwo}, the {\it probable} members that appear as phase-space outliers correspond to faint stars.}
\item {\clustersix}: similarly to the {\clustertwo}, this cluster is clearly separated from the field in phase-space. Again, some of the {\it probable} members appear as outliers particularly in proper motion space, but they correspond to \editfinal{faint} stars with large astrometric uncertainties. The small number of {\it possible} members is again a reflection of the large kinematic difference between the cluster and field.
\item {\clusterseven}: this cluster has a similar proper motion in RA to that of the field, and therefore the parallax and proper motion in Dec are the variables that allow us to better distinguish the {\it probable} cluster members from the {\it non} members. However, given its small parallax value (compared to other clusters), and similarly to {\clusterthree}, {\clusterseven} also has a large number of stars classified as {\it possible} members that overlap with the {\it non} members in phase-space (and moreover, for magnitudes fainter than apparent $G \simeq$ 19 mag, the astrometric errors are so large that no star can be reliably classified as a {\it probable} member).
\end{itemize}
\section{Comparison of our Masses and Temperatures with Literature values}
\label{appendix:app_mass_and_Teff_comparison}
\restartappendixnumbering

\begin{figure}[h]
\epsscale{1.05}  
\plotone{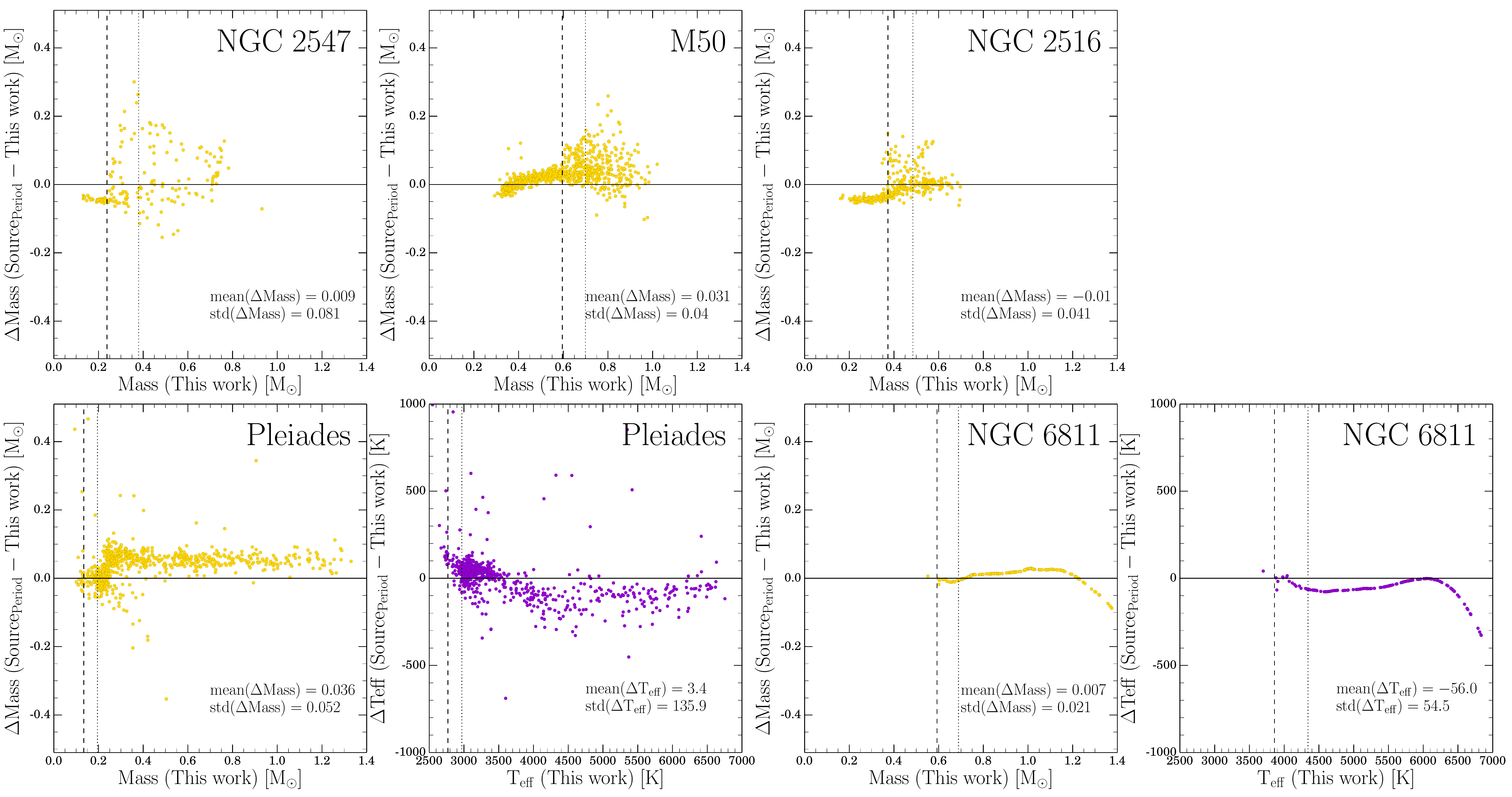}
\caption{Comparison of our mass and {\Teff} estimates with literature values (analogous to Figure \ref{fig:Figure_working_example_cluster_mass_comparison}). The mass comparisons are shown in yellow, while the {\Teff} comparisons are shown in purple. The axis scale is the same for all the mass and {\Teff} comparisons. Each panel indicates the respective cluster name in the top right corner and the mean and standard deviation of the mass or {\Teff} difference in the bottom right corner. The dotted and dashed lines correspond to the apparent $G=17.5$ and 18.5 mag limits from Figure \ref{fig:Figure_all_clusters_apparent_CMD} translated to mass and {\Teff} coordinates. The comparisons suggest that our masses and temperatures are subject to systematic uncertainties of $\sim 0.05 \textendash 0.1 M_{\odot}$ and $\sim 150$ K, which are modest considering the different methods, photometric data, and underlying models employed.}
\label{fig:Figure_all_clusters_mass_teff_comparison}
\end{figure}

Analogous to the comparison made in \S\ref{subsec:properties_stellar} for {\clusterone}, some of the other clusters we study have mass (and sometimes {\Teff}) estimates available in the references that we use as sources for the period information. In particular, \citet{irwin08a}, \citet{irwin09}, and \citet{irwin07b} report masses for the {\clusterone}, {\clusterthree}, and {\clusterfour} stars, and \citet{stauffer16} and \citet{curtis19a} report mass and {\Teff} for the {\clustertwo} and {\clusterseven} stars. We show the comparison of these estimates with ours in Figure \ref{fig:Figure_all_clusters_mass_teff_comparison}. Note that for {\clusterfive} and {\clustersix} the references used as sources for the period information did not report mass or temperature estimates, and these clusters are missing from the comparison.

The mass comparisons with {\clusterone}, {\clusterthree}, and {\clusterfour} are topologically similar to each other: in the {\it faint regime} (masses below the dashed lines), where we do not account for the contribution of photometric binaries, we see mostly constant differences of order $\approx -0.05 M_{\odot}$. For masses higher than this limit, the scatter increases and we observe that many stars have positive mass differences, which is due to the magnitude projections we use in the {\it intermediate} and {\it bright} regimes. 

For the {\clustertwo}, the mass and temperature comparisons with \citet{stauffer16} are in agreement with our expectations: while we employed the projection technique to account for unresolved binaries in our {\it bright} regime and partially in our {\it intermediate} regime, \citet{stauffer16} used this approach for their entire sample. Accordingly, we observe an approximately constant offset trend in $\Delta$Mass ($\approx +0.05 M_{\odot}$) and $\Delta T_{\text{eff}}$ ($\approx -150$ K) for stars more massive/hotter than the dotted line, which then turns to a mass/{\Teff}-dependent trend for stars in the {\it intermediate} and {\it faint} regimes.

For {\clusterseven}, \citet{curtis19a} derived temperatures by generating their own color-{\Teff} relations, and translated these to masses by interpolating tabulated spectral energy distributions. The mass and {\Teff} comparisons show an excellent agreement, with differences being $\lesssim 0.025 M_{\odot}$ and $\lesssim 100$ K for most of the mass/{\Teff} range. Given that both our and their calculations are based only on the {\gaia} DR2 photometry (unlike the previously described comparisons), this result is consistent with the expectations. 

Ultimately, all the mass and temperature comparisons previously described are subject to different assumptions regarding the cluster properties, stellar evolutionary models (and their specific input physics; e.g., \citealt{tayar20}), and considerations for the contribution of unresolved photometric binaries. Furthermore, in most cases, the photometry employed by the references that supply the period information is different from that of {\gaia}. In spite of this, we observe a good overall agreement with the literature estimates, and these comparisons suggest possible systematic uncertainties of $\sim 0.05 \textendash 0.1 M_{\odot}$ and $\sim 150$ K in our masses and temperatures.
\section{Data Table for the Periodic Samples}
\label{appendix:app_tables_rotation_period_sample}
\restartappendixnumbering

Table \ref{tab:table_periodic_samples} reports the main information and properties for the periodic samples of the clusters we study. We list all stars that were reported in the catalogs used as sources for the periods (see \S\ref{sec:sample_selection_and_period_data}), regardless of their membership classification, or if we found them when crossmatching with the {\gaia} DR2 data. We extend the same cautionary note given in Table \ref{tab:table_membership_information_probandposs_members} to Table \ref{tab:table_periodic_samples}, regarding that the $M_{G_{0,\text{proj}}}$, mass, and {\Teff} values are calculated assuming as if all the stars are true main-sequence cluster members (and are all at the same respective global cluster parallax).

\begin{table}
\scriptsize
\caption{Main information and properties for the periodic samples. \label{tab:table_periodic_samples}}
\begin{tabular}{lll}
\hline
\hline
Column & Source & Description\\
\hline 
Cluster & This paper & Cluster with respect to which this star's membership was calculated\\
Source ID & Period Source Catalog & ID in the catalog used as source of the period information\\
Period & Period Source Catalog & Period \\
{\gaia} DR2 & {\gaia} DR2 Catalog & {\gaia} DR2 Source ID\\
$\alpha$ & {\gaia} DR2 Catalog & RA coordinate\\
$\delta$ & {\gaia} DR2 Catalog & Dec coordinate\\
$\varpi$ & {\gaia} DR2 Catalog & Parallax (not including zero-point value)\\
$\sigma_{\varpi}$ & {\gaia} DR2 Catalog & Parallax error\\
$\mu_{\alpha}$ & {\gaia} DR2 Catalog & Proper motion in RA  \\
$\sigma_{\mu_{\alpha}}$ & {\gaia} DR2 Catalog & Proper motion in RA error\\
$\mu_{\delta}$ & {\gaia} DR2 Catalog & Proper motion in Dec  \\
$\sigma_{\mu_{\delta}}$ & {\gaia} DR2 Catalog & Proper motion in Dec error\\
$G$ & {\gaia} DR2 Catalog & $G$ band magnitude \\
$G_{\text{BP}}$ & {\gaia} DR2 Catalog & $G_{\text{BP}}$ band magnitude \\
$G_{\text{RP}}$ & {\gaia} DR2 Catalog & $G_{\text{RP}}$ band magnitude \\
$G_{\text{BP}}-G_{\text{RP}}$ & {\gaia} DR2 Catalog & $G_{\text{BP}}-G_{\text{RP}}$ color\\
$\Delta$ & This paper & $\Delta$ quantity (as defined in Equation (\ref{eq:definition_Delta}))\\
Probability & This paper & Membership probability (as defined in Equation (\ref{eq:membership_probability}))\\
Classification & This paper & Membership classification\\
$M_{G_{0,\text{proj}}}$ & This paper & Projected absolute and dereddened $G$ band magnitude\\
Mass & This paper & Mass obtained by interpolating the best-fit cluster model\\
{\Teff} & This paper & Effective temperature obtained by interpolating the best-fit cluster model\\
Flag Mass \& {\Teff} & This paper & Regime in which the $M_{G_{0,\text{proj}}}$, mass, and {\Teff} values were calculated\\
\hline
\\
\end{tabular}
\tablecomments{(The full table is available online in machine-readable format.) From the catalogs used as source of the period measurements, we report the IDs used in them (which vary on a cluster by cluster basis) and the periods. We note that although many of the {\clustertwo} and {\clustersix} stars had more than one period reported, this table only lists their main value, i.e., $P1$ from \citet{rebull16a,rebull16b} and \citet{rebull17}, respectively. From {\gaia} DR2, we report the main astrometric and photometric information, and the parallaxes are reported as they appear in {\gaia} DR2 and do not include the zero-point values we have used elsewhere. From our analysis we report the membership probability and classification regardless of their category (i.e., {\it probable}, {\it possible} or {\it non} member, or {\it no info} star). We also report our derived projected absolute and dereddened $G$ band magnitudes ($M_{G_{0,\text{proj}}}$), as well as the respective mass and temperature values obtained by interpolating the models, and the regime in which they were calculated (see \S\ref{subsec:properties_stellar} for details).}
\end{table}

\section{Percentiles of the Period Distributions}
\label{appendix:app_table_percentiles}
\restartappendixnumbering

Table \ref{tab:rotation_percentiles} reports the percentiles of the distribution of rotation periods as a function of mass and age. Note that only the stars that survive our membership analysis have been considered (i.e., the stars of the revised rotational sequences, see Figure \ref{fig:period_mass_evolution_aftergaiadr2}). These values can be used to calibrate and test future models of angular momentum evolution. 

\begin{longrotatetable}
\begin{deluxetable*}{lc|c|c|ccc|ccc|ccc|ccc|ccc}
\tablecaption{Percentiles of the distribution of rotation periods as a function of mass and age.\label{tab:rotation_percentiles}}
\tablewidth{700pt}
\tabletypesize{\scriptsize}
\startdata
Cluster & Age & Mass & N$_{\star}$ & $\text{P}_{10}$ & MAD$_{10}$ &$\sigma_{10}$ & $\text{P}_{25}$ & MAD$_{25}$ & $\sigma_{25}$ & $\text{P}_{50}$ & MAD$_{50}$ & $\sigma_{50}$ & $\text{P}_{75}$ & MAD$_{75}$ & $\sigma_{75}$ & $\text{P}_{90}$ & MAD$_{90}$ & $\sigma_{90}$ \\ 
- & [Myr] & [$M_{\odot}$] & - & [days] & [days] & [days] & [days] & [days] & [days] & [days] & [days] & [days] & [days] & [days] & [days] & [days] & [days] & [days] \\ \hline
        NGC 2547 & 35 & 0.2 & 34 & 0.318 & 0.091 & 0.035 & 0.471 & 0.101 & 0.024 & 0.628 & 0.212 & 0.036 & 0.886 & 0.13 & 0.032 & 1.161 & 0.218 & 0.084 \\ 
        NGC 2547 & 35 & 0.3 & 29 & 0.393 & 0.04 & 0.017 & 0.586 & 0.193 & 0.051 & 1.063 & 0.344 & 0.064 & 1.369 & 0.161 & 0.042 & 2.408 & 0.81 & 0.336 \\ 
        NGC 2547 & 35 & 0.4 & 24 & 0.39 & 0.152 & 0.069 & 0.78 & 0.317 & 0.092 & 1.447 & 0.747 & 0.152 & 5.024 & 2.873 & 0.829 & 8.045 & 0.747 & 0.341 \\ 
        NGC 2547 & 35 & 0.55 & 24 & 0.807 & 0.409 & 0.187 & 1.591 & 0.433 & 0.125 & 3.146 & 1.61 & 0.329 & 5.111 & 1.463 & 0.422 & 7.287 & 1.43 & 0.652 \\ 
        NGC 2547 & 35 & 0.8 & 21 & 0.287 & 0.06 & 0.029 & 1.219 & 0.922 & 0.285 & 4.41 & 1.197 & 0.261 & 5.221 & 0.268 & 0.083 & 5.56 & 0.192 & 0.094 \\ \hline
        Pleiades & 125 & 0.2 & 268 & 0.238 & 0.041 & 0.006 & 0.308 & 0.051 & 0.004 & 0.445 & 0.148 & 0.009 & 0.653 & 0.134 & 0.012 & 0.909 & 0.161 & 0.022 \\ 
        Pleiades & 125 & 0.3 & 132 & 0.457 & 0.068 & 0.013 & 0.611 & 0.139 & 0.017 & 0.837 & 0.283 & 0.025 & 1.241 & 0.322 & 0.04 & 2.011 & 0.262 & 0.051 \\ 
        Pleiades & 125 & 0.4 & 70 & 0.288 & 0.032 & 0.009 & 0.476 & 0.207 & 0.035 & 1.134 & 0.709 & 0.085 & 1.996 & 0.462 & 0.078 & 4.041 & 1.584 & 0.423 \\ 
        Pleiades & 125 & 0.5 & 40 & 0.327 & 0.022 & 0.008 & 0.597 & 0.277 & 0.062 & 2.153 & 1.56 & 0.247 & 3.734 & 1.333 & 0.298 & 7.434 & 0.674 & 0.238 \\ 
        Pleiades & 125 & 0.6 & 54 & 0.399 & 0.094 & 0.029 & 1.661 & 1.304 & 0.251 & 3.887 & 2.48 & 0.337 & 6.641 & 1.83 & 0.352 & 9.292 & 0.667 & 0.203 \\
        Pleiades & 125 & 0.7 & 35 & 0.394 & 0.048 & 0.018 & 0.56 & 0.199 & 0.047 & 5.002 & 3.235 & 0.547 & 7.479 & 0.889 & 0.212 & 8.44 & 0.158 & 0.06 \\ 
        Pleiades & 125 & 0.8 & 23 & 0.54 & 0.226 & 0.105 & 6.13 & 0.627 & 0.185 & 6.813 & 0.742 & 0.155 & 7.65 & 0.336 & 0.099 & 8.249 & 0.302 & 0.141 \\ 
        Pleiades & 125 & 0.9 & 28 & 0.831 & 0.142 & 0.06 & 4.031 & 1.374 & 0.367 & 5.444 & 0.548 & 0.103 & 5.933 & 0.22 & 0.059 & 6.154 & 0.005 & 0.002 \\ 
        Pleiades & 125 & 1 & 20 & 1.476 & 0.765 & 0.382 & 3.297 & 0.211 & 0.067 & 3.991 & 0.658 & 0.147 & 4.549 & 0.465 & 0.147 & 5.137 & 0.575 & 0.287 \\ 
        Pleiades & 125 & 1.1 & 18 & 1.116 & 0.279 & 0.147 & 2.138 & 0.259 & 0.086 & 2.453 & 0.455 & 0.107 & 3.081 & 0.552 & 0.184 & 3.907 & 0.334 & 0.176 \\
        Pleiades & 125 & 1.2 & 14 & 0.856 & 0.026 & 0.016 & 1.172 & 0.334 & 0.126 & 1.908 & 0.933 & 0.249 & 3.095 & 0.952 & 0.36 & 6.792 & 0.07 & 0.042 \\ \hline
        NGC 2516 & 150 & 0.2 & 27 & 0.291 & 0.038 & 0.017 & 0.37 & 0.027 & 0.007 & 0.413 & 0.068 & 0.013 & 0.561 & 0.101 & 0.027 & 2.136 & 1.487 & 0.64 \\ 
        NGC 2516 & 150 & 0.3 & 86 & 0.34 & 0.029 & 0.007 & 0.46 & 0.09 & 0.014 & 0.599 & 0.185 & 0.02 & 0.829 & 0.181 & 0.028 & 1.194 & 0.169 & 0.041 \\ 
        NGC 2516 & 150 & 0.4 & 110 & 0.364 & 0.056 & 0.012 & 0.605 & 0.236 & 0.032 & 1.179 & 0.541 & 0.052 & 1.685 & 0.414 & 0.056 & 2.987 & 0.709 & 0.151 \\ 
        NGC 2516 & 150 & 0.5 & 72 & 0.334 & 0.052 & 0.014 & 0.558 & 0.257 & 0.043 & 2.235 & 1.759 & 0.207 & 4.879 & 1.853 & 0.309 & 6.694 & 0.974 & 0.257 \\ 
        NGC 2516 & 150 & 0.6 & 49 & 0.446 & 0.039 & 0.012 & 1.282 & 0.95 & 0.192 & 4.971 & 3.227 & 0.461 & 7.756 & 1.268 & 0.256 & 9.606 & 0.908 & 0.29 \\ \hline
        M50 & 150 & 0.6 & 54 & 0.272 & 0.024 & 0.007 & 0.53 & 0.3 & 0.058 & 3.452 & 2.828 & 0.385 & 6.417 & 1.754 & 0.338 & 10.327 & 2.364 & 0.719 \\ 
        M50 & 150 & 0.7 & 106 & 0.456 & 0.111 & 0.024 & 1.116 & 0.844 & 0.116 & 5.627 & 3.191 & 0.31 & 7.753 & 1.104 & 0.152 & 9.522 & 0.666 & 0.145 \\ 
        M50 & 150 & 0.8 & 57 & 0.723 & 0.161 & 0.048 & 3.48 & 1.941 & 0.363 & 6.188 & 1.594 & 0.211 & 7.47 & 0.481 & 0.09 & 8.483 & 0.651 & 0.193 \\ 
        M50 & 150 & 0.9 & 46 & 1.29 & 0.854 & 0.282 & 3.633 & 1.137 & 0.237 & 4.992 & 1.541 & 0.227 & 6.723 & 0.923 & 0.192 & 7.868 & 0.626 & 0.206 \\ \hline
        M37 & 500 & 0.5 & 67 & 0.579 & 0.063 & 0.017 & 0.779 & 0.316 & 0.055 & 2.273 & 1.782 & 0.218 & 13.758 & 3.185 & 0.55 & 16.609 & 0.857 & 0.234 \\ 
        M37 & 500 & 0.6 & 78 & 1.292 & 0.352 & 0.089 & 2.208 & 1.073 & 0.172 & 8.665 & 5.491 & 0.622 & 13.801 & 1.147 & 0.184 & 15.055 & 0.473 & 0.12 \\ 
        M37 & 500 & 0.7 & 63 & 5.134 & 0.775 & 0.218 & 7.418 & 2.248 & 0.4 & 10.435 & 1.211 & 0.153 & 11.474 & 0.601 & 0.107 & 12.388 & 0.23 & 0.065 \\
        M37 & 500 & 0.8 & 58 & 7.512 & 0.401 & 0.118 & 8.242 & 0.429 & 0.08 & 9.06 & 0.821 & 0.108 & 9.902 & 0.377 & 0.07 & 10.303 & 0.189 & 0.056 \\ 
        M37 & 500 & 0.9 & 39 & 6.829 & 0.168 & 0.06 & 7.162 & 0.178 & 0.04 & 7.507 & 0.51 & 0.082 & 8.249 & 0.486 & 0.11 & 9.731 & 1.246 & 0.446 \\ 
        M37 & 500 & 1 & 18 & 5.372 & 0.356 & 0.188 & 6.37 & 0.354 & 0.118 & 6.817 & 0.272 & 0.064 & 7.01 & 0.175 & 0.058 & 8.514 & 1.166 & 0.615 \\ 
        M37 & 500 & 1.1 & 18 & 3.225 & 0.049 & 0.026 & 3.531 & 0.383 & 0.128 & 4.632 & 1.182 & 0.279 & 6.22 & 0.609 & 0.203 & 6.776 & 0.46 & 0.243 \\ \hline
        Praesepe & 700 & 0.2 & 132 & 0.254 & 0.026 & 0.005 & 0.331 & 0.09 & 0.011 & 0.635 & 0.331 & 0.029 & 1.138 & 0.364 & 0.045 & 1.806 & 0.194 & 0.038 \\ 
        Praesepe & 700 & 0.3 & 186 & 0.858 & 0.251 & 0.041 & 1.268 & 0.25 & 0.026 & 1.878 & 0.779 & 0.057 & 5.09 & 2.958 & 0.307 & 16.447 & 3.984 & 0.653 \\ 
        Praesepe & 700 & 0.4 & 86 & 0.902 & 0.122 & 0.029 & 3.25 & 2.352 & 0.359 & 12.682 & 8.758 & 0.944 & 20.605 & 4.221 & 0.644 & 25.071 & 1.267 & 0.306 \\
        Praesepe & 700 & 0.5 & 53 & 2.076 & 1.037 & 0.319 & 16.739 & 2.013 & 0.391 & 18.873 & 2.064 & 0.283 & 20.857 & 1.141 & 0.222 & 23.111 & 1.049 & 0.322 \\ 
        Praesepe & 700 & 0.6 & 63 & 7.529 & 3.363 & 0.948 & 14.365 & 0.846 & 0.151 & 15.454 & 1.133 & 0.143 & 16.732 & 0.682 & 0.121 & 18.07 & 0.645 & 0.182 \\ 
        Praesepe & 700 & 0.7 & 55 & 11.743 & 0.174 & 0.053 & 12.199 & 0.384 & 0.073 & 12.969 & 0.667 & 0.09 & 13.607 & 0.371 & 0.071 & 14.511 & 0.27 & 0.081 \\ 
        Praesepe & 700 & 0.8 & 29 & 9.368 & 0.807 & 0.335 & 10.421 & 0.249 & 0.066 & 10.886 & 0.552 & 0.103 & 11.556 & 0.26 & 0.068 & 12.073 & 0.257 & 0.107 \\ 
        Praesepe & 700 & 0.9 & 34 & 8.059 & 0.259 & 0.099 & 9.147 & 0.239 & 0.058 & 9.585 & 0.424 & 0.073 & 10.049 & 0.393 & 0.095 & 11.816 & 0.903 & 0.346 \\ 
        Praesepe & 700 & 1 & 30 & 4.489 & 1.513 & 0.618 & 8.048 & 0.557 & 0.144 & 8.899 & 0.472 & 0.086 & 9.088 & 0.107 & 0.028 & 10.268 & 0.642 & 0.262 \\ 
        Praesepe & 700 & 1.1 & 22 & 4.399 & 0.336 & 0.16 & 5.928 & 0.826 & 0.249 & 6.845 & 0.923 & 0.197 & 7.897 & 0.334 & 0.101 & 8.548 & 0.561 & 0.267 \\ 
        Praesepe & 700 & 1.2 & 14 & --- & --- & --- & 3.122 & 0.497 & 0.188 & 3.745 & 0.969 & 0.259 & 5.077 & 0.443 & 0.168 & --- & --- & --- \\ 
        Praesepe & 700 & 1.3 & 12 & --- & --- & --- & 1.008 & 0.282 & 0.115 & 2.686 & 1.178 & 0.34 & 3.657 & 0.377 & 0.154 & --- & --- & --- \\
        Praesepe & 700 & 1.4 & 8 & --- & --- & --- & 0.71 & 0.214 & 0.107 & 1.092 & 0.452 & 0.16 & 1.919 & 0.565 & 0.283 & --- & --- & --- \\ \hline
        NGC 6811 & 950 & 0.6 & 12 & --- & --- & --- & 13.983 & 0.403 & 0.164 & 14.43 & 0.375 & 0.108 & 14.97 & 0.54 & 0.22 & --- & --- & --- \\ 
        NGC 6811 & 950 & 0.7 & 51 & 11.48 & 0.233 & 0.073 & 11.917 & 0.175 & 0.035 & 12.597 & 0.679 & 0.095 & 13.203 & 0.422 & 0.084 & 15.841 & 2.391 & 0.749 \\ 
        NGC 6811 & 950 & 0.8 & 36 & 7.82 & 2.626 & 0.978 & 10.591 & 0.256 & 0.06 & 10.966 & 0.447 & 0.074 & 11.672 & 0.705 & 0.166 & 12.859 & 0.349 & 0.13 \\ 
        NGC 6811 & 950 & 0.9 & 35 & 9.747 & 0.136 & 0.051 & 10.47 & 0.345 & 0.082 & 10.89 & 0.371 & 0.063 & 11.26 & 0.369 & 0.088 & 12.376 & 0.623 & 0.235 \\ 
        NGC 6811 & 950 & 1 & 25 & 9.555 & 0.009 & 0.004 & 9.918 & 0.305 & 0.086 & 10.303 & 0.588 & 0.118 & 10.929 & 0.482 & 0.136 & 13.007 & 0.253 & 0.113 \\ 
        NGC 6811 & 950 & 1.1 & 26 & 4.785 & 0.518 & 0.227 & 6.204 & 0.689 & 0.191 & 7.066 & 1.834 & 0.36 & 9.35 & 0.602 & 0.167 & 10.842 & 0.889 & 0.39 \\ 
        NGC 6811 & 950 & 1.2 & 26 & 1.472 & 0.122 & 0.054 & 2.022 & 0.278 & 0.077 & 2.828 & 0.824 & 0.162 & 3.726 & 0.763 & 0.212 & 7.434 & 2.693 & 1.181 \\
        NGC 6811 & 950 & 1.35 & 14 & --- & --- & --- & 0.987 & 0.067 & 0.025 & 1.601 & 0.547 & 0.146 & 2.145 & 0.199 & 0.075 & --- & --- & --- \\ 
\enddata
\tablecomments{(The table is also available online in machine-readable format.) The clusters are sorted by their revised age values, and only the stars that survive our membership analysis are considered. The data are binned in $0.1 M_{\odot}$ bins (unless otherwise indicated), then rank ordered from short to long periods. For each cluster, we report the 10th, 25th, 50th, 75th, and 90th percentiles in the period distribution, along with their median absolute deviations (MAD). Standard errors of the medians are converted from MADs using the assumption of a normal distribution and root N statistics. The 10th and 90th percentiles are not presented for bins with less than 20 entries. The 0.55 and 0.8 $M_{\odot}$ bins for {\clusterone} and the 1.35 $M_{\odot}$ bin for {\clusterseven} are binned in broader bins ($0.45\textendash0.70$, $0.7\textendash0.9$, and $1.25\textendash1.45 M_{\odot}$, respectively) because of small sample sizes.}
\end{deluxetable*}
\end{longrotatetable}
\bibliography{bibliography.bib}{}
\bibliographystyle{aasjournal}



\end{document}